\documentclass[%
superscriptaddress,
 amsmath,amssymb,
 aps,
 twocolumn,
prl,
]{revtex4-2}

\usepackage{xcolor}
\usepackage{graphicx}
\usepackage{comment}
\usepackage{mathtools}
\usepackage{dcolumn}
\usepackage{bm}
\usepackage{graphicx} 
\usepackage{amsmath}
\usepackage{bm}
\usepackage[section]{placeins}
\usepackage[normalem]{ulem} 
\usepackage[colorlinks=true,
            linkcolor=red,
            urlcolor=blue,
            citecolor=blue]{hyperref}

\newcommand{\NTT}{NTT Basic Research Laboratories, NTT Corporation, 3-1 Morinosato Wakamiya, Atsugi, Kanagawa, 243-0198, Japan}

\newcommand{\UEC}{Department of Engineering Science, University of Electro-Communications, Tokyo 182-8585, Japan}

\newcommand{\TQP}{NTT Research Center for Theoretical Quantum Information, NTT Corporation, 3-1 Morinosato Wakamiya, Atsugi, Kanagawa, 243-0198, Japan}

\newcommand{\OIST}{Quantum Systems Unit, Okinawa Institute of Science and Technology Graduate University, Onna-son, Okinawa 904-0495, Japan}

\begin{document}

\preprint{APS/123-QED}

\title{Strong Dipole-Dipole Interactions via Enhanced Light-Matter Coupling in Composite Nanofiber Waveguides}

\author{Kritika Jain}
\email{kritika.jain@oist.jp}

\affiliation{\OIST}

\author{Lewis Ruks}
\email{lewis.ruks@ntt.com} 
\affiliation{\NTT}
\affiliation{\TQP}

\author{Fam le Kien}
\affiliation{\UEC}

\author{Thomas Busch}
\affiliation{\OIST}

\date{\today}

\begin{abstract}
\noindent We study the interaction of emitters with a composite waveguide formed from two parallel optical nanofibers in currently unexplored regimes of experimental importance for atomic gases or solid-state emitters. Using the exact dyadic Green's function we comprehensively investigate the coupling efficiency and the fiber-induced Lamb shift accounting for variations in emitter positions and fiber configurations. This reveals coupling efficiencies and Purcell factors that are enhanced considerably beyond those using a single fiber waveguide, and robustness in the figures of merit. We finally investigate resonant dipole-dipole interactions and the generation of entanglement between two emitters mediated through the composite waveguide under excitation. We show that the concurrence can be enhanced for two fiber systems, such that entanglement may be present even in cases where it is zero for a single fiber. All-fiber systems are simple in construction and benefit from a wealth of existing telecommunications technologies, whilst enjoying strong couplings to emitters and offering novel light-matter functionalities specific to slot waveguides.

\end{abstract}

\maketitle

\paragraph{Introduction.}
Emitters coupled to the evanescent field of nanophotonic optical waveguides realise a powerful light-matter platform with potential for applications in sensing~\cite{sensing_1,okaba2014lamb,sensing}, photonic state control~\cite{tudela_abitrary,mahmoodian_nonclassical,chang_transistor,corzo_atoms}, and quantum information processing~\cite{waveguide_quantum,Paulisch_2016}. The performance of such hybrid devices is dictated by the emission rate and coupling efficiency into the waveguides. Recognizing the modest coupling to low-loss dielectric waveguides, current implementations featuring atomic gases~\cite{ritter2018coupling,lightcage2021,okaba2014lamb} and solid-state quantum emitters~\cite{high-purcell,q_emitter_slot} seek to improve the efficiency chiefly by confining the field further into a fabricated void containing the emitters~\cite{wenfang_void}, or extending the light-matter interaction time using slow light inherent to photonic crystals~\cite{goban2015superradiance,near-unity}. In addition, precision operation requires a detailed consideration of the accompanying Lamb shift and cooperative interactions mediated through the structure~\cite{sheremet2023waveguide}, which in turn may entail further variations of the waveguide. Such complex waveguide designs are challenging to fabricate, whilst facing hurdles in integration with existing telecommunications technologies. They are further difficult and costly to accurately simulate in conjunction with emitters~\cite{ana_exponential}, which has restricted theoretical investigation.

\indent On the other hand, the optical nanofiber (ONF) enjoys a simple structure enabling precise theoretical modelling and experimental fabrication~\cite{nieddu2016optical,Nayak_onf_review,solano2017optical,zhang2024optical}, with natural integration into well-established fiber-based technologies. ONF platforms integrating neutral atoms~\cite{vetsch2012nanofiber,solano_superradiance,corzo_atoms} or solid state emitters~\cite{yalla2012efficient,diamond_nanofiber} are routinely demonstrated experimentally, and a well-established toolbox to control such systems exits~\cite{morrissey_2013}. To overcome the relatively small coupling efficiency of single fibers whilst retaining simplicity of fabrication, pairs of parallel ONFs -- or couplers~\cite{Ding_2019,yu2022highly,Li:18} -- have been proposed to enable enhanced coupling~\cite{shao2022twin} to surface-bound quantum dots. However, a study of the figures of merit beyond the coupling efficiency in ONFs more generally, and in regimes relevant to near-term implementations using atomic gases, is so far lacking.

\indent  In this work, we provide a comprehensive study of emitter coupling to and through waveguides formed simply by two parallel ONFs, and propose their use as part of an inexpensive light-matter platform enabling appreciable figures of merit, most notably in the coupling efficiency. We exploit the full Green's function providing an exact description of dipolar emission, which may be efficiently calculated for an arbitrary configuration of parallel cylinders~\cite{fussell2004three} compared to  computationally prohibitive FDTD methods frequently suffering from significant numerical error~\cite{almokhtar2014numerical,solano2019alignment}. We show first how the realistic slot geometry formed from only two ONFs can robustly enhance coupling efficiency beyond the single ONF by more than 60$\%$ for nearby atoms, and may enhance the Purcell factor by nearly an order of magnitude for surface-bound solid-state emitters. The two fiber scheme is optimal -- in the sense that adding more fibers \textit{reduces} the maximal coupling efficiency -- and does not rely on intricate refractive index modulation of photonic crystals~\cite{near-unity} or lossy plasmonic resonances~\cite{chang_transistor}. With the full Green's function, we are able to further obtain exact Lamb shifts, induced by the composite waveguide, whose precise knowledge is vital in precision metrology~\cite{okaba2014lamb}, near-surface atom trapping schemes~\cite{chang2014}, and nanophotonic Rydberg atom platforms~\cite{rajasree2020,stourm2020spontaneous}. Finally, we use the Green's function to investigate collective emission and the accompanying establishing of entanglement amongst two emitters interfaced with the ONFs. Whilst a difference in coupling efficiency is typically regarded as a quantitative change, we show that this change can result in the complete destruction of steady-state concurrence -- and thus distillable entanglement -- between two atoms subject to a waveguide driving. 
\paragraph{Model}
\begin{figure*}[t!]
   	\centering	
    \includegraphics[width=0.99\linewidth]{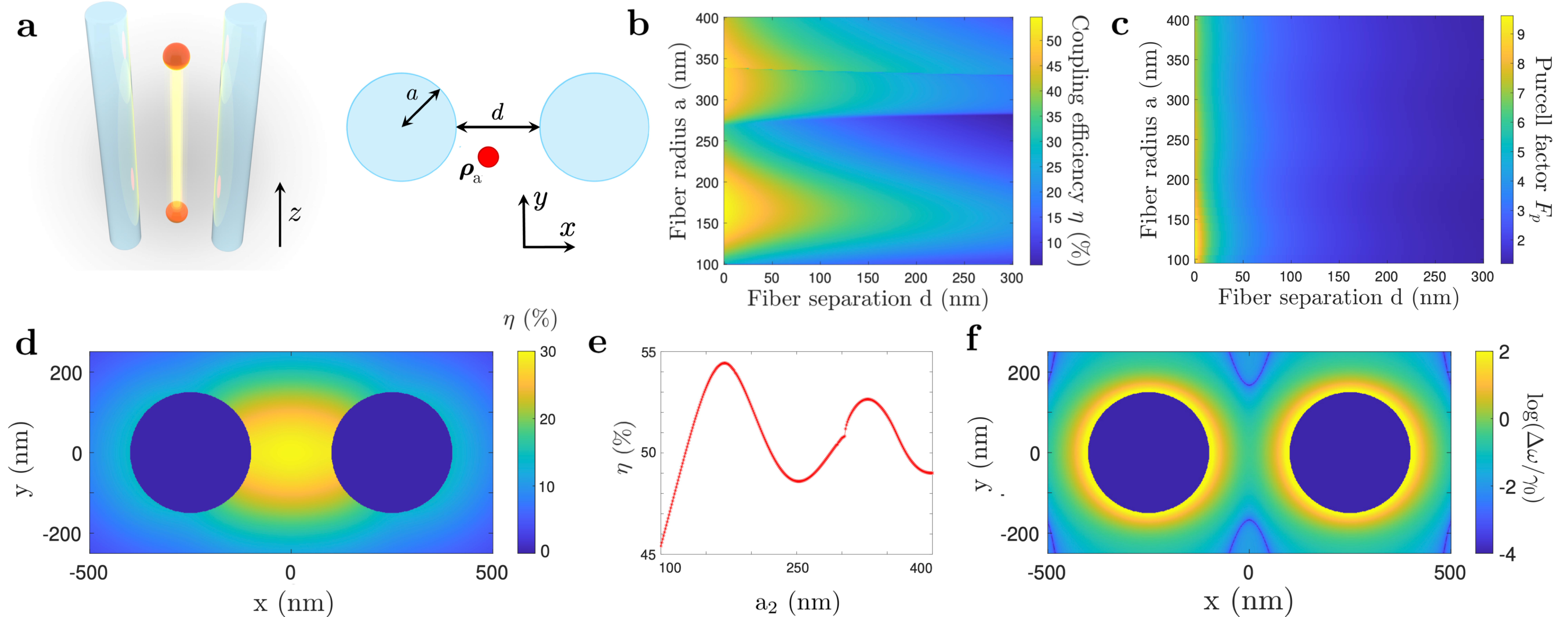}
    \caption{{(a) (left) Schematic of two emitters interfaced through the guided mode of a pair of identical parallel ONFs, with radii $a$ and separation $d$, and (right) configuration in the transverse plane, for transverse emitter  position $\bm{\rho}_{\text{a}}$. (b) Coupling efficiency $\eta$, and (c) Purcell factor $F_{p}$ for each emitter at the center of the two-fiber system, as $a$ and $d$ are varied. (d) Coupling efficiency for varying emitter position in the transverse plane for fiber parameters $(a, d) = (150$ nm, $200$ nm). (e) Variation of coupling efficiency with the radius $a_2$ of the second fiber, when the radius of first fiber $a_1 =175$ nm for touching fibers, $d=0$. (f) Lamb shift $\Delta \omega$, normalized to the free-space linewidth $\gamma_{0},$ on a log-scale for varying emitter position in the transverse plane. Fiber parameters are the same as in (d).}}
    \label{fig:single-atom main}
\end{figure*}
We consider a geometry formed by $N$ optical nanofibers extending parallel along the $z$-axis in vacuum. Fiber $l$ has a radius of $a_{l}$ and is centered at the transverse position $\bm{c}_{l}$ in the $x$-$y$ plane with local coordinates $(\rho_{l},\phi_{l})$. The refractive index $n(\textbf{r})$, for position $\textbf{r} = (\bm{\rho},z)$, takes the value $n_{-}$ within each fiber and $n_{+} = 1$ in the surrounding vacuum. We consider two emitters with identical transverse position $\bm{\rho}_{\mathrm{a}}$, transition frequency $\omega = ck$ and (oscillating) dipole moment $\textbf{d}$, such that the (Dyadic) Green's function, or Green's tensor~\cite{novotny2012principles}, $\mathbf{G}(\textbf{r},\textbf{r}') = \mathbf{G}(\textbf{r},\textbf{r}';\omega)$ satisfying,
\begin{equation} 
    \boldsymbol{\nabla} \times \boldsymbol{\nabla} \times \textbf{G}(\textbf{r},\textbf{r}') - k^2n(\textbf{r})^2\textbf{G}\left(\textbf{r}, \textbf{r}^{\prime}\right)=\textbf{I} \delta\left(\textbf{r} -\textbf{r}^{\prime}\right),
    \label{eq:green-de}
\end{equation}
then gives the electric field $\frac{k^2}{\epsilon_0}\textbf{G}(\textbf{r}_{1},\textbf{r}_{2}) \cdot \textbf{d}$ observed by the dipole at $\textbf{r}_{1} = (\bm{\rho}_{\mathrm{a}},z_{1})$ and radiated by the emitter at position $\textbf{r}_{2} = (\bm{\rho}_{\mathrm{a}},z_{2})$. The Green's function then completely describes a system of dipoles interacting through resonant excitation exchange in the presence of an arbitrary dielectric environment. Exploiting translational invariance in the $z$-direction allows us to consider the 2D problem for the Fourier transform $\widetilde{\textbf{G}}(\bm{\rho},\bm{\rho}',\beta)$ satisfying $\textbf{G}(\textbf{r},\textbf{r}')= \mathbf{G}(\bm{\rho},\bm{\rho}',z-z') =  \frac{1}{2\pi} \int_{-\infty}^{\infty}  e^{i\beta(z-z^\prime)} \widetilde{\textbf{G}} (\bm{\rho},\bm{\rho}',\beta)d\beta$. Cylindrical symmetry then permits the global field expansion~\cite{fussell2004three,white2002multipole}
\begin{align}
     \widetilde{\mathbf{G}}= 
     \widetilde{\textbf{G}}_{0} +\sum_{l=1}^{N} \sum_{m=-\infty}^{\infty} \widetilde{\bm{\mathcal{B}}}{}^{l}_{m} H_m^{(1)}\left(k_\rho \rho_l\right) e^{i m \phi_l},
     \label{eq:global-field}
\end{align}
valid outside all the fibers, with Hankel function $H_{m}^{(1)}$ of the first kind for the azimuthal mode number $m$, and transverse wavenumber $k_{\rho} = \sqrt{k^2 n_{+}^2  - \beta^2}$ outside of the fibers. Equation~\eqref{eq:global-field} decomposes the Green's function via the superposition principle~\cite{chew1999waves} into vacuum component $\widetilde{\textbf{G}}_{0}$, and circular waves scattered from cylinder $l$ with amplitude $\widetilde{\bm{\mathcal{B}}}_{m}^{l}$ to be determined. This expression includes multiple scattering between fibers, and is thus exact. As detailed in the SM, Graf's addition theorem allows the field \eqref{eq:global-field} to be expressed as cylindrical waves locally around each fiber $l$, as a function of $(\rho_{l},\phi_{l})$ only. Here, the boundary conditions readily allow for the $\widetilde{\bm{\mathcal{B}}}_{m}^{l}$ to be expressed in terms of $\widetilde{\mathbf{G}}_{0}$, so that $\textbf{G}$ is obtained in real space upon Fourier inversion. In the following, we have verified our calculations against existing results for single-fiber~\cite{klimov2004spontaneous} and two-fiber~\cite{le2021spatial} geometries (see Supplemental Material).

\indent \paragraph{Single-emitter figures of merit.}
In the following, we focus on the case of two identical ONFs with separation $d$ and, unless otherwise specified, radius $a$, as depicted in Fig.~\ref{fig:single-atom main}(a). In the following we concretely take the emitter transition wavelength $\lambda = 2\pi/k = 780$ nm, corresponding to the D2 line of rubidium. Once we fix the fiber refractive index, the Green's function is determined (neglecting dispersion) by the ratios $a/\lambda$ and $d/\lambda$, so that our conclusions may be extended to general emission wavelengths. As the transverse positions and dipoles of the emitters are identical, they share the same coupling efficiency,
\begin{equation}
    \eta = \frac{\Gamma_{\text{1D}}}{\Gamma},
\end{equation}
which is calculated using the emitter decay rate $\Gamma_{\text{1D}}$  into the guided modes of the fibers and the total decay rate $\Gamma$. These are expressed via the imaginary part of the diagonal Green's function elements~\cite{stourm2020spontaneous} as $\Gamma = \frac{2k^2}{\hbar\epsilon_{0}} \textbf{d} \cdot \Im [\textbf{G}(\bm{\rho}_{\mathrm{a}},\bm{\rho}_{\mathrm{a}},0)] \cdot \textbf{d}^{*},$ and $\Gamma_{\text{1D}} = \frac{2k^2}{\hbar\epsilon_{0}} \textbf{d} \cdot \Im [\textbf{G}_g(\bm{\rho}_{\mathrm{a}},\bm{\rho}_{\mathrm{a}},0)] \cdot \textbf{d}^{*}$ respectively. The guided mode contribution $\textbf{G}_g$ to the Green's function is calculated by restricting the Fourier inversion integral of $\tilde{\mathbf{G}}$ to the region $k n_{+} <|\beta| < kn_{-}$ \cite{marcuse1982light,fussell2004three}. In the following, we focus our attention on $x$-polarized dipoles that maximize the coupling efficiency to the single TE-like mode for emitters  at the center of the two ONFs~\cite{le2021spatial} (see SM). The exact calculation of the Green's function  gives coupling efficiencies over a wide range of parameters, as can be seen in Fig.~\ref{fig:single-atom main}(b), including for smaller separations $d \approx 0$ suitable for interfacing with surface-bound quantum dots~\cite{shao2022twin}, and for larger separations $d \gtrapprox 200$ nm more relevant to centrally trapped neutral atoms~\cite{daly2014nanostructured, nieddu2016optical}. Systems with atoms trapped at the center between the fibers with larger separations $d \sim 300$ nm can show coupling efficiencies around $22$\% compared with the single-ONF value of $15\%$, whilst the coupling efficiency may increase beyond $50\%$~\cite{shao2022twin} for small ONF separations, compared to the single-ONF maximum of $32\%$~\cite{yalla2012efficient,Kien2008}. In the Supplement, we further show that coupling efficiencies to two-ONFs may also outperform that of more complex geometries featuring additional ONFs, showing that the simplest multi-coupler scheme is the most effective. 
The discontinuities observed in coupling efficiency represent cutoffs of $\mathrm{TE}$-like modes, which reflects the fact that poles of $\widetilde{\textbf{G}}$ are exactly the propagation constants of guided modes. We note that our calculation is exact even in the case of $a,d\lessapprox \lambda$, where the commonly used coupled-mode theory may give erroneous results~\cite{le2021spatial}. Beyond the cutoffs, emitters continue to enjoy appreciable total emission including into the higher-order modes, which could benefit multimode~\cite{Maeda:23} fiber-based ~\cite{kato2015strong,kato2019observation} cavity QED schemes similarly to in standard Fabry-Perot resonators~\cite{vaidya_2018,multimodecqed_robb}. 
\begin{figure}[t!]
   	\centering	
    \includegraphics[width=0.99\linewidth]{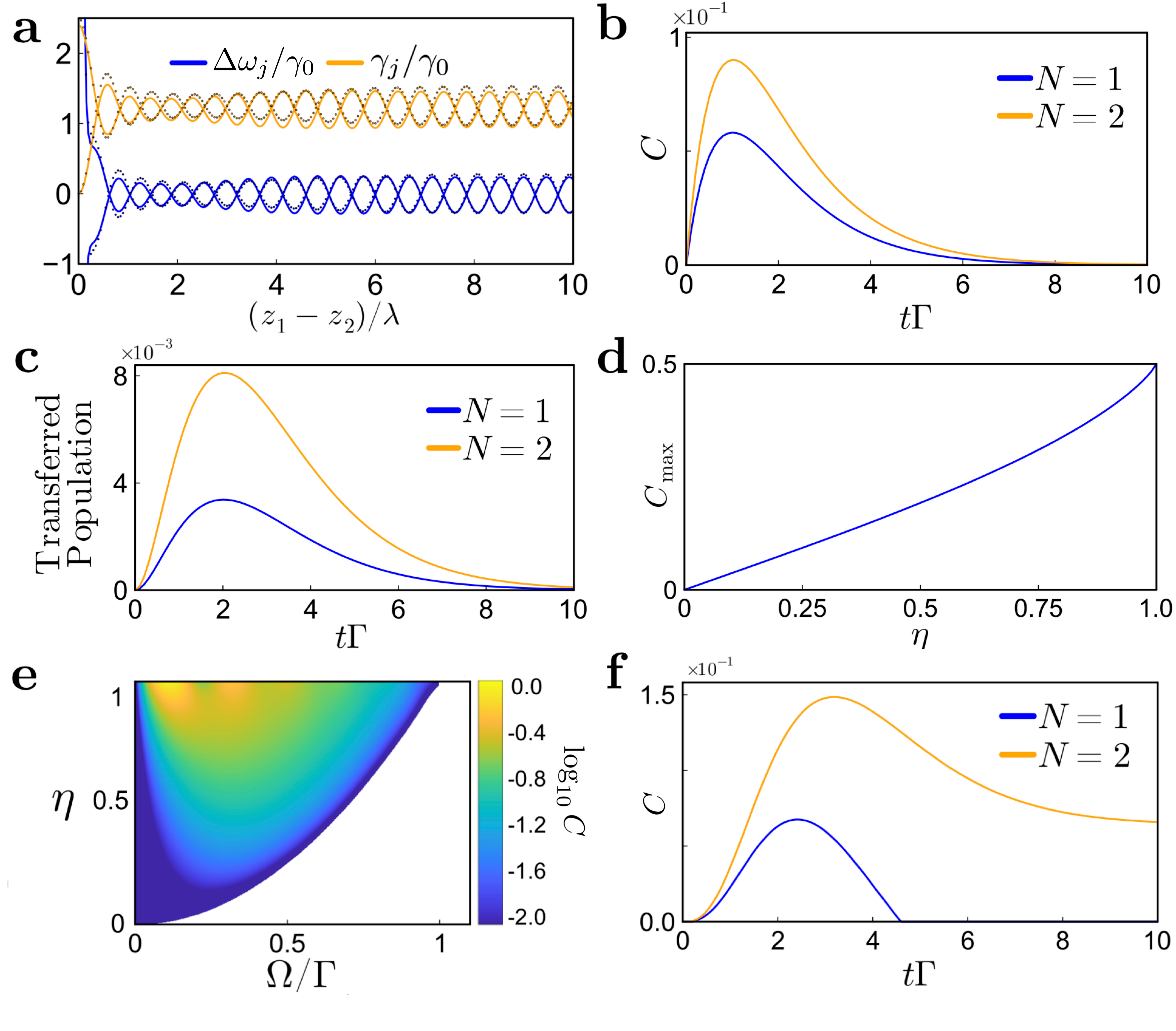}
	\caption{(a) Normalized linewidth $\gamma_{j}/\gamma_{0}$ and lineshifts $\Delta\omega_{j}/\gamma_{0}$ of the two collective resonances formed by two emitters located at the center of two ONFs in the configuration of Fig.~\ref{fig:single-atom main}(a). The dots give the numbers obtained using the approximate Green's function elements \eqref{eq:long-range}, whilst the solid lines give the values obtained from the exact Green's function.   (b) Evolution of concurrence $C$ and (c) population $\langle \sigma_{2}^{+}\sigma_{2}\rangle$ transferred  when evolving the master equation Eq. \eqref{eq:master-equation} from the initial state with $\langle \sigma_{2}^{+}\sigma_{2}\rangle = 0, \langle \sigma_{1}^{+}\sigma_{1}\rangle = 1$. (d) Maximal concurrence observed between two atoms for the same configuration as (b,c) but allowing the  coupling coefficient $\eta$ to vary. Note that this value is independent of $\Gamma$. (e) Steady-state concurrence (log-scale) observed under the continuous driving of strength $\Omega$ for coupling efficiency $\eta$. The white region denotes $C = 0.$ (f) Evolution of concurrence over time for two atoms initially in the ground state.}
	\label{fig:two-atom}
\end{figure}
In addition to coupling efficiency, the Purcell factor $F_{p} = \frac{\Gamma}{\Gamma_{0}}$, for free-space decay rate $\Gamma_{0},$ is a key figure of merit when considering efficient on-chip single-photon sources~\cite{purcell2018}, with large $F_{p} \gg 1$ preferable. Whilst single-ONF systems typically observe $F_{p} \approx 1.5$~\cite{solano2019alignment,le2005spontaneous}, Fig. \ref{fig:single-atom main}(c) reveals Purcell factors on the order of 10 for touching ONFs over a range of $100$ nm $\lessapprox a \lessapprox 175$ nm of fiber radii. The regions of large $F_{p}$ crucially overlap with regions of larger coupling efficiency, which is a result of the TE-like slot mode guided between the two fibers, and implied by the coupling efficiency profile observed for varying $\bm{\rho}_{a}$ (Fig.~\ref{fig:single-atom main}(d)). Further detail is presented in the SM. In particular, the coupling efficiency remains appreciable away from the center of the two ONFs. Although operation commonly envisions fixed emitters trapped at the two-ONF center, this result suggests that considerable position fluctuations may be tolerated by neutral atoms in the near term without considerably sacrificing large waveguide couplings. The coupling efficiency is further robust to small changes in the radius of either ONF (Fig.~\ref{fig:single-atom main}(e), see SM) on the order of tens of nanometers~\cite{Fatemi:17}, so that fiber-based cavities~\cite{kato2019observation,kato2015strong} can reliably observe a cavity-emitter coupling $g \propto \sqrt{F_p \eta}$~\cite{nayak2019real} enjoying increased contributions from both $\eta$ and $F_p$.

Our calculation of the full Green's function further yields the exact single-emitter Lamb shift, defined as,
\begin{equation}
  \Delta \omega= \frac{k^2}{\hbar\epsilon_{0}} \textbf{d} \cdot \Re [\textbf{G}(\bm{\rho}_{\mathrm{a}},\bm{\rho}_{\mathrm{a}},0)] \cdot \textbf{d}^{*},   
\end{equation}
where the divergent vacuum contribution is understood to be contained within the bare atomic frequency. The Lamb shift can generate a repulsive potential for excited state atoms and exploited to counteract strong Van der Waals forces close proximity to the surface~\cite{chang2014}. On the other hand, it must be accounted for in precision metrology~\cite{okaba2014lamb} once $\frac{\Delta \omega}{\gamma_{0}}\gtrapprox 1$, for the free-space linewidth $\gamma_{0} = \Gamma_{0}/2$, and for atoms in highly-excited Rydberg states~\cite{rajasree2020,stourm2020spontaneous} where the shift may be extremely large. In Fig. \ref{fig:single-atom main}(f), we find that the Lamb shift for each emitter may be small, and notably even zero on a curve lying between the two ONFs. The zero occurs in a region of appreciable coupling efficiency ($\eta \sim 20$\%), so that the effects of single-emitter shifts can be eliminated with a trapping scheme above the slot similar to~\cite{goban2015superradiance} without significantly compromising coupling. The shift diverges for shorter atom-surface distances, with the metrologically relevant regime of $\Delta\omega/\gamma_{0} \gtrapprox 1$ occurring once the atom-surface distance $d/2$ drops below approximately 75 nm. We find this bound holds independently of the value of $a$ (see SM). 

\indent \paragraph{Collective resonances of emitters.}
Key functionalities of waveguide-based light-matter platforms are enabled by the collective resonances and interactions mediated through waveguided photons. For linearly responding emitters, the eigenvalues $\lambda_{j} = \Delta\omega_{j} + i\gamma_{j}$ of the coupling matrix $G_{mn} = \Omega_{mn} + \frac{i\Gamma_{mn}}{2}$ formed by the Green's function elements
\begin{align}
    \Omega_{mn} &= \frac{k^2}{\hbar\epsilon_{0}}\textbf{d}\cdot \mathfrak{R}\left[\textbf{G}(\textbf{r}_{m},\textbf{r}_{n})\right]\cdot \textbf{d}^{*}, \label{eq:om}\\
    \Gamma_{mn} &= \frac{2k^2}{\hbar\epsilon_{0}}\textbf{d}\cdot \mathfrak{I}\left[\textbf{G}(\textbf{r}_{m},\textbf{r}_{n})\right]\cdot \textbf{d}^{*} \label{eq:gam},
\end{align}
define lineshifts $\Delta\omega_{j}$ and linewidths $\gamma_{j}$ of collective resonances under waveguide driving. The infinite-range interaction induced by the 1D geometry is apparent upon inspection of the asymptotic limit of $|z - z'|/\lambda \to \infty$ for a 1D waveguide~\cite{mrao_2007} homogeneous along the $z$-direction
\begin{align}
    &\textbf{G}_{\text{1D}}(\textbf{r},\textbf{r}') = \textbf{G}_{\text{1D}}(\bm{\rho},\bm{\rho}',z-z') \nonumber\\
    &\sim 
    \frac{ic^2}{2\omega}\sum_{\mu}\frac{d\beta_{\mu}(\omega)}{d\omega} \times
    \begin{cases}
        \textbf{E}_{\mu}(\textbf{r})\otimes \textbf{E}_{\mu}^{*}(\textbf{r}') & z - z' > 0 \\
        \textbf{E}_{\mu}^{*}(\textbf{r})\otimes \textbf{E}_{\mu}(\textbf{r}') & z - z' < 0,
    \end{cases}
    \label{eq:long-range}
\end{align}
plus $O(\lambda|z-z'|^{-1})$ corrections. Eq. \eqref{eq:long-range} formalises the intuition that only guided modes $\mu$ with normalized field amplitudes $\textbf{E}_{\mu}(\textbf{r}) = \bm{\mathcal{E}}_{\mu}(\bm{\rho})e^{i\beta_{\mu}(\omega)z}$ (such that $\int d^2 \bm{\rho} n|\textbf{E}_{\mu}|^2 = 1$) and propagation constants $\beta_{\mu}(\omega) \ (> 0)$ contribute significantly to atom-atom interactions for large on-axis  separations. For two $x$-polarized emitters at the center of a two-ONF system, oscillations~\cite{fam_2005} of collective lineshifts and linewidths are shown in Fig.~\ref{fig:two-atom}(a) for $(a,d) = (180\text{nm},300\text{nm})$. Owing to the exclusive coupling to the TE-like mode at the center, the persistent oscillations have a wavenumber $\beta_{\text{TE}} \approx 1.08 \times k$ -- the propagation constant of the TE-like mode coupling to the dipoles. Even for separations on the order of a wavelength the discrepancy between the exact calculations and the typical approximation obtained using \eqref{eq:long-range} may be significant, which in turn results in appreciable differences in transmission of weak waveguided fields through atomic chains when comparing the exact and approximate expressions~\cite{svendsen2023modified}. The linewidth of the most superradiant (subradiant) resonance for large separations is given by $\Gamma(1\pm \eta)$, where $\eta \approx 0.2$ for two ONFs compared with the maximally obtainable $\eta = 0.13$ for the emitter with identical emitter-surface distance for one ONF. The reduction of the emission rate $\Gamma(1-\eta)$ outside of the waveguide enables higher fidelity generation of collective emitter states via superradiance and subradiance~\cite{tudela_abitrary}, including those exhibiting nonclassical correlations.
\paragraph{Entanglement generation between quantum emitters.}
Waveguide coupling efficiency also plays a key role in the dynamics of \textit{quantum} emitters. More generally, $\Omega_{mn}$ and $\Gamma_{mn}$ mediate the coherent and incoherent interactions respectively in the master equation $\dot{\rho} = -(i/\hbar)\left[H,\rho\right] + \mathcal{L}[\rho]$ for two-level quantum emitters with density matrix $\rho$. Assuming identical two-level-systems and a common, real coherent driving $\Omega$, the effective Hamiltonian and dissipator in the frame rotating at the emitters' frequency read
\begin{align}
    H &= \sum_{j}\frac{\Omega}{2}\left(\sigma_{j} + \sigma_{j}^{+}\right) + \sum_{jk}\Omega_{jk}\sigma_{j}^{+}\sigma_{k}, \nonumber\\
    \mathcal{L}[\rho] &= \sum_{jk}\frac{\Gamma_{jk}}{2}\left(\left[\sigma_{k}\rho,\sigma_{j}^{+}\right] + \left[\sigma_{k},\rho\sigma_{j}^{+}\right]\right),
    \label{eq:master-equation}
\end{align}
for the raising (lowering) operator $\sigma_{j}^{+} (\sigma_{j})$ of emitter $j$. 
For sufficiently high coupling efficiencies, a 1D waveguide can mediate appreciable entanglement between emitters with arbitrarily large separation along the ONF~\cite{zheng2013entanglement,tudela2011entanglement}. In particular, the concurrence~\cite{wooters1998entanglement} $C$ (see SM) has been shown to dictate the amount of entanglement detectable between the two quantum emitters under a single photon input~\cite{ballestero2014entanglement}, and here we investigate the generation of entanglement in the parameter regime of modest coupling efficiencies typical of ONFs. For centrally trapped atoms $100$ nm away from the ONF of radius $200$ nm, we solve~\cite{kramer_qoptics} the master equation Eq. \eqref{eq:master-equation} for two distinct excitations schemes. Assuming spacing $|z_{1}-z_{2}| = 10\times 2\pi/\beta_{\text{TE}}$ commensurate with the guided mode wavelength, dominant emitter-emitter interactions are due to collective dissipation, and we neglect the small contributions of $\Omega_{jk}$ at this separation. The system dynamics according to Eq.~\eqref{eq:master-equation} are then characterized by $\Gamma,$ $\eta,$ and $\Omega$. In Fig.~\ref{fig:two-atom}(b), we present the transient concurrence generated by collective emission in the case where emitter one is initially excited and emitter two de-excited, with $\Omega = 0$. We compare the evolution with that of two atoms in the same geometry, but with one of the ONFs absent. For reference, the concurrence of the maximally entangled two-atom state is $C=1$, whilst $C=0$ for non-entangled states. Provided $C > 0$, then many copies of the state can be used to distill high-fidelity Bell states~\cite{distill1996}. The maximum concurrence for the two-ONF system is enhanced compared to the single ONF, which is reflected in the linear growth of $C$ with $\eta$ (Fig.~\ref{fig:two-atom}(d)). For $\eta = 1$, half of the initial excitation is coupled into the long-lived subradiant Bell state, giving $C_{\text{max}} = 1/2$.  Similarly in Fig.~\ref{fig:two-atom}(c), the population $\langle \sigma_{2}^{+}\sigma_{2}\rangle$ transferred to the second atom is significantly enhanced by the presence of the extra ONF. In Fig.~\ref{fig:two-atom}(e) we also present the steady-state concurrence generated under a continuous driving for both atoms initially deexcited, whilst allowing $\eta$ to vary freely. Remarkably, for each driving strength there exists a threshold efficiency beyond which concurrence is identically zero, and thus where entanglement distillation is impossible. At a driving strength of $\Omega = 0.45\Gamma,$ for example, the two-ONF system with $\eta \approx 0.24$ may exhibit non-zero steady-state concurrence absent in the single-ONF system ($\eta \approx 0.16)$. The time evolution Fig.~\ref{fig:two-atom}(f) highlights the eventually loss of all concurrence for the single ONF due to the combined effects of atomic saturation and increased losses.  Whilst small differences in concurrence may be tolerable for $\eta$ close to unity~\cite{tudela2011entanglement}, the more pronounced difference for lower efficiencies suggests two-ONF geometries as a possible first step towards the observation of entanglement two waveguide-trapped atoms.
\paragraph{Conclusion.}
We have demonstrated enhanced photon coupling efficiencies in a simple geometry of only two ONFs and for experimental parameters typical of realistic light-matter systems including atomic gases and quantum dots.
On the theoretical side, our results have qualitative predictive power in a wide array of existing systems, and the Green's function can be readily exploited -- in general multiple-cylinder geometries~\cite{lightcage} comprising various dielectrics~\cite{nanowire} -- to further study, for example, F{\"o}rster resonance energy transfer~\cite{blum2012nanophotonic}, optical forces~\cite{fang2021optical,janus2023,fiberforces2009}, and higher order multipole transitions ~\cite{Ray_2020} that play a key role in fiber-based Rydberg atom platforms~\cite{stourm2020spontaneous,rajasree2020}. On the practical side, the attractive figures of merit demonstrated are realizable in simple geometries of ONFs using existing technologies~\cite{yu2022highly}, where the enhanced coupling efficiencies, Purcell factors and accompanying emitter-emitter entanglement points towards their use as tools for photonic state generation and distributed quantum computing within all-fiber~\cite{kato2019observation,fiber_Cavity_dark} networks.

\section{Acknowledgments}
We are grateful for fruitful discussions with Maki Maeda, Zohreh Shahrabifarahani and Dylan Brown. We acknowledge the support provided by the Scientific Computing
and Data Analysis section of Research Support Division at OIST. LR acknowledges the support, in part, from Moonshot R\&D, JST JPMJMS2061. K.J. and L.R. contributed equally to this work.

\bibliography{ms}

\providecommand{\noopsort}[1]{}\providecommand{\singleletter}[1]{#1}%
\begin{thebibliography}{73}%
\makeatletter
\providecommand \@ifxundefined [1]{%
 \@ifx{#1\undefined}
}%
\providecommand \@ifnum [1]{%
 \ifnum #1\expandafter \@firstoftwo
 \else \expandafter \@secondoftwo
 \fi
}%
\providecommand \@ifx [1]{%
 \ifx #1\expandafter \@firstoftwo
 \else \expandafter \@secondoftwo
 \fi
}%
\providecommand \natexlab [1]{#1}%
\providecommand \enquote  [1]{``#1''}%
\providecommand \bibnamefont  [1]{#1}%
\providecommand \bibfnamefont [1]{#1}%
\providecommand \citenamefont [1]{#1}%
\providecommand \href@noop [0]{\@secondoftwo}%
\providecommand \href [0]{\begingroup \@sanitize@url \@href}%
\providecommand \@href[1]{\@@startlink{#1}\@@href}%
\providecommand \@@href[1]{\endgroup#1\@@endlink}%
\providecommand \@sanitize@url [0]{\catcode `\\12\catcode `\$12\catcode `\&12\catcode `\#12\catcode `\^12\catcode `\_12\catcode `\%12\relax}%
\providecommand \@@startlink[1]{}%
\providecommand \@@endlink[0]{}%
\providecommand \url  [0]{\begingroup\@sanitize@url \@url }%
\providecommand \@url [1]{\endgroup\@href {#1}{\urlprefix }}%
\providecommand \urlprefix  [0]{URL }%
\providecommand \Eprint [0]{\href }%
\providecommand \doibase [0]{https://doi.org/}%
\providecommand \selectlanguage [0]{\@gobble}%
\providecommand \bibinfo  [0]{\@secondoftwo}%
\providecommand \bibfield  [0]{\@secondoftwo}%
\providecommand \translation [1]{[#1]}%
\providecommand \BibitemOpen [0]{}%
\providecommand \bibitemStop [0]{}%
\providecommand \bibitemNoStop [0]{.\EOS\space}%
\providecommand \EOS [0]{\spacefactor3000\relax}%
\providecommand \BibitemShut  [1]{\csname bibitem#1\endcsname}%
\let\auto@bib@innerbib\@empty
\bibitem [{\citenamefont {Paulisch}\ \emph {et~al.}(2019)\citenamefont {Paulisch}, \citenamefont {Perarnau-Llobet}, \citenamefont {Gonz\'alez-Tudela},\ and\ \citenamefont {Cirac}}]{sensing_1}%
  \BibitemOpen
  \bibfield  {author} {\bibinfo {author} {\bibfnamefont {V.}~\bibnamefont {Paulisch}}, \bibinfo {author} {\bibfnamefont {M.}~\bibnamefont {Perarnau-Llobet}}, \bibinfo {author} {\bibfnamefont {A.}~\bibnamefont {Gonz\'alez-Tudela}},\ and\ \bibinfo {author} {\bibfnamefont {J.~I.}\ \bibnamefont {Cirac}},\ }\bibfield  {title} {\bibinfo {title} {Quantum metrology with one-dimensional superradiant photonic states},\ }\href {https://doi.org/10.1103/PhysRevA.99.043807} {\bibfield  {journal} {\bibinfo  {journal} {Phys. Rev. A}\ }\textbf {\bibinfo {volume} {99}},\ \bibinfo {pages} {043807} (\bibinfo {year} {2019})}\BibitemShut {NoStop}%
\bibitem [{\citenamefont {Okaba}\ \emph {et~al.}(2014)\citenamefont {Okaba}, \citenamefont {Takano}, \citenamefont {Benabid}, \citenamefont {Bradley}, \citenamefont {Vincetti}, \citenamefont {Maizelis}, \citenamefont {Yampol'Skii}, \citenamefont {Nori},\ and\ \citenamefont {Katori}}]{okaba2014lamb}%
  \BibitemOpen
  \bibfield  {author} {\bibinfo {author} {\bibfnamefont {S.}~\bibnamefont {Okaba}}, \bibinfo {author} {\bibfnamefont {T.}~\bibnamefont {Takano}}, \bibinfo {author} {\bibfnamefont {F.}~\bibnamefont {Benabid}}, \bibinfo {author} {\bibfnamefont {T.}~\bibnamefont {Bradley}}, \bibinfo {author} {\bibfnamefont {L.}~\bibnamefont {Vincetti}}, \bibinfo {author} {\bibfnamefont {Z.}~\bibnamefont {Maizelis}}, \bibinfo {author} {\bibfnamefont {V.}~\bibnamefont {Yampol'Skii}}, \bibinfo {author} {\bibfnamefont {F.}~\bibnamefont {Nori}},\ and\ \bibinfo {author} {\bibfnamefont {H.}~\bibnamefont {Katori}},\ }\bibfield  {title} {\bibinfo {title} {Lamb-dicke spectroscopy of atoms in a hollow-core photonic crystal fibre},\ }\href {https://doi.org/10.1038/ncomms5096} {\bibfield  {journal} {\bibinfo  {journal} {Nature communications}\ }\textbf {\bibinfo {volume} {5}},\ \bibinfo {pages} {4096} (\bibinfo {year} {2014})}\BibitemShut {NoStop}%
\bibitem [{\citenamefont {Yu}\ \emph {et~al.}(2018)\citenamefont {Yu}, \citenamefont {Zhi}, \citenamefont {Tang}, \citenamefont {Li}, \citenamefont {Gong}, \citenamefont {Qiu},\ and\ \citenamefont {Xiao}}]{sensing}%
  \BibitemOpen
  \bibfield  {author} {\bibinfo {author} {\bibfnamefont {X.-C.}\ \bibnamefont {Yu}}, \bibinfo {author} {\bibfnamefont {Y.}~\bibnamefont {Zhi}}, \bibinfo {author} {\bibfnamefont {S.-J.}\ \bibnamefont {Tang}}, \bibinfo {author} {\bibfnamefont {B.-B.}\ \bibnamefont {Li}}, \bibinfo {author} {\bibfnamefont {Q.}~\bibnamefont {Gong}}, \bibinfo {author} {\bibfnamefont {C.-W.}\ \bibnamefont {Qiu}},\ and\ \bibinfo {author} {\bibfnamefont {Y.-F.}\ \bibnamefont {Xiao}},\ }\bibfield  {title} {\bibinfo {title} {Optically sizing single atmospheric particulates with a 10-nm resolution using a strong evanescent field},\ }\href {https://doi.org/10.1038/lsa.2018.3} {\bibfield  {journal} {\bibinfo  {journal} {Light: Science \& Applications}\ }\textbf {\bibinfo {volume} {7}},\ \bibinfo {pages} {18003} (\bibinfo {year} {2018})}\BibitemShut {NoStop}%
\bibitem [{\citenamefont {Gonz\'alez-Tudela}\ \emph {et~al.}(2015)\citenamefont {Gonz\'alez-Tudela}, \citenamefont {Paulisch}, \citenamefont {Chang}, \citenamefont {Kimble},\ and\ \citenamefont {Cirac}}]{tudela_abitrary}%
  \BibitemOpen
  \bibfield  {author} {\bibinfo {author} {\bibfnamefont {A.}~\bibnamefont {Gonz\'alez-Tudela}}, \bibinfo {author} {\bibfnamefont {V.}~\bibnamefont {Paulisch}}, \bibinfo {author} {\bibfnamefont {D.~E.}\ \bibnamefont {Chang}}, \bibinfo {author} {\bibfnamefont {H.~J.}\ \bibnamefont {Kimble}},\ and\ \bibinfo {author} {\bibfnamefont {J.~I.}\ \bibnamefont {Cirac}},\ }\bibfield  {title} {\bibinfo {title} {Deterministic generation of arbitrary photonic states assisted by dissipation},\ }\href {https://doi.org/10.1103/PhysRevLett.115.163603} {\bibfield  {journal} {\bibinfo  {journal} {Phys. Rev. Lett.}\ }\textbf {\bibinfo {volume} {115}},\ \bibinfo {pages} {163603} (\bibinfo {year} {2015})}\BibitemShut {NoStop}%
\bibitem [{\citenamefont {Mahmoodian}\ \emph {et~al.}(2020)\citenamefont {Mahmoodian}, \citenamefont {Calaj\'o}, \citenamefont {Chang}, \citenamefont {Hammerer},\ and\ \citenamefont {S\o{}rensen}}]{mahmoodian_nonclassical}%
  \BibitemOpen
  \bibfield  {author} {\bibinfo {author} {\bibfnamefont {S.}~\bibnamefont {Mahmoodian}}, \bibinfo {author} {\bibfnamefont {G.}~\bibnamefont {Calaj\'o}}, \bibinfo {author} {\bibfnamefont {D.~E.}\ \bibnamefont {Chang}}, \bibinfo {author} {\bibfnamefont {K.}~\bibnamefont {Hammerer}},\ and\ \bibinfo {author} {\bibfnamefont {A.~S.}\ \bibnamefont {S\o{}rensen}},\ }\bibfield  {title} {\bibinfo {title} {Dynamics of many-body photon bound states in chiral waveguide {QED}},\ }\href {https://doi.org/10.1103/PhysRevX.10.031011} {\bibfield  {journal} {\bibinfo  {journal} {Phys. Rev. X}\ }\textbf {\bibinfo {volume} {10}},\ \bibinfo {pages} {031011} (\bibinfo {year} {2020})}\BibitemShut {NoStop}%
\bibitem [{\citenamefont {Chang}\ \emph {et~al.}(2007)\citenamefont {Chang}, \citenamefont {S{\o}rensen}, \citenamefont {Demler},\ and\ \citenamefont {Lukin}}]{chang_transistor}%
  \BibitemOpen
  \bibfield  {author} {\bibinfo {author} {\bibfnamefont {D.~E.}\ \bibnamefont {Chang}}, \bibinfo {author} {\bibfnamefont {A.~S.}\ \bibnamefont {S{\o}rensen}}, \bibinfo {author} {\bibfnamefont {E.~A.}\ \bibnamefont {Demler}},\ and\ \bibinfo {author} {\bibfnamefont {M.~D.}\ \bibnamefont {Lukin}},\ }\bibfield  {title} {\bibinfo {title} {A single-photon transistor using nanoscale surface plasmons},\ }\href {https://doi.org/10.1038/nphys708} {\bibfield  {journal} {\bibinfo  {journal} {Nature Physics}\ }\textbf {\bibinfo {volume} {3}},\ \bibinfo {pages} {807} (\bibinfo {year} {2007})}\BibitemShut {NoStop}%
\bibitem [{\citenamefont {Corzo}\ \emph {et~al.}(2019)\citenamefont {Corzo}, \citenamefont {Raskop}, \citenamefont {Chandra}, \citenamefont {Sheremet}, \citenamefont {Gouraud},\ and\ \citenamefont {Laurat}}]{corzo_atoms}%
  \BibitemOpen
  \bibfield  {author} {\bibinfo {author} {\bibfnamefont {N.~V.}\ \bibnamefont {Corzo}}, \bibinfo {author} {\bibfnamefont {J.}~\bibnamefont {Raskop}}, \bibinfo {author} {\bibfnamefont {A.}~\bibnamefont {Chandra}}, \bibinfo {author} {\bibfnamefont {A.~S.}\ \bibnamefont {Sheremet}}, \bibinfo {author} {\bibfnamefont {B.}~\bibnamefont {Gouraud}},\ and\ \bibinfo {author} {\bibfnamefont {J.}~\bibnamefont {Laurat}},\ }\bibfield  {title} {\bibinfo {title} {Waveguide-coupled single collective excitation of atomic arrays},\ }\href {https://doi.org/10.1038/s41586-019-0902-3} {\bibfield  {journal} {\bibinfo  {journal} {Nature}\ }\textbf {\bibinfo {volume} {566}},\ \bibinfo {pages} {359} (\bibinfo {year} {2019})}\BibitemShut {NoStop}%
\bibitem [{\citenamefont {Zheng}\ \emph {et~al.}(2013)\citenamefont {Zheng}, \citenamefont {Gauthier},\ and\ \citenamefont {Baranger}}]{waveguide_quantum}%
  \BibitemOpen
  \bibfield  {author} {\bibinfo {author} {\bibfnamefont {H.}~\bibnamefont {Zheng}}, \bibinfo {author} {\bibfnamefont {D.~J.}\ \bibnamefont {Gauthier}},\ and\ \bibinfo {author} {\bibfnamefont {H.~U.}\ \bibnamefont {Baranger}},\ }\bibfield  {title} {\bibinfo {title} {Waveguide-{QED}-based photonic quantum computation},\ }\href {https://doi.org/10.1103/PhysRevLett.111.090502} {\bibfield  {journal} {\bibinfo  {journal} {Phys. Rev. Lett.}\ }\textbf {\bibinfo {volume} {111}},\ \bibinfo {pages} {090502} (\bibinfo {year} {2013})}\BibitemShut {NoStop}%
\bibitem [{\citenamefont {Paulisch}\ \emph {et~al.}(2016)\citenamefont {Paulisch}, \citenamefont {Kimble},\ and\ \citenamefont {González-Tudela}}]{Paulisch_2016}%
  \BibitemOpen
  \bibfield  {author} {\bibinfo {author} {\bibfnamefont {V.}~\bibnamefont {Paulisch}}, \bibinfo {author} {\bibfnamefont {H.~J.}\ \bibnamefont {Kimble}},\ and\ \bibinfo {author} {\bibfnamefont {A.}~\bibnamefont {González-Tudela}},\ }\bibfield  {title} {\bibinfo {title} {Universal quantum computation in waveguide {QED} using decoherence free subspaces},\ }\href {https://doi.org/10.1088/1367-2630/18/4/043041} {\bibfield  {journal} {\bibinfo  {journal} {New Journal of Physics}\ }\textbf {\bibinfo {volume} {18}},\ \bibinfo {pages} {043041} (\bibinfo {year} {2016})}\BibitemShut {NoStop}%
\bibitem [{\citenamefont {Ritter}\ \emph {et~al.}(2018)\citenamefont {Ritter}, \citenamefont {Gruhler}, \citenamefont {Dobbertin}, \citenamefont {K{\"u}bler}, \citenamefont {Scheel}, \citenamefont {Pernice}, \citenamefont {Pfau},\ and\ \citenamefont {L{\"o}w}}]{ritter2018coupling}%
  \BibitemOpen
  \bibfield  {author} {\bibinfo {author} {\bibfnamefont {R.}~\bibnamefont {Ritter}}, \bibinfo {author} {\bibfnamefont {N.}~\bibnamefont {Gruhler}}, \bibinfo {author} {\bibfnamefont {H.}~\bibnamefont {Dobbertin}}, \bibinfo {author} {\bibfnamefont {H.}~\bibnamefont {K{\"u}bler}}, \bibinfo {author} {\bibfnamefont {S.}~\bibnamefont {Scheel}}, \bibinfo {author} {\bibfnamefont {W.}~\bibnamefont {Pernice}}, \bibinfo {author} {\bibfnamefont {T.}~\bibnamefont {Pfau}},\ and\ \bibinfo {author} {\bibfnamefont {R.}~\bibnamefont {L{\"o}w}},\ }\bibfield  {title} {\bibinfo {title} {Coupling thermal atomic vapor to slot waveguides},\ }\href {https://doi.org/10.1103/physrevx.8.021032} {\bibfield  {journal} {\bibinfo  {journal} {Phys. Rev. X}\ }\textbf {\bibinfo {volume} {8}},\ \bibinfo {pages} {021032} (\bibinfo {year} {2018})}\BibitemShut {NoStop}%
\bibitem [{\citenamefont {Davidson-Marquis}\ \emph {et~al.}(2021{\natexlab{a}})\citenamefont {Davidson-Marquis}, \citenamefont {Gargiulo}, \citenamefont {G{\'o}mez-L{\'o}pez}, \citenamefont {Jang}, \citenamefont {Kroh}, \citenamefont {M{\"u}ller}, \citenamefont {Ziegler}, \citenamefont {Maier}, \citenamefont {K{\"u}bler}, \citenamefont {Schmidt},\ and\ \citenamefont {Benson}}]{lightcage2021}%
  \BibitemOpen
  \bibfield  {author} {\bibinfo {author} {\bibfnamefont {F.}~\bibnamefont {Davidson-Marquis}}, \bibinfo {author} {\bibfnamefont {J.}~\bibnamefont {Gargiulo}}, \bibinfo {author} {\bibfnamefont {E.}~\bibnamefont {G{\'o}mez-L{\'o}pez}}, \bibinfo {author} {\bibfnamefont {B.}~\bibnamefont {Jang}}, \bibinfo {author} {\bibfnamefont {T.}~\bibnamefont {Kroh}}, \bibinfo {author} {\bibfnamefont {C.}~\bibnamefont {M{\"u}ller}}, \bibinfo {author} {\bibfnamefont {M.}~\bibnamefont {Ziegler}}, \bibinfo {author} {\bibfnamefont {S.~A.}\ \bibnamefont {Maier}}, \bibinfo {author} {\bibfnamefont {H.}~\bibnamefont {K{\"u}bler}}, \bibinfo {author} {\bibfnamefont {M.~A.}\ \bibnamefont {Schmidt}},\ and\ \bibinfo {author} {\bibfnamefont {O.}~\bibnamefont {Benson}},\ }\bibfield  {title} {\bibinfo {title} {Coherent interaction of atoms with a beam of light confined in a light cage},\ }\href {https://doi.org/10.1038/s41377-021-00556-z} {\bibfield  {journal} {\bibinfo  {journal} {Light: Science \& Applications}\ }\textbf {\bibinfo {volume}
  {10}},\ \bibinfo {pages} {114} (\bibinfo {year} {2021}{\natexlab{a}})}\BibitemShut {NoStop}%
\bibitem [{\citenamefont {Kolchin}\ \emph {et~al.}(2015)\citenamefont {Kolchin}, \citenamefont {Pholchai}, \citenamefont {Mikkelsen}, \citenamefont {Oh}, \citenamefont {Ota}, \citenamefont {Islam}, \citenamefont {Yin},\ and\ \citenamefont {Zhang}}]{high-purcell}%
  \BibitemOpen
  \bibfield  {author} {\bibinfo {author} {\bibfnamefont {P.}~\bibnamefont {Kolchin}}, \bibinfo {author} {\bibfnamefont {N.}~\bibnamefont {Pholchai}}, \bibinfo {author} {\bibfnamefont {M.~H.}\ \bibnamefont {Mikkelsen}}, \bibinfo {author} {\bibfnamefont {J.}~\bibnamefont {Oh}}, \bibinfo {author} {\bibfnamefont {S.}~\bibnamefont {Ota}}, \bibinfo {author} {\bibfnamefont {M.~S.}\ \bibnamefont {Islam}}, \bibinfo {author} {\bibfnamefont {X.}~\bibnamefont {Yin}},\ and\ \bibinfo {author} {\bibfnamefont {X.}~\bibnamefont {Zhang}},\ }\bibfield  {title} {\bibinfo {title} {High {P}urcell factor due to coupling of a single emitter to a dielectric slot waveguide},\ }\href {https://doi.org/10.1021/nl5037808} {\bibfield  {journal} {\bibinfo  {journal} {Nano Letters}\ }\textbf {\bibinfo {volume} {15}},\ \bibinfo {pages} {464} (\bibinfo {year} {2015})}\BibitemShut {NoStop}%
\bibitem [{\citenamefont {Blauth}\ \emph {et~al.}(2018)\citenamefont {Blauth}, \citenamefont {Jürgensen}, \citenamefont {Vest}, \citenamefont {Hartwig}, \citenamefont {Prechtl}, \citenamefont {Cerne}, \citenamefont {Finley},\ and\ \citenamefont {Kaniber}}]{q_emitter_slot}%
  \BibitemOpen
  \bibfield  {author} {\bibinfo {author} {\bibfnamefont {M.}~\bibnamefont {Blauth}}, \bibinfo {author} {\bibfnamefont {M.}~\bibnamefont {Jürgensen}}, \bibinfo {author} {\bibfnamefont {G.}~\bibnamefont {Vest}}, \bibinfo {author} {\bibfnamefont {O.}~\bibnamefont {Hartwig}}, \bibinfo {author} {\bibfnamefont {M.}~\bibnamefont {Prechtl}}, \bibinfo {author} {\bibfnamefont {J.}~\bibnamefont {Cerne}}, \bibinfo {author} {\bibfnamefont {J.~J.}\ \bibnamefont {Finley}},\ and\ \bibinfo {author} {\bibfnamefont {M.}~\bibnamefont {Kaniber}},\ }\bibfield  {title} {\bibinfo {title} {Coupling single photons from discrete quantum emitters in {WS}e2 to lithographically defined plasmonic slot waveguides},\ }\href {https://doi.org/10.1021/acs.nanolett.8b02687} {\bibfield  {journal} {\bibinfo  {journal} {Nano Letters}\ }\textbf {\bibinfo {volume} {18}},\ \bibinfo {pages} {6812} (\bibinfo {year} {2018})}\BibitemShut {NoStop}%
\bibitem [{\citenamefont {Li}\ \emph {et~al.}(2018{\natexlab{a}})\citenamefont {Li}, \citenamefont {Du},\ and\ \citenamefont {Chormaic}}]{wenfang_void}%
  \BibitemOpen
  \bibfield  {author} {\bibinfo {author} {\bibfnamefont {W.}~\bibnamefont {Li}}, \bibinfo {author} {\bibfnamefont {J.}~\bibnamefont {Du}},\ and\ \bibinfo {author} {\bibfnamefont {S.~N.}\ \bibnamefont {Chormaic}},\ }\bibfield  {title} {\bibinfo {title} {Tailoring a nanofiber for enhanced photon emission and coupling efficiency from single quantum emitters},\ }\href {https://doi.org/10.1364/OL.43.001674} {\bibfield  {journal} {\bibinfo  {journal} {Opt. Lett.}\ }\textbf {\bibinfo {volume} {43}},\ \bibinfo {pages} {1674} (\bibinfo {year} {2018}{\natexlab{a}})}\BibitemShut {NoStop}%
\bibitem [{\citenamefont {Goban}\ \emph {et~al.}(2015)\citenamefont {Goban}, \citenamefont {Hung}, \citenamefont {Hood}, \citenamefont {Yu}, \citenamefont {Muniz}, \citenamefont {Painter},\ and\ \citenamefont {Kimble}}]{goban2015superradiance}%
  \BibitemOpen
  \bibfield  {author} {\bibinfo {author} {\bibfnamefont {A.}~\bibnamefont {Goban}}, \bibinfo {author} {\bibfnamefont {C.-L.}\ \bibnamefont {Hung}}, \bibinfo {author} {\bibfnamefont {J.}~\bibnamefont {Hood}}, \bibinfo {author} {\bibfnamefont {S.-P.}\ \bibnamefont {Yu}}, \bibinfo {author} {\bibfnamefont {J.}~\bibnamefont {Muniz}}, \bibinfo {author} {\bibfnamefont {O.}~\bibnamefont {Painter}},\ and\ \bibinfo {author} {\bibfnamefont {H.}~\bibnamefont {Kimble}},\ }\bibfield  {title} {\bibinfo {title} {Superradiance for atoms trapped along a photonic crystal waveguide},\ }\href {https://doi.org/10.1103/PhysRevLett.115.063601} {\bibfield  {journal} {\bibinfo  {journal} {Physical review letters}\ }\textbf {\bibinfo {volume} {115}},\ \bibinfo {pages} {063601} (\bibinfo {year} {2015})}\BibitemShut {NoStop}%
\bibitem [{\citenamefont {Arcari}\ \emph {et~al.}(2014)\citenamefont {Arcari}, \citenamefont {S\"ollner}, \citenamefont {Javadi}, \citenamefont {Lindskov~Hansen}, \citenamefont {Mahmoodian}, \citenamefont {Liu}, \citenamefont {Thyrrestrup}, \citenamefont {Lee}, \citenamefont {Song}, \citenamefont {Stobbe},\ and\ \citenamefont {Lodahl}}]{near-unity}%
  \BibitemOpen
  \bibfield  {author} {\bibinfo {author} {\bibfnamefont {M.}~\bibnamefont {Arcari}}, \bibinfo {author} {\bibfnamefont {I.}~\bibnamefont {S\"ollner}}, \bibinfo {author} {\bibfnamefont {A.}~\bibnamefont {Javadi}}, \bibinfo {author} {\bibfnamefont {S.}~\bibnamefont {Lindskov~Hansen}}, \bibinfo {author} {\bibfnamefont {S.}~\bibnamefont {Mahmoodian}}, \bibinfo {author} {\bibfnamefont {J.}~\bibnamefont {Liu}}, \bibinfo {author} {\bibfnamefont {H.}~\bibnamefont {Thyrrestrup}}, \bibinfo {author} {\bibfnamefont {E.~H.}\ \bibnamefont {Lee}}, \bibinfo {author} {\bibfnamefont {J.~D.}\ \bibnamefont {Song}}, \bibinfo {author} {\bibfnamefont {S.}~\bibnamefont {Stobbe}},\ and\ \bibinfo {author} {\bibfnamefont {P.}~\bibnamefont {Lodahl}},\ }\bibfield  {title} {\bibinfo {title} {Near-unity coupling efficiency of a quantum emitter to a photonic crystal waveguide},\ }\href {https://doi.org/10.1103/PhysRevLett.113.093603} {\bibfield  {journal} {\bibinfo  {journal} {Phys. Rev. Lett.}\ }\textbf {\bibinfo {volume} {113}},\ \bibinfo
  {pages} {093603} (\bibinfo {year} {2014})}\BibitemShut {NoStop}%
\bibitem [{\citenamefont {Sheremet}\ \emph {et~al.}(2023)\citenamefont {Sheremet}, \citenamefont {Petrov}, \citenamefont {Iorsh}, \citenamefont {Poshakinskiy},\ and\ \citenamefont {Poddubny}}]{sheremet2023waveguide}%
  \BibitemOpen
  \bibfield  {author} {\bibinfo {author} {\bibfnamefont {A.~S.}\ \bibnamefont {Sheremet}}, \bibinfo {author} {\bibfnamefont {M.~I.}\ \bibnamefont {Petrov}}, \bibinfo {author} {\bibfnamefont {I.~V.}\ \bibnamefont {Iorsh}}, \bibinfo {author} {\bibfnamefont {A.~V.}\ \bibnamefont {Poshakinskiy}},\ and\ \bibinfo {author} {\bibfnamefont {A.~N.}\ \bibnamefont {Poddubny}},\ }\bibfield  {title} {\bibinfo {title} {Waveguide quantum electrodynamics: collective radiance and photon-photon correlations},\ }\href {https://doi.org/10.1103/revmodphys.95.015002} {\bibfield  {journal} {\bibinfo  {journal} {Rev. of Mod. Phys.}\ }\textbf {\bibinfo {volume} {95}},\ \bibinfo {pages} {015002} (\bibinfo {year} {2023})}\BibitemShut {NoStop}%
\bibitem [{\citenamefont {Asenjo-Garcia}\ \emph {et~al.}(2017)\citenamefont {Asenjo-Garcia}, \citenamefont {Moreno-Cardoner}, \citenamefont {Albrecht}, \citenamefont {Kimble},\ and\ \citenamefont {Chang}}]{ana_exponential}%
  \BibitemOpen
  \bibfield  {author} {\bibinfo {author} {\bibfnamefont {A.}~\bibnamefont {Asenjo-Garcia}}, \bibinfo {author} {\bibfnamefont {M.}~\bibnamefont {Moreno-Cardoner}}, \bibinfo {author} {\bibfnamefont {A.}~\bibnamefont {Albrecht}}, \bibinfo {author} {\bibfnamefont {H.~J.}\ \bibnamefont {Kimble}},\ and\ \bibinfo {author} {\bibfnamefont {D.~E.}\ \bibnamefont {Chang}},\ }\bibfield  {title} {\bibinfo {title} {Exponential improvement in photon storage fidelities using subradiance and ``selective radiance'' in atomic arrays},\ }\href {https://doi.org/10.1103/PhysRevX.7.031024} {\bibfield  {journal} {\bibinfo  {journal} {Phys. Rev. X}\ }\textbf {\bibinfo {volume} {7}},\ \bibinfo {pages} {031024} (\bibinfo {year} {2017})}\BibitemShut {NoStop}%
\bibitem [{\citenamefont {Nieddu}\ \emph {et~al.}(2016)\citenamefont {Nieddu}, \citenamefont {Gokhroo},\ and\ \citenamefont {Chormaic}}]{nieddu2016optical}%
  \BibitemOpen
  \bibfield  {author} {\bibinfo {author} {\bibfnamefont {T.}~\bibnamefont {Nieddu}}, \bibinfo {author} {\bibfnamefont {V.}~\bibnamefont {Gokhroo}},\ and\ \bibinfo {author} {\bibfnamefont {S.~N.}\ \bibnamefont {Chormaic}},\ }\bibfield  {title} {\bibinfo {title} {Optical nanofibres and neutral atoms},\ }\href {https://iopscience.iop.org/article/10.1088/2040-8978/18/5/053001} {\bibfield  {journal} {\bibinfo  {journal} {Journal of Optics}\ }\textbf {\bibinfo {volume} {18}},\ \bibinfo {pages} {053001} (\bibinfo {year} {2016})}\BibitemShut {NoStop}%
\bibitem [{\citenamefont {Nayak}\ \emph {et~al.}(2018)\citenamefont {Nayak}, \citenamefont {Sadgrove}, \citenamefont {Yalla}, \citenamefont {Kien},\ and\ \citenamefont {Hakuta}}]{Nayak_onf_review}%
  \BibitemOpen
  \bibfield  {author} {\bibinfo {author} {\bibfnamefont {K.~P.}\ \bibnamefont {Nayak}}, \bibinfo {author} {\bibfnamefont {M.}~\bibnamefont {Sadgrove}}, \bibinfo {author} {\bibfnamefont {R.}~\bibnamefont {Yalla}}, \bibinfo {author} {\bibfnamefont {F.~L.}\ \bibnamefont {Kien}},\ and\ \bibinfo {author} {\bibfnamefont {K.}~\bibnamefont {Hakuta}},\ }\bibfield  {title} {\bibinfo {title} {Nanofiber quantum photonics},\ }\href {https://doi.org/10.1088/2040-8986/aac35e} {\bibfield  {journal} {\bibinfo  {journal} {Journal of Optics}\ }\textbf {\bibinfo {volume} {20}},\ \bibinfo {pages} {073001} (\bibinfo {year} {2018})}\BibitemShut {NoStop}%
\bibitem [{\citenamefont {Solano}\ \emph {et~al.}(2017{\natexlab{a}})\citenamefont {Solano}, \citenamefont {Grover}, \citenamefont {Hoffman}, \citenamefont {Ravets}, \citenamefont {Fatemi}, \citenamefont {Orozco},\ and\ \citenamefont {Rolston}}]{solano2017optical}%
  \BibitemOpen
  \bibfield  {author} {\bibinfo {author} {\bibfnamefont {P.}~\bibnamefont {Solano}}, \bibinfo {author} {\bibfnamefont {J.~A.}\ \bibnamefont {Grover}}, \bibinfo {author} {\bibfnamefont {J.~E.}\ \bibnamefont {Hoffman}}, \bibinfo {author} {\bibfnamefont {S.}~\bibnamefont {Ravets}}, \bibinfo {author} {\bibfnamefont {F.~K.}\ \bibnamefont {Fatemi}}, \bibinfo {author} {\bibfnamefont {L.~A.}\ \bibnamefont {Orozco}},\ and\ \bibinfo {author} {\bibfnamefont {S.~L.}\ \bibnamefont {Rolston}},\ }\bibfield  {title} {\bibinfo {title} {Optical nanofibers: a new platform for quantum optics},\ }in\ \href {https://doi.org/10.1016/bs.aamop.2017.02.003} {\emph {\bibinfo {booktitle} {Advances In Atomic, Molecular, and Optical Physics}}},\ Vol.~\bibinfo {volume} {66}\ (\bibinfo  {publisher} {Elsevier},\ \bibinfo {year} {2017})\ pp.\ \bibinfo {pages} {439--505}\BibitemShut {NoStop}%
\bibitem [{\citenamefont {Zhang}\ \emph {et~al.}(2024)\citenamefont {Zhang}, \citenamefont {Fang}, \citenamefont {Wang}, \citenamefont {Fang}, \citenamefont {Zhang}, \citenamefont {Guo},\ and\ \citenamefont {Tong}}]{zhang2024optical}%
  \BibitemOpen
  \bibfield  {author} {\bibinfo {author} {\bibfnamefont {J.}~\bibnamefont {Zhang}}, \bibinfo {author} {\bibfnamefont {H.}~\bibnamefont {Fang}}, \bibinfo {author} {\bibfnamefont {P.}~\bibnamefont {Wang}}, \bibinfo {author} {\bibfnamefont {W.}~\bibnamefont {Fang}}, \bibinfo {author} {\bibfnamefont {L.}~\bibnamefont {Zhang}}, \bibinfo {author} {\bibfnamefont {X.}~\bibnamefont {Guo}},\ and\ \bibinfo {author} {\bibfnamefont {L.}~\bibnamefont {Tong}},\ }\bibfield  {title} {\bibinfo {title} {Optical microfiber or nanofiber: a miniature fiber-optic platform for nanophotonics},\ }\href {https://www.spiedigitallibrary.org/journals/photonics-insights/volume-3/issue-1/R02/Optical-microfiber-or-nanofiber--a-miniature-fiber-optic-platform/10.3788/PI.2024.R02.short} {\bibfield  {journal} {\bibinfo  {journal} {Photonics Insights}\ }\textbf {\bibinfo {volume} {3}},\ \bibinfo {pages} {R02} (\bibinfo {year} {2024})}\BibitemShut {NoStop}%
\bibitem [{\citenamefont {Vetsch}\ \emph {et~al.}(2012)\citenamefont {Vetsch}, \citenamefont {Dawkins}, \citenamefont {Mitsch}, \citenamefont {Reitz}, \citenamefont {Schneeweiss},\ and\ \citenamefont {Rauschenbeutel}}]{vetsch2012nanofiber}%
  \BibitemOpen
  \bibfield  {author} {\bibinfo {author} {\bibfnamefont {E.}~\bibnamefont {Vetsch}}, \bibinfo {author} {\bibfnamefont {S.~T.}\ \bibnamefont {Dawkins}}, \bibinfo {author} {\bibfnamefont {R.}~\bibnamefont {Mitsch}}, \bibinfo {author} {\bibfnamefont {D.}~\bibnamefont {Reitz}}, \bibinfo {author} {\bibfnamefont {P.}~\bibnamefont {Schneeweiss}},\ and\ \bibinfo {author} {\bibfnamefont {A.}~\bibnamefont {Rauschenbeutel}},\ }\bibfield  {title} {\bibinfo {title} {Nanofiber-based optical trapping of cold neutral atoms},\ }\href {https://doi.org/10.1109/JSTQE.2012.2196025} {\bibfield  {journal} {\bibinfo  {journal} {IEEE Journal of Selected Topics in Quantum Electronics}\ }\textbf {\bibinfo {volume} {18}},\ \bibinfo {pages} {1763} (\bibinfo {year} {2012})}\BibitemShut {NoStop}%
\bibitem [{\citenamefont {Solano}\ \emph {et~al.}(2017{\natexlab{b}})\citenamefont {Solano}, \citenamefont {Barberis-Blostein}, \citenamefont {Fatemi}, \citenamefont {Orozco},\ and\ \citenamefont {Rolston}}]{solano_superradiance}%
  \BibitemOpen
  \bibfield  {author} {\bibinfo {author} {\bibfnamefont {P.}~\bibnamefont {Solano}}, \bibinfo {author} {\bibfnamefont {P.}~\bibnamefont {Barberis-Blostein}}, \bibinfo {author} {\bibfnamefont {F.~K.}\ \bibnamefont {Fatemi}}, \bibinfo {author} {\bibfnamefont {L.~A.}\ \bibnamefont {Orozco}},\ and\ \bibinfo {author} {\bibfnamefont {S.~L.}\ \bibnamefont {Rolston}},\ }\bibfield  {title} {\bibinfo {title} {Super-radiance reveals infinite-range dipole interactions through a nanofiber},\ }\href {https://doi.org/10.1038/s41467-017-01994-3} {\bibfield  {journal} {\bibinfo  {journal} {Nature Communications}\ }\textbf {\bibinfo {volume} {8}},\ \bibinfo {pages} {1857} (\bibinfo {year} {2017}{\natexlab{b}})}\BibitemShut {NoStop}%
\bibitem [{\citenamefont {Yalla}\ \emph {et~al.}(2012)\citenamefont {Yalla}, \citenamefont {Le~Kien}, \citenamefont {Morinaga},\ and\ \citenamefont {Hakuta}}]{yalla2012efficient}%
  \BibitemOpen
  \bibfield  {author} {\bibinfo {author} {\bibfnamefont {R.}~\bibnamefont {Yalla}}, \bibinfo {author} {\bibfnamefont {F.}~\bibnamefont {Le~Kien}}, \bibinfo {author} {\bibfnamefont {M.}~\bibnamefont {Morinaga}},\ and\ \bibinfo {author} {\bibfnamefont {K.}~\bibnamefont {Hakuta}},\ }\bibfield  {title} {\bibinfo {title} {Efficient channeling of fluorescence photons from single quantum dots into guided modes of optical nanofiber},\ }\href {https://link.aps.org/doi/10.1103/PhysRevLett.109.063602} {\bibfield  {journal} {\bibinfo  {journal} {Physical review letters}\ }\textbf {\bibinfo {volume} {109}},\ \bibinfo {pages} {063602} (\bibinfo {year} {2012})}\BibitemShut {NoStop}%
\bibitem [{\citenamefont {Liebermeister}\ \emph {et~al.}(2014)\citenamefont {Liebermeister}, \citenamefont {Petersen}, \citenamefont {Münchow}, \citenamefont {Burchardt}, \citenamefont {Hermelbracht}, \citenamefont {Tashima}, \citenamefont {Schell}, \citenamefont {Benson}, \citenamefont {Meinhardt}, \citenamefont {Krueger}, \citenamefont {Stiebeiner}, \citenamefont {Rauschenbeutel}, \citenamefont {Weinfurter},\ and\ \citenamefont {Weber}}]{diamond_nanofiber}%
  \BibitemOpen
  \bibfield  {author} {\bibinfo {author} {\bibfnamefont {L.}~\bibnamefont {Liebermeister}}, \bibinfo {author} {\bibfnamefont {F.}~\bibnamefont {Petersen}}, \bibinfo {author} {\bibfnamefont {A.~v.}\ \bibnamefont {Münchow}}, \bibinfo {author} {\bibfnamefont {D.}~\bibnamefont {Burchardt}}, \bibinfo {author} {\bibfnamefont {J.}~\bibnamefont {Hermelbracht}}, \bibinfo {author} {\bibfnamefont {T.}~\bibnamefont {Tashima}}, \bibinfo {author} {\bibfnamefont {A.~W.}\ \bibnamefont {Schell}}, \bibinfo {author} {\bibfnamefont {O.}~\bibnamefont {Benson}}, \bibinfo {author} {\bibfnamefont {T.}~\bibnamefont {Meinhardt}}, \bibinfo {author} {\bibfnamefont {A.}~\bibnamefont {Krueger}}, \bibinfo {author} {\bibfnamefont {A.}~\bibnamefont {Stiebeiner}}, \bibinfo {author} {\bibfnamefont {A.}~\bibnamefont {Rauschenbeutel}}, \bibinfo {author} {\bibfnamefont {H.}~\bibnamefont {Weinfurter}},\ and\ \bibinfo {author} {\bibfnamefont {M.}~\bibnamefont {Weber}},\ }\bibfield  {title} {\bibinfo {title} {{Tapered fiber coupling of single
  photons emitted by a deterministically positioned single nitrogen vacancy center}},\ }\href {https://doi.org/10.1063/1.4862207} {\bibfield  {journal} {\bibinfo  {journal} {Applied Physics Letters}\ }\textbf {\bibinfo {volume} {104}},\ \bibinfo {pages} {031101} (\bibinfo {year} {2014})}\BibitemShut {NoStop}%
\bibitem [{\citenamefont {Morrissey}\ \emph {et~al.}(2013)\citenamefont {Morrissey}, \citenamefont {Deasy}, \citenamefont {Frawley}, \citenamefont {Kumar}, \citenamefont {Prel}, \citenamefont {Russell}, \citenamefont {Truong},\ and\ \citenamefont {Nic~Chormaic}}]{morrissey_2013}%
  \BibitemOpen
  \bibfield  {author} {\bibinfo {author} {\bibfnamefont {M.~J.}\ \bibnamefont {Morrissey}}, \bibinfo {author} {\bibfnamefont {K.}~\bibnamefont {Deasy}}, \bibinfo {author} {\bibfnamefont {M.}~\bibnamefont {Frawley}}, \bibinfo {author} {\bibfnamefont {R.}~\bibnamefont {Kumar}}, \bibinfo {author} {\bibfnamefont {E.}~\bibnamefont {Prel}}, \bibinfo {author} {\bibfnamefont {L.}~\bibnamefont {Russell}}, \bibinfo {author} {\bibfnamefont {V.~G.}\ \bibnamefont {Truong}},\ and\ \bibinfo {author} {\bibfnamefont {S.}~\bibnamefont {Nic~Chormaic}},\ }\bibfield  {title} {\bibinfo {title} {Spectroscopy, manipulation and trapping of neutral atoms, molecules, and other particles using optical nanofibers: A review},\ }\href {https://doi.org/10.3390/s130810449} {\bibfield  {journal} {\bibinfo  {journal} {Sensors}\ }\textbf {\bibinfo {volume} {13}},\ \bibinfo {pages} {10449} (\bibinfo {year} {2013})}\BibitemShut {NoStop}%
\bibitem [{\citenamefont {Ding}\ \emph {et~al.}(2019)\citenamefont {Ding}, \citenamefont {Loo}, \citenamefont {Pigeon}, \citenamefont {Gautier}, \citenamefont {Joos}, \citenamefont {Wu}, \citenamefont {Giacobino}, \citenamefont {Bramati},\ and\ \citenamefont {Glorieux}}]{Ding_2019}%
  \BibitemOpen
  \bibfield  {author} {\bibinfo {author} {\bibfnamefont {C.}~\bibnamefont {Ding}}, \bibinfo {author} {\bibfnamefont {V.}~\bibnamefont {Loo}}, \bibinfo {author} {\bibfnamefont {S.}~\bibnamefont {Pigeon}}, \bibinfo {author} {\bibfnamefont {R.}~\bibnamefont {Gautier}}, \bibinfo {author} {\bibfnamefont {M.}~\bibnamefont {Joos}}, \bibinfo {author} {\bibfnamefont {E.}~\bibnamefont {Wu}}, \bibinfo {author} {\bibfnamefont {E.}~\bibnamefont {Giacobino}}, \bibinfo {author} {\bibfnamefont {A.}~\bibnamefont {Bramati}},\ and\ \bibinfo {author} {\bibfnamefont {Q.}~\bibnamefont {Glorieux}},\ }\bibfield  {title} {\bibinfo {title} {Fabrication and characterization of optical nanofiber interferometer and resonator for the visible range},\ }\href {https://doi.org/10.1088/1367-2630/ab31cc} {\bibfield  {journal} {\bibinfo  {journal} {New Journal of Physics}\ }\textbf {\bibinfo {volume} {21}},\ \bibinfo {pages} {073060} (\bibinfo {year} {2019})}\BibitemShut {NoStop}%
\bibitem [{\citenamefont {Yu}\ \emph {et~al.}(2022)\citenamefont {Yu}, \citenamefont {Yao}, \citenamefont {Pan}, \citenamefont {Fang}, \citenamefont {Li}, \citenamefont {Tong},\ and\ \citenamefont {Zhang}}]{yu2022highly}%
  \BibitemOpen
  \bibfield  {author} {\bibinfo {author} {\bibfnamefont {W.}~\bibnamefont {Yu}}, \bibinfo {author} {\bibfnamefont {N.}~\bibnamefont {Yao}}, \bibinfo {author} {\bibfnamefont {J.}~\bibnamefont {Pan}}, \bibinfo {author} {\bibfnamefont {W.}~\bibnamefont {Fang}}, \bibinfo {author} {\bibfnamefont {X.}~\bibnamefont {Li}}, \bibinfo {author} {\bibfnamefont {L.}~\bibnamefont {Tong}},\ and\ \bibinfo {author} {\bibfnamefont {L.}~\bibnamefont {Zhang}},\ }\bibfield  {title} {\bibinfo {title} {Highly sensitive and fast response strain sensor based on evanescently coupled micro/nanofibers},\ }\href {https://doi.org/10.29026/oea.2022.210101} {\bibfield  {journal} {\bibinfo  {journal} {Opto-Electronic Advances}\ }\textbf {\bibinfo {volume} {5}},\ \bibinfo {pages} {210101} (\bibinfo {year} {2022})}\BibitemShut {NoStop}%
\bibitem [{\citenamefont {Li}\ \emph {et~al.}(2018{\natexlab{b}})\citenamefont {Li}, \citenamefont {Zhang}, \citenamefont {Zhang}, \citenamefont {Liu}, \citenamefont {Zhang},\ and\ \citenamefont {Wei}}]{Li:18}%
  \BibitemOpen
  \bibfield  {author} {\bibinfo {author} {\bibfnamefont {K.}~\bibnamefont {Li}}, \bibinfo {author} {\bibfnamefont {N.}~\bibnamefont {Zhang}}, \bibinfo {author} {\bibfnamefont {N.~M.~Y.}\ \bibnamefont {Zhang}}, \bibinfo {author} {\bibfnamefont {G.}~\bibnamefont {Liu}}, \bibinfo {author} {\bibfnamefont {T.}~\bibnamefont {Zhang}},\ and\ \bibinfo {author} {\bibfnamefont {L.}~\bibnamefont {Wei}},\ }\bibfield  {title} {\bibinfo {title} {Ultrasensitive measurement of gas refractive index using an optical nanofiber coupler},\ }\href {https://doi.org/10.1364/OL.43.000679} {\bibfield  {journal} {\bibinfo  {journal} {Opt. Lett.}\ }\textbf {\bibinfo {volume} {43}},\ \bibinfo {pages} {679} (\bibinfo {year} {2018}{\natexlab{b}})}\BibitemShut {NoStop}%
\bibitem [{\citenamefont {Shao}\ \emph {et~al.}(2022)\citenamefont {Shao}, \citenamefont {Wu}, \citenamefont {Fang},\ and\ \citenamefont {Tong}}]{shao2022twin}%
  \BibitemOpen
  \bibfield  {author} {\bibinfo {author} {\bibfnamefont {L.}~\bibnamefont {Shao}}, \bibinfo {author} {\bibfnamefont {H.}~\bibnamefont {Wu}}, \bibinfo {author} {\bibfnamefont {W.}~\bibnamefont {Fang}},\ and\ \bibinfo {author} {\bibfnamefont {L.}~\bibnamefont {Tong}},\ }\bibfield  {title} {\bibinfo {title} {Twin-nanofiber structure for a highly efficient single-photon collection},\ }\href {https://doi.org/10.1364/OE.454616} {\bibfield  {journal} {\bibinfo  {journal} {Optics Express}\ }\textbf {\bibinfo {volume} {30}},\ \bibinfo {pages} {9147} (\bibinfo {year} {2022})}\BibitemShut {NoStop}%
\bibitem [{\citenamefont {Fussell}\ \emph {et~al.}(2004)\citenamefont {Fussell}, \citenamefont {McPhedran},\ and\ \citenamefont {de~Sterke}}]{fussell2004three}%
  \BibitemOpen
  \bibfield  {author} {\bibinfo {author} {\bibfnamefont {D.}~\bibnamefont {Fussell}}, \bibinfo {author} {\bibfnamefont {R.}~\bibnamefont {McPhedran}},\ and\ \bibinfo {author} {\bibfnamefont {C.~M.}\ \bibnamefont {de~Sterke}},\ }\bibfield  {title} {\bibinfo {title} {Three-dimensional green’s tensor, local density of states, and spontaneous emission in finite two-dimensional photonic crystals composed of cylinders},\ }\href {https://doi.org/10.1103/PhysRevE.70.066608} {\bibfield  {journal} {\bibinfo  {journal} {Physical Review E}\ }\textbf {\bibinfo {volume} {70}},\ \bibinfo {pages} {066608} (\bibinfo {year} {2004})}\BibitemShut {NoStop}%
\bibitem [{\citenamefont {Almokhtar}\ \emph {et~al.}(2014)\citenamefont {Almokhtar}, \citenamefont {Fujiwara}, \citenamefont {Takashima},\ and\ \citenamefont {Takeuchi}}]{almokhtar2014numerical}%
  \BibitemOpen
  \bibfield  {author} {\bibinfo {author} {\bibfnamefont {M.}~\bibnamefont {Almokhtar}}, \bibinfo {author} {\bibfnamefont {M.}~\bibnamefont {Fujiwara}}, \bibinfo {author} {\bibfnamefont {H.}~\bibnamefont {Takashima}},\ and\ \bibinfo {author} {\bibfnamefont {S.}~\bibnamefont {Takeuchi}},\ }\bibfield  {title} {\bibinfo {title} {Numerical simulations of nanodiamond nitrogen-vacancy centers coupled with tapered optical fibers as hybrid quantum nanophotonic devices},\ }\href {https://doi.org/10.1364/OE.22.020045} {\bibfield  {journal} {\bibinfo  {journal} {Optics express}\ }\textbf {\bibinfo {volume} {22}},\ \bibinfo {pages} {20045} (\bibinfo {year} {2014})}\BibitemShut {NoStop}%
\bibitem [{\citenamefont {Solano}\ \emph {et~al.}(2019)\citenamefont {Solano}, \citenamefont {Grover}, \citenamefont {Xu}, \citenamefont {Barberis-Blostein}, \citenamefont {Munday}, \citenamefont {Orozco}, \citenamefont {Phillips},\ and\ \citenamefont {Rolston}}]{solano2019alignment}%
  \BibitemOpen
  \bibfield  {author} {\bibinfo {author} {\bibfnamefont {P.}~\bibnamefont {Solano}}, \bibinfo {author} {\bibfnamefont {J.~A.}\ \bibnamefont {Grover}}, \bibinfo {author} {\bibfnamefont {Y.}~\bibnamefont {Xu}}, \bibinfo {author} {\bibfnamefont {P.}~\bibnamefont {Barberis-Blostein}}, \bibinfo {author} {\bibfnamefont {J.~N.}\ \bibnamefont {Munday}}, \bibinfo {author} {\bibfnamefont {L.~A.}\ \bibnamefont {Orozco}}, \bibinfo {author} {\bibfnamefont {W.~D.}\ \bibnamefont {Phillips}},\ and\ \bibinfo {author} {\bibfnamefont {S.~L.}\ \bibnamefont {Rolston}},\ }\bibfield  {title} {\bibinfo {title} {Alignment-dependent decay rate of an atomic dipole near an optical nanofiber},\ }\href {https://doi.org/10.1103/PhysRevA.99.013822} {\bibfield  {journal} {\bibinfo  {journal} {Physical Review A}\ }\textbf {\bibinfo {volume} {99}},\ \bibinfo {pages} {013822} (\bibinfo {year} {2019})}\BibitemShut {NoStop}%
\bibitem [{\citenamefont {Chang}\ \emph {et~al.}(2014)\citenamefont {Chang}, \citenamefont {Sinha}, \citenamefont {Taylor},\ and\ \citenamefont {Kimble}}]{chang2014}%
  \BibitemOpen
  \bibfield  {author} {\bibinfo {author} {\bibfnamefont {D.~E.}\ \bibnamefont {Chang}}, \bibinfo {author} {\bibfnamefont {K.}~\bibnamefont {Sinha}}, \bibinfo {author} {\bibfnamefont {J.~M.}\ \bibnamefont {Taylor}},\ and\ \bibinfo {author} {\bibfnamefont {H.~J.}\ \bibnamefont {Kimble}},\ }\bibfield  {title} {\bibinfo {title} {Trapping atoms using nanoscale quantum vacuum forces},\ }\href {https://doi.org/10.1038/ncomms5343} {\bibfield  {journal} {\bibinfo  {journal} {Nature Communications}\ }\textbf {\bibinfo {volume} {5}},\ \bibinfo {pages} {4343} (\bibinfo {year} {2014})}\BibitemShut {NoStop}%
\bibitem [{\citenamefont {Rajasree}\ \emph {et~al.}(2020)\citenamefont {Rajasree}, \citenamefont {Ray}, \citenamefont {Karlsson}, \citenamefont {Everett},\ and\ \citenamefont {Chormaic}}]{rajasree2020}%
  \BibitemOpen
  \bibfield  {author} {\bibinfo {author} {\bibfnamefont {K.~S.}\ \bibnamefont {Rajasree}}, \bibinfo {author} {\bibfnamefont {T.}~\bibnamefont {Ray}}, \bibinfo {author} {\bibfnamefont {K.}~\bibnamefont {Karlsson}}, \bibinfo {author} {\bibfnamefont {J.~L.}\ \bibnamefont {Everett}},\ and\ \bibinfo {author} {\bibfnamefont {S.~N.}\ \bibnamefont {Chormaic}},\ }\bibfield  {title} {\bibinfo {title} {Generation of cold {R}ydberg atoms at submicron distances from an optical nanofiber},\ }\href {https://doi.org/10.1103/PhysRevResearch.2.012038} {\bibfield  {journal} {\bibinfo  {journal} {Phys. Rev. Res.}\ }\textbf {\bibinfo {volume} {2}},\ \bibinfo {pages} {012038} (\bibinfo {year} {2020})}\BibitemShut {NoStop}%
\bibitem [{\citenamefont {Stourm}\ \emph {et~al.}(2020)\citenamefont {Stourm}, \citenamefont {Lepers}, \citenamefont {Robert}, \citenamefont {Chormaic}, \citenamefont {M{\o}lmer},\ and\ \citenamefont {Brion}}]{stourm2020spontaneous}%
  \BibitemOpen
  \bibfield  {author} {\bibinfo {author} {\bibfnamefont {E.}~\bibnamefont {Stourm}}, \bibinfo {author} {\bibfnamefont {M.}~\bibnamefont {Lepers}}, \bibinfo {author} {\bibfnamefont {J.}~\bibnamefont {Robert}}, \bibinfo {author} {\bibfnamefont {S.~N.}\ \bibnamefont {Chormaic}}, \bibinfo {author} {\bibfnamefont {K.}~\bibnamefont {M{\o}lmer}},\ and\ \bibinfo {author} {\bibfnamefont {E.}~\bibnamefont {Brion}},\ }\bibfield  {title} {\bibinfo {title} {Spontaneous emission and energy shifts of a {R}ydberg rubidium atom close to an optical nanofiber},\ }\href {https://doi.org/10.1103/PhysRevA.101.052508} {\bibfield  {journal} {\bibinfo  {journal} {Physical Review A}\ }\textbf {\bibinfo {volume} {101}},\ \bibinfo {pages} {052508} (\bibinfo {year} {2020})}\BibitemShut {NoStop}%
\bibitem [{\citenamefont {Novotny}\ and\ \citenamefont {Hecht}(2012)}]{novotny2012principles}%
  \BibitemOpen
  \bibfield  {author} {\bibinfo {author} {\bibfnamefont {L.}~\bibnamefont {Novotny}}\ and\ \bibinfo {author} {\bibfnamefont {B.}~\bibnamefont {Hecht}},\ }\href {https://doi.org/10.1017/CBO9780511813535} {\emph {\bibinfo {title} {Principles of nano-optics}}}\ (\bibinfo  {publisher} {Cambridge university press},\ \bibinfo {year} {2012})\BibitemShut {NoStop}%
\bibitem [{\citenamefont {White}\ \emph {et~al.}(2002)\citenamefont {White}, \citenamefont {Kuhlmey}, \citenamefont {McPhedran}, \citenamefont {Maystre}, \citenamefont {Renversez}, \citenamefont {De~Sterke},\ and\ \citenamefont {Botten}}]{white2002multipole}%
  \BibitemOpen
  \bibfield  {author} {\bibinfo {author} {\bibfnamefont {T.}~\bibnamefont {White}}, \bibinfo {author} {\bibfnamefont {B.}~\bibnamefont {Kuhlmey}}, \bibinfo {author} {\bibfnamefont {R.}~\bibnamefont {McPhedran}}, \bibinfo {author} {\bibfnamefont {D.}~\bibnamefont {Maystre}}, \bibinfo {author} {\bibfnamefont {G.}~\bibnamefont {Renversez}}, \bibinfo {author} {\bibfnamefont {C.~M.}\ \bibnamefont {De~Sterke}},\ and\ \bibinfo {author} {\bibfnamefont {L.}~\bibnamefont {Botten}},\ }\bibfield  {title} {\bibinfo {title} {Multipole method for microstructured optical fibers. {I}. formulation},\ }\href {https://doi.org/10.1364/JOSAB.19.002322} {\bibfield  {journal} {\bibinfo  {journal} {JOSA B}\ }\textbf {\bibinfo {volume} {19}},\ \bibinfo {pages} {2322} (\bibinfo {year} {2002})}\BibitemShut {NoStop}%
\bibitem [{\citenamefont {Chew}(1999)}]{chew1999waves}%
  \BibitemOpen
  \bibfield  {author} {\bibinfo {author} {\bibfnamefont {W.~C.}\ \bibnamefont {Chew}},\ }\href {https://ieeexplore.ieee.org/servlet/opac?bknumber=5270998} {\emph {\bibinfo {title} {Waves and fields in inhomogenous media}}},\ Vol.~\bibinfo {volume} {16}\ (\bibinfo  {publisher} {John Wiley \& Sons},\ \bibinfo {year} {1999})\BibitemShut {NoStop}%
\bibitem [{\citenamefont {Klimov}\ and\ \citenamefont {Ducloy}(2004)}]{klimov2004spontaneous}%
  \BibitemOpen
  \bibfield  {author} {\bibinfo {author} {\bibfnamefont {V.~V.}\ \bibnamefont {Klimov}}\ and\ \bibinfo {author} {\bibfnamefont {M.}~\bibnamefont {Ducloy}},\ }\bibfield  {title} {\bibinfo {title} {Spontaneous emission rate of an excited atom placed near a nanofiber},\ }\href {https://doi.org/10.1103/PhysRevA.69.013812} {\bibfield  {journal} {\bibinfo  {journal} {Physical Review A}\ }\textbf {\bibinfo {volume} {69}},\ \bibinfo {pages} {013812} (\bibinfo {year} {2004})}\BibitemShut {NoStop}%
\bibitem [{\citenamefont {Le~Kien}\ \emph {et~al.}(2021)\citenamefont {Le~Kien}, \citenamefont {Ruks}, \citenamefont {Chormaic},\ and\ \citenamefont {Busch}}]{le2021spatial}%
  \BibitemOpen
  \bibfield  {author} {\bibinfo {author} {\bibfnamefont {F.}~\bibnamefont {Le~Kien}}, \bibinfo {author} {\bibfnamefont {L.}~\bibnamefont {Ruks}}, \bibinfo {author} {\bibfnamefont {S.~N.}\ \bibnamefont {Chormaic}},\ and\ \bibinfo {author} {\bibfnamefont {T.}~\bibnamefont {Busch}},\ }\bibfield  {title} {\bibinfo {title} {Spatial distributions of the fields in guided normal modes of two coupled parallel optical nanofibers},\ }\href {https://iopscience.iop.org/article/10.1088/1367-2630/abea44/meta} {\bibfield  {journal} {\bibinfo  {journal} {New Journal of Physics}\ }\textbf {\bibinfo {volume} {23}},\ \bibinfo {pages} {043006} (\bibinfo {year} {2021})}\BibitemShut {NoStop}%
\bibitem [{\citenamefont {Marcuse}(1982)}]{marcuse1982light}%
  \BibitemOpen
  \bibfield  {author} {\bibinfo {author} {\bibfnamefont {D.}~\bibnamefont {Marcuse}},\ }\bibfield  {title} {\bibinfo {title} {Light transmission optics},\ }\href@noop {} {\bibfield  {journal} {\bibinfo  {journal} {New York}\ } (\bibinfo {year} {1982})}\BibitemShut {NoStop}%
\bibitem [{\citenamefont {Daly}\ \emph {et~al.}(2014)\citenamefont {Daly}, \citenamefont {Truong}, \citenamefont {Phelan}, \citenamefont {Deasy},\ and\ \citenamefont {Chormaic}}]{daly2014nanostructured}%
  \BibitemOpen
  \bibfield  {author} {\bibinfo {author} {\bibfnamefont {M.}~\bibnamefont {Daly}}, \bibinfo {author} {\bibfnamefont {V.~G.}\ \bibnamefont {Truong}}, \bibinfo {author} {\bibfnamefont {C.}~\bibnamefont {Phelan}}, \bibinfo {author} {\bibfnamefont {K.}~\bibnamefont {Deasy}},\ and\ \bibinfo {author} {\bibfnamefont {S.~N.}\ \bibnamefont {Chormaic}},\ }\bibfield  {title} {\bibinfo {title} {Nanostructured optical nanofibres for atom trapping},\ }\href {https://iopscience.iop.org/article/10.1088/1367-2630/16/5/053052/meta} {\bibfield  {journal} {\bibinfo  {journal} {New Journal of Physics}\ }\textbf {\bibinfo {volume} {16}},\ \bibinfo {pages} {053052} (\bibinfo {year} {2014})}\BibitemShut {NoStop}%
\bibitem [{\citenamefont {Kien}\ and\ \citenamefont {Hakuta}(2008)}]{Kien2008}%
  \BibitemOpen
  \bibfield  {author} {\bibinfo {author} {\bibfnamefont {F.~L.}\ \bibnamefont {Kien}}\ and\ \bibinfo {author} {\bibfnamefont {K.}~\bibnamefont {Hakuta}},\ }\bibfield  {title} {\bibinfo {title} {Cooperative enhancement of channeling of emission from atoms into a nanofiber},\ }\href {https://doi.org/10.1103/physreva.77.013801} {\bibfield  {journal} {\bibinfo  {journal} {Phys. Rev. A}\ }\textbf {\bibinfo {volume} {77}} (\bibinfo {year} {2008})}\BibitemShut {NoStop}%
\bibitem [{\citenamefont {Maeda}\ \emph {et~al.}(2023)\citenamefont {Maeda}, \citenamefont {Keloth},\ and\ \citenamefont {Chormaic}}]{Maeda:23}%
  \BibitemOpen
  \bibfield  {author} {\bibinfo {author} {\bibfnamefont {M.}~\bibnamefont {Maeda}}, \bibinfo {author} {\bibfnamefont {J.}~\bibnamefont {Keloth}},\ and\ \bibinfo {author} {\bibfnamefont {S.~N.}\ \bibnamefont {Chormaic}},\ }\bibfield  {title} {\bibinfo {title} {Manipulation of polarization topology using a {F}abry-{P}\'erot fiber cavity with a higher-order mode optical nanofiber},\ }\href {https://doi.org/10.1364/PRJ.486373} {\bibfield  {journal} {\bibinfo  {journal} {Photon. Res.}\ }\textbf {\bibinfo {volume} {11}},\ \bibinfo {pages} {1029} (\bibinfo {year} {2023})}\BibitemShut {NoStop}%
\bibitem [{\citenamefont {Kato}\ and\ \citenamefont {Aoki}(2015)}]{kato2015strong}%
  \BibitemOpen
  \bibfield  {author} {\bibinfo {author} {\bibfnamefont {S.}~\bibnamefont {Kato}}\ and\ \bibinfo {author} {\bibfnamefont {T.}~\bibnamefont {Aoki}},\ }\bibfield  {title} {\bibinfo {title} {Strong coupling between a trapped single atom and an all-fiber cavity},\ }\href {https://doi.org/10.1103/PhysRevLett.115.093603} {\bibfield  {journal} {\bibinfo  {journal} {Physical review letters}\ }\textbf {\bibinfo {volume} {115}},\ \bibinfo {pages} {093603} (\bibinfo {year} {2015})}\BibitemShut {NoStop}%
\bibitem [{\citenamefont {Kato}\ \emph {et~al.}(2019)\citenamefont {Kato}, \citenamefont {N{\'e}met}, \citenamefont {Senga}, \citenamefont {Mizukami}, \citenamefont {Huang}, \citenamefont {Parkins},\ and\ \citenamefont {Aoki}}]{kato2019observation}%
  \BibitemOpen
  \bibfield  {author} {\bibinfo {author} {\bibfnamefont {S.}~\bibnamefont {Kato}}, \bibinfo {author} {\bibfnamefont {N.}~\bibnamefont {N{\'e}met}}, \bibinfo {author} {\bibfnamefont {K.}~\bibnamefont {Senga}}, \bibinfo {author} {\bibfnamefont {S.}~\bibnamefont {Mizukami}}, \bibinfo {author} {\bibfnamefont {X.}~\bibnamefont {Huang}}, \bibinfo {author} {\bibfnamefont {S.}~\bibnamefont {Parkins}},\ and\ \bibinfo {author} {\bibfnamefont {T.}~\bibnamefont {Aoki}},\ }\bibfield  {title} {\bibinfo {title} {Observation of dressed states of distant atoms with delocalized photons in coupled-cavities quantum electrodynamics},\ }\href {https://doi.org/10.1038/s41467-019-08975-8} {\bibfield  {journal} {\bibinfo  {journal} {Nature communications}\ }\textbf {\bibinfo {volume} {10}},\ \bibinfo {pages} {1160} (\bibinfo {year} {2019})}\BibitemShut {NoStop}%
\bibitem [{\citenamefont {Vaidya}\ \emph {et~al.}(2018)\citenamefont {Vaidya}, \citenamefont {Guo}, \citenamefont {Kroeze}, \citenamefont {Ballantine}, \citenamefont {Koll\'ar}, \citenamefont {Keeling},\ and\ \citenamefont {Lev}}]{vaidya_2018}%
  \BibitemOpen
  \bibfield  {author} {\bibinfo {author} {\bibfnamefont {V.~D.}\ \bibnamefont {Vaidya}}, \bibinfo {author} {\bibfnamefont {Y.}~\bibnamefont {Guo}}, \bibinfo {author} {\bibfnamefont {R.~M.}\ \bibnamefont {Kroeze}}, \bibinfo {author} {\bibfnamefont {K.~E.}\ \bibnamefont {Ballantine}}, \bibinfo {author} {\bibfnamefont {A.~J.}\ \bibnamefont {Koll\'ar}}, \bibinfo {author} {\bibfnamefont {J.}~\bibnamefont {Keeling}},\ and\ \bibinfo {author} {\bibfnamefont {B.~L.}\ \bibnamefont {Lev}},\ }\bibfield  {title} {\bibinfo {title} {Tunable-range, photon-mediated atomic interactions in multimode cavity {QED}},\ }\href {https://doi.org/10.1103/PhysRevX.8.011002} {\bibfield  {journal} {\bibinfo  {journal} {Phys. Rev. X}\ }\textbf {\bibinfo {volume} {8}},\ \bibinfo {pages} {011002} (\bibinfo {year} {2018})}\BibitemShut {NoStop}%
\bibitem [{\citenamefont {Wickenbrock}\ \emph {et~al.}(2013)\citenamefont {Wickenbrock}, \citenamefont {Hemmerling}, \citenamefont {Robb}, \citenamefont {Emary},\ and\ \citenamefont {Renzoni}}]{multimodecqed_robb}%
  \BibitemOpen
  \bibfield  {author} {\bibinfo {author} {\bibfnamefont {A.}~\bibnamefont {Wickenbrock}}, \bibinfo {author} {\bibfnamefont {M.}~\bibnamefont {Hemmerling}}, \bibinfo {author} {\bibfnamefont {G.~R.~M.}\ \bibnamefont {Robb}}, \bibinfo {author} {\bibfnamefont {C.}~\bibnamefont {Emary}},\ and\ \bibinfo {author} {\bibfnamefont {F.}~\bibnamefont {Renzoni}},\ }\bibfield  {title} {\bibinfo {title} {Collective strong coupling in multimode cavity {QED}},\ }\href {https://doi.org/10.1103/PhysRevA.87.043817} {\bibfield  {journal} {\bibinfo  {journal} {Phys. Rev. A}\ }\textbf {\bibinfo {volume} {87}},\ \bibinfo {pages} {043817} (\bibinfo {year} {2013})}\BibitemShut {NoStop}%
\bibitem [{\citenamefont {Liu}\ \emph {et~al.}(2018)\citenamefont {Liu}, \citenamefont {Brash}, \citenamefont {O'Hara}, \citenamefont {Martins}, \citenamefont {Phillips}, \citenamefont {Coles}, \citenamefont {Royall}, \citenamefont {Clarke}, \citenamefont {Bentham}, \citenamefont {Prtljaga}, \citenamefont {Itskevich}, \citenamefont {Wilson}, \citenamefont {Skolnick},\ and\ \citenamefont {Fox}}]{purcell2018}%
  \BibitemOpen
  \bibfield  {author} {\bibinfo {author} {\bibfnamefont {F.}~\bibnamefont {Liu}}, \bibinfo {author} {\bibfnamefont {A.~J.}\ \bibnamefont {Brash}}, \bibinfo {author} {\bibfnamefont {J.}~\bibnamefont {O'Hara}}, \bibinfo {author} {\bibfnamefont {L.~M. P.~P.}\ \bibnamefont {Martins}}, \bibinfo {author} {\bibfnamefont {C.~L.}\ \bibnamefont {Phillips}}, \bibinfo {author} {\bibfnamefont {R.~J.}\ \bibnamefont {Coles}}, \bibinfo {author} {\bibfnamefont {B.}~\bibnamefont {Royall}}, \bibinfo {author} {\bibfnamefont {E.}~\bibnamefont {Clarke}}, \bibinfo {author} {\bibfnamefont {C.}~\bibnamefont {Bentham}}, \bibinfo {author} {\bibfnamefont {N.}~\bibnamefont {Prtljaga}}, \bibinfo {author} {\bibfnamefont {I.~E.}\ \bibnamefont {Itskevich}}, \bibinfo {author} {\bibfnamefont {L.~R.}\ \bibnamefont {Wilson}}, \bibinfo {author} {\bibfnamefont {M.~S.}\ \bibnamefont {Skolnick}},\ and\ \bibinfo {author} {\bibfnamefont {A.~M.}\ \bibnamefont {Fox}},\ }\bibfield  {title} {\bibinfo {title} {High {P}urcell factor generation of
  indistinguishable on-chip single photons},\ }\href {https://doi.org/10.1038/s41565-018-0188-x} {\bibfield  {journal} {\bibinfo  {journal} {Nature Nanotechnology}\ }\textbf {\bibinfo {volume} {13}},\ \bibinfo {pages} {835} (\bibinfo {year} {2018})}\BibitemShut {NoStop}%
\bibitem [{\citenamefont {Le~Kien}\ \emph {et~al.}(2005{\natexlab{a}})\citenamefont {Le~Kien}, \citenamefont {Gupta}, \citenamefont {Balykin},\ and\ \citenamefont {Hakuta}}]{le2005spontaneous}%
  \BibitemOpen
  \bibfield  {author} {\bibinfo {author} {\bibfnamefont {F.}~\bibnamefont {Le~Kien}}, \bibinfo {author} {\bibfnamefont {S.~D.}\ \bibnamefont {Gupta}}, \bibinfo {author} {\bibfnamefont {V.}~\bibnamefont {Balykin}},\ and\ \bibinfo {author} {\bibfnamefont {K.}~\bibnamefont {Hakuta}},\ }\bibfield  {title} {\bibinfo {title} {Spontaneous emission of a cesium atom near a nanofiber: Efficient coupling of light to guided modes},\ }\href {https://doi.org/10.1103/PhysRevA.72.032509} {\bibfield  {journal} {\bibinfo  {journal} {Physical Review A}\ }\textbf {\bibinfo {volume} {72}},\ \bibinfo {pages} {032509} (\bibinfo {year} {2005}{\natexlab{a}})}\BibitemShut {NoStop}%
\bibitem [{\citenamefont {Fatemi}\ \emph {et~al.}(2017)\citenamefont {Fatemi}, \citenamefont {Hoffman}, \citenamefont {Solano}, \citenamefont {Fenton}, \citenamefont {Beadie}, \citenamefont {Rolston},\ and\ \citenamefont {Orozco}}]{Fatemi:17}%
  \BibitemOpen
  \bibfield  {author} {\bibinfo {author} {\bibfnamefont {F.~K.}\ \bibnamefont {Fatemi}}, \bibinfo {author} {\bibfnamefont {J.~E.}\ \bibnamefont {Hoffman}}, \bibinfo {author} {\bibfnamefont {P.}~\bibnamefont {Solano}}, \bibinfo {author} {\bibfnamefont {E.~F.}\ \bibnamefont {Fenton}}, \bibinfo {author} {\bibfnamefont {G.}~\bibnamefont {Beadie}}, \bibinfo {author} {\bibfnamefont {S.~L.}\ \bibnamefont {Rolston}},\ and\ \bibinfo {author} {\bibfnamefont {L.~A.}\ \bibnamefont {Orozco}},\ }\bibfield  {title} {\bibinfo {title} {Modal interference in optical nanofibers for sub-angstrom radius sensitivity},\ }\href {https://doi.org/10.1364/OPTICA.4.000157} {\bibfield  {journal} {\bibinfo  {journal} {Optica}\ }\textbf {\bibinfo {volume} {4}},\ \bibinfo {pages} {157} (\bibinfo {year} {2017})}\BibitemShut {NoStop}%
\bibitem [{\citenamefont {Nayak}\ \emph {et~al.}(2019)\citenamefont {Nayak}, \citenamefont {Wang},\ and\ \citenamefont {Keloth}}]{nayak2019real}%
  \BibitemOpen
  \bibfield  {author} {\bibinfo {author} {\bibfnamefont {K.~P.}\ \bibnamefont {Nayak}}, \bibinfo {author} {\bibfnamefont {J.}~\bibnamefont {Wang}},\ and\ \bibinfo {author} {\bibfnamefont {J.}~\bibnamefont {Keloth}},\ }\bibfield  {title} {\bibinfo {title} {Real-time observation of single atoms trapped and interfaced to a nanofiber cavity},\ }\href {https://doi.org/10.1103/PhysRevLett.123.213602} {\bibfield  {journal} {\bibinfo  {journal} {Physical Review Letters}\ }\textbf {\bibinfo {volume} {123}},\ \bibinfo {pages} {213602} (\bibinfo {year} {2019})}\BibitemShut {NoStop}%
\bibitem [{\citenamefont {Manga~Rao}\ and\ \citenamefont {Hughes}(2007)}]{mrao_2007}%
  \BibitemOpen
  \bibfield  {author} {\bibinfo {author} {\bibfnamefont {V.~S.~C.}\ \bibnamefont {Manga~Rao}}\ and\ \bibinfo {author} {\bibfnamefont {S.}~\bibnamefont {Hughes}},\ }\bibfield  {title} {\bibinfo {title} {Single quantum-dot purcell factor and $\ensuremath{\beta}$ factor in a photonic crystal waveguide},\ }\href {https://doi.org/10.1103/PhysRevB.75.205437} {\bibfield  {journal} {\bibinfo  {journal} {Phys. Rev. B}\ }\textbf {\bibinfo {volume} {75}},\ \bibinfo {pages} {205437} (\bibinfo {year} {2007})}\BibitemShut {NoStop}%
\bibitem [{\citenamefont {Le~Kien}\ \emph {et~al.}(2005{\natexlab{b}})\citenamefont {Le~Kien}, \citenamefont {Gupta}, \citenamefont {Nayak},\ and\ \citenamefont {Hakuta}}]{fam_2005}%
  \BibitemOpen
  \bibfield  {author} {\bibinfo {author} {\bibfnamefont {F.}~\bibnamefont {Le~Kien}}, \bibinfo {author} {\bibfnamefont {S.~D.}\ \bibnamefont {Gupta}}, \bibinfo {author} {\bibfnamefont {K.~P.}\ \bibnamefont {Nayak}},\ and\ \bibinfo {author} {\bibfnamefont {K.}~\bibnamefont {Hakuta}},\ }\bibfield  {title} {\bibinfo {title} {Nanofiber-mediated radiative transfer between two distant atoms},\ }\href {https://doi.org/10.1103/PhysRevA.72.063815} {\bibfield  {journal} {\bibinfo  {journal} {Phys. Rev. A}\ }\textbf {\bibinfo {volume} {72}},\ \bibinfo {pages} {063815} (\bibinfo {year} {2005}{\natexlab{b}})}\BibitemShut {NoStop}%
\bibitem [{\citenamefont {Svendsen}\ and\ \citenamefont {Olmos}(2023)}]{svendsen2023modified}%
  \BibitemOpen
  \bibfield  {author} {\bibinfo {author} {\bibfnamefont {M.~B.}\ \bibnamefont {Svendsen}}\ and\ \bibinfo {author} {\bibfnamefont {B.}~\bibnamefont {Olmos}},\ }\bibfield  {title} {\bibinfo {title} {Modified dipole-dipole interactions in the presence of a nanophotonic waveguide},\ }\href {https://doi.org/10.22331/q-2023-08-22-1091} {\bibfield  {journal} {\bibinfo  {journal} {Quantum}\ }\textbf {\bibinfo {volume} {7}},\ \bibinfo {pages} {1091} (\bibinfo {year} {2023})}\BibitemShut {NoStop}%
\bibitem [{\citenamefont {Zheng}\ and\ \citenamefont {Baranger}(2013)}]{zheng2013entanglement}%
  \BibitemOpen
  \bibfield  {author} {\bibinfo {author} {\bibfnamefont {H.}~\bibnamefont {Zheng}}\ and\ \bibinfo {author} {\bibfnamefont {H.~U.}\ \bibnamefont {Baranger}},\ }\bibfield  {title} {\bibinfo {title} {Persistent quantum beats and long-distance entanglement from waveguide-mediated interactions},\ }\href {https://doi.org/10.1103/PhysRevLett.110.113601} {\bibfield  {journal} {\bibinfo  {journal} {Phys. Rev. Lett.}\ }\textbf {\bibinfo {volume} {110}},\ \bibinfo {pages} {113601} (\bibinfo {year} {2013})}\BibitemShut {NoStop}%
\bibitem [{\citenamefont {Gonzalez-Tudela}\ \emph {et~al.}(2011)\citenamefont {Gonzalez-Tudela}, \citenamefont {Martin-Cano}, \citenamefont {Moreno}, \citenamefont {Martin-Moreno}, \citenamefont {Tejedor},\ and\ \citenamefont {Garcia-Vidal}}]{tudela2011entanglement}%
  \BibitemOpen
  \bibfield  {author} {\bibinfo {author} {\bibfnamefont {A.}~\bibnamefont {Gonzalez-Tudela}}, \bibinfo {author} {\bibfnamefont {D.}~\bibnamefont {Martin-Cano}}, \bibinfo {author} {\bibfnamefont {E.}~\bibnamefont {Moreno}}, \bibinfo {author} {\bibfnamefont {L.}~\bibnamefont {Martin-Moreno}}, \bibinfo {author} {\bibfnamefont {C.}~\bibnamefont {Tejedor}},\ and\ \bibinfo {author} {\bibfnamefont {F.~J.}\ \bibnamefont {Garcia-Vidal}},\ }\bibfield  {title} {\bibinfo {title} {Entanglement of two qubits mediated by one-dimensional plasmonic waveguides},\ }\href {https://doi.org/10.1103/PhysRevLett.106.020501} {\bibfield  {journal} {\bibinfo  {journal} {Phys. Rev. Lett.}\ }\textbf {\bibinfo {volume} {106}},\ \bibinfo {pages} {020501} (\bibinfo {year} {2011})}\BibitemShut {NoStop}%
\bibitem [{\citenamefont {Wootters}(1998)}]{wooters1998entanglement}%
  \BibitemOpen
  \bibfield  {author} {\bibinfo {author} {\bibfnamefont {W.~K.}\ \bibnamefont {Wootters}},\ }\bibfield  {title} {\bibinfo {title} {Entanglement of formation of an arbitrary state of two qubits},\ }\href {https://doi.org/10.1103/PhysRevLett.80.2245} {\bibfield  {journal} {\bibinfo  {journal} {Phys. Rev. Lett.}\ }\textbf {\bibinfo {volume} {80}},\ \bibinfo {pages} {2245} (\bibinfo {year} {1998})}\BibitemShut {NoStop}%
\bibitem [{\citenamefont {Gonzalez-Ballestero}\ \emph {et~al.}(2014)\citenamefont {Gonzalez-Ballestero}, \citenamefont {Moreno},\ and\ \citenamefont {Garcia-Vidal}}]{ballestero2014entanglement}%
  \BibitemOpen
  \bibfield  {author} {\bibinfo {author} {\bibfnamefont {C.}~\bibnamefont {Gonzalez-Ballestero}}, \bibinfo {author} {\bibfnamefont {E.}~\bibnamefont {Moreno}},\ and\ \bibinfo {author} {\bibfnamefont {F.~J.}\ \bibnamefont {Garcia-Vidal}},\ }\bibfield  {title} {\bibinfo {title} {Generation, manipulation, and detection of two-qubit entanglement in waveguide {QED}},\ }\href {https://doi.org/10.1103/PhysRevA.89.042328} {\bibfield  {journal} {\bibinfo  {journal} {Phys. Rev. A}\ }\textbf {\bibinfo {volume} {89}},\ \bibinfo {pages} {042328} (\bibinfo {year} {2014})}\BibitemShut {NoStop}%
\bibitem [{\citenamefont {Kr{\"a}mer}\ \emph {et~al.}(2018)\citenamefont {Kr{\"a}mer}, \citenamefont {Plankensteiner}, \citenamefont {Ostermann},\ and\ \citenamefont {Ritsch}}]{kramer_qoptics}%
  \BibitemOpen
  \bibfield  {author} {\bibinfo {author} {\bibfnamefont {S.}~\bibnamefont {Kr{\"a}mer}}, \bibinfo {author} {\bibfnamefont {D.}~\bibnamefont {Plankensteiner}}, \bibinfo {author} {\bibfnamefont {L.}~\bibnamefont {Ostermann}},\ and\ \bibinfo {author} {\bibfnamefont {H.}~\bibnamefont {Ritsch}},\ }\bibfield  {title} {\bibinfo {title} {Quantumoptics. jl: A {J}ulia framework for simulating open quantum systems},\ }\href {https://doi.org/10.1016/j.cpc.2018.02.004} {\bibfield  {journal} {\bibinfo  {journal} {Computer Physics Communications}\ }\textbf {\bibinfo {volume} {227}},\ \bibinfo {pages} {109} (\bibinfo {year} {2018})}\BibitemShut {NoStop}%
\bibitem [{\citenamefont {Bennett}\ \emph {et~al.}(1996)\citenamefont {Bennett}, \citenamefont {Bernstein}, \citenamefont {Popescu},\ and\ \citenamefont {Schumacher}}]{distill1996}%
  \BibitemOpen
  \bibfield  {author} {\bibinfo {author} {\bibfnamefont {C.~H.}\ \bibnamefont {Bennett}}, \bibinfo {author} {\bibfnamefont {H.~J.}\ \bibnamefont {Bernstein}}, \bibinfo {author} {\bibfnamefont {S.}~\bibnamefont {Popescu}},\ and\ \bibinfo {author} {\bibfnamefont {B.}~\bibnamefont {Schumacher}},\ }\bibfield  {title} {\bibinfo {title} {Concentrating partial entanglement by local operations},\ }\href {https://doi.org/10.1103/PhysRevA.53.2046} {\bibfield  {journal} {\bibinfo  {journal} {Phys. Rev. A}\ }\textbf {\bibinfo {volume} {53}},\ \bibinfo {pages} {2046} (\bibinfo {year} {1996})}\BibitemShut {NoStop}%
\bibitem [{\citenamefont {Davidson-Marquis}\ \emph {et~al.}(2021{\natexlab{b}})\citenamefont {Davidson-Marquis}, \citenamefont {Gargiulo}, \citenamefont {G{\'o}mez-L{\'o}pez}, \citenamefont {Jang}, \citenamefont {Kroh}, \citenamefont {M{\"u}ller}, \citenamefont {Ziegler}, \citenamefont {Maier}, \citenamefont {K{\"u}bler}, \citenamefont {Schmidt},\ and\ \citenamefont {Benson}}]{lightcage}%
  \BibitemOpen
  \bibfield  {author} {\bibinfo {author} {\bibfnamefont {F.}~\bibnamefont {Davidson-Marquis}}, \bibinfo {author} {\bibfnamefont {J.}~\bibnamefont {Gargiulo}}, \bibinfo {author} {\bibfnamefont {E.}~\bibnamefont {G{\'o}mez-L{\'o}pez}}, \bibinfo {author} {\bibfnamefont {B.}~\bibnamefont {Jang}}, \bibinfo {author} {\bibfnamefont {T.}~\bibnamefont {Kroh}}, \bibinfo {author} {\bibfnamefont {C.}~\bibnamefont {M{\"u}ller}}, \bibinfo {author} {\bibfnamefont {M.}~\bibnamefont {Ziegler}}, \bibinfo {author} {\bibfnamefont {S.~A.}\ \bibnamefont {Maier}}, \bibinfo {author} {\bibfnamefont {H.}~\bibnamefont {K{\"u}bler}}, \bibinfo {author} {\bibfnamefont {M.~A.}\ \bibnamefont {Schmidt}},\ and\ \bibinfo {author} {\bibfnamefont {O.}~\bibnamefont {Benson}},\ }\bibfield  {title} {\bibinfo {title} {Coherent interaction of atoms with a beam of light confined in a light cage},\ }\href {https://doi.org/10.1038/s41377-021-00556-z} {\bibfield  {journal} {\bibinfo  {journal} {Light: Science \& Applications}\ }\textbf {\bibinfo {volume}
  {10}},\ \bibinfo {pages} {114} (\bibinfo {year} {2021}{\natexlab{b}})}\BibitemShut {NoStop}%
\bibitem [{\citenamefont {Dzsotjan}\ \emph {et~al.}(2010)\citenamefont {Dzsotjan}, \citenamefont {S\o{}rensen},\ and\ \citenamefont {Fleischhauer}}]{nanowire}%
  \BibitemOpen
  \bibfield  {author} {\bibinfo {author} {\bibfnamefont {D.}~\bibnamefont {Dzsotjan}}, \bibinfo {author} {\bibfnamefont {A.~S.}\ \bibnamefont {S\o{}rensen}},\ and\ \bibinfo {author} {\bibfnamefont {M.}~\bibnamefont {Fleischhauer}},\ }\bibfield  {title} {\bibinfo {title} {Quantum emitters coupled to surface plasmons of a nanowire: A {G}reen's function approach},\ }\href {https://doi.org/10.1103/PhysRevB.82.075427} {\bibfield  {journal} {\bibinfo  {journal} {Phys. Rev. B}\ }\textbf {\bibinfo {volume} {82}},\ \bibinfo {pages} {075427} (\bibinfo {year} {2010})}\BibitemShut {NoStop}%
\bibitem [{\citenamefont {Blum}\ \emph {et~al.}(2012)\citenamefont {Blum}, \citenamefont {Zijlstra}, \citenamefont {Lagendijk}, \citenamefont {Wubs}, \citenamefont {Mosk}, \citenamefont {Subramaniam},\ and\ \citenamefont {Vos}}]{blum2012nanophotonic}%
  \BibitemOpen
  \bibfield  {author} {\bibinfo {author} {\bibfnamefont {C.}~\bibnamefont {Blum}}, \bibinfo {author} {\bibfnamefont {N.}~\bibnamefont {Zijlstra}}, \bibinfo {author} {\bibfnamefont {A.}~\bibnamefont {Lagendijk}}, \bibinfo {author} {\bibfnamefont {M.}~\bibnamefont {Wubs}}, \bibinfo {author} {\bibfnamefont {A.~P.}\ \bibnamefont {Mosk}}, \bibinfo {author} {\bibfnamefont {V.}~\bibnamefont {Subramaniam}},\ and\ \bibinfo {author} {\bibfnamefont {W.~L.}\ \bibnamefont {Vos}},\ }\bibfield  {title} {\bibinfo {title} {Nanophotonic control of the f{\"o}rster resonance energy transfer efficiency},\ }\href {https://doi.org/10.1103/PhysRevLett.109.203601} {\bibfield  {journal} {\bibinfo  {journal} {Physical review letters}\ }\textbf {\bibinfo {volume} {109}},\ \bibinfo {pages} {203601} (\bibinfo {year} {2012})}\BibitemShut {NoStop}%
\bibitem [{\citenamefont {Fang}\ and\ \citenamefont {Wang}(2021)}]{fang2021optical}%
  \BibitemOpen
  \bibfield  {author} {\bibinfo {author} {\bibfnamefont {L.}~\bibnamefont {Fang}}\ and\ \bibinfo {author} {\bibfnamefont {J.}~\bibnamefont {Wang}},\ }\bibfield  {title} {\bibinfo {title} {Optical trapping separation of chiral nanoparticles by subwavelength slot waveguides},\ }\href {https://doi.org/10.1103/PhysRevLett.127.233902} {\bibfield  {journal} {\bibinfo  {journal} {Physical Review Letters}\ }\textbf {\bibinfo {volume} {127}},\ \bibinfo {pages} {233902} (\bibinfo {year} {2021})}\BibitemShut {NoStop}%
\bibitem [{\citenamefont {Tkachenko}\ \emph {et~al.}(2023)\citenamefont {Tkachenko}, \citenamefont {Truong}, \citenamefont {Esporlas}, \citenamefont {Sanskriti},\ and\ \citenamefont {Nic~Chormaic}}]{janus2023}%
  \BibitemOpen
  \bibfield  {author} {\bibinfo {author} {\bibfnamefont {G.}~\bibnamefont {Tkachenko}}, \bibinfo {author} {\bibfnamefont {V.~G.}\ \bibnamefont {Truong}}, \bibinfo {author} {\bibfnamefont {C.~L.}\ \bibnamefont {Esporlas}}, \bibinfo {author} {\bibfnamefont {I.}~\bibnamefont {Sanskriti}},\ and\ \bibinfo {author} {\bibfnamefont {S.}~\bibnamefont {Nic~Chormaic}},\ }\bibfield  {title} {\bibinfo {title} {Evanescent field trapping and propulsion of {J}anus particles along optical nanofibers},\ }\href {https://doi.org/10.1038/s41467-023-37448-2} {\bibfield  {journal} {\bibinfo  {journal} {Nature Communications}\ }\textbf {\bibinfo {volume} {14}},\ \bibinfo {pages} {1691} (\bibinfo {year} {2023})}\BibitemShut {NoStop}%
\bibitem [{\citenamefont {Li}\ \emph {et~al.}(2009)\citenamefont {Li}, \citenamefont {Pernice},\ and\ \citenamefont {Tang}}]{fiberforces2009}%
  \BibitemOpen
  \bibfield  {author} {\bibinfo {author} {\bibfnamefont {M.}~\bibnamefont {Li}}, \bibinfo {author} {\bibfnamefont {W.~H.~P.}\ \bibnamefont {Pernice}},\ and\ \bibinfo {author} {\bibfnamefont {H.~X.}\ \bibnamefont {Tang}},\ }\bibfield  {title} {\bibinfo {title} {Tunable bipolar optical interactions between guided lightwaves},\ }\href {https://doi.org/10.1038/nphoton.2009.116} {\bibfield  {journal} {\bibinfo  {journal} {Nature Photonics}\ }\textbf {\bibinfo {volume} {3}},\ \bibinfo {pages} {464} (\bibinfo {year} {2009})}\BibitemShut {NoStop}%
\bibitem [{\citenamefont {Ray}\ \emph {et~al.}(2020)\citenamefont {Ray}, \citenamefont {Gupta}, \citenamefont {Gokhroo}, \citenamefont {Everett}, \citenamefont {Nieddu}, \citenamefont {Rajasree},\ and\ \citenamefont {Chormaic}}]{Ray_2020}%
  \BibitemOpen
  \bibfield  {author} {\bibinfo {author} {\bibfnamefont {T.}~\bibnamefont {Ray}}, \bibinfo {author} {\bibfnamefont {R.~K.}\ \bibnamefont {Gupta}}, \bibinfo {author} {\bibfnamefont {V.}~\bibnamefont {Gokhroo}}, \bibinfo {author} {\bibfnamefont {J.~L.}\ \bibnamefont {Everett}}, \bibinfo {author} {\bibfnamefont {T.}~\bibnamefont {Nieddu}}, \bibinfo {author} {\bibfnamefont {K.~S.}\ \bibnamefont {Rajasree}},\ and\ \bibinfo {author} {\bibfnamefont {S.~N.}\ \bibnamefont {Chormaic}},\ }\bibfield  {title} {\bibinfo {title} {Observation of the 87{R}b 5{S}1/2 to 4{D}3/2 electric quadrupole transition at 516.6 nm mediated via an optical nanofibre},\ }\href {https://doi.org/10.1088/1367-2630/ab8265} {\bibfield  {journal} {\bibinfo  {journal} {New Journal of Physics}\ }\textbf {\bibinfo {volume} {22}},\ \bibinfo {pages} {062001} (\bibinfo {year} {2020})}\BibitemShut {NoStop}%
\bibitem [{\citenamefont {White}\ \emph {et~al.}(2019)\citenamefont {White}, \citenamefont {Kato}, \citenamefont {N\'emet}, \citenamefont {Parkins},\ and\ \citenamefont {Aoki}}]{fiber_Cavity_dark}%
  \BibitemOpen
  \bibfield  {author} {\bibinfo {author} {\bibfnamefont {D.~H.}\ \bibnamefont {White}}, \bibinfo {author} {\bibfnamefont {S.}~\bibnamefont {Kato}}, \bibinfo {author} {\bibfnamefont {N.}~\bibnamefont {N\'emet}}, \bibinfo {author} {\bibfnamefont {S.}~\bibnamefont {Parkins}},\ and\ \bibinfo {author} {\bibfnamefont {T.}~\bibnamefont {Aoki}},\ }\bibfield  {title} {\bibinfo {title} {Cavity dark mode of distant coupled atom-cavity systems},\ }\href {https://doi.org/10.1103/PhysRevLett.122.253603} {\bibfield  {journal} {\bibinfo  {journal} {Phys. Rev. Lett.}\ }\textbf {\bibinfo {volume} {122}},\ \bibinfo {pages} {253603} (\bibinfo {year} {2019})}\BibitemShut {NoStop}%
\bibitem [{\citenamefont {Yasumoto}(2018)}]{yasumoto2018electromagnetic}%
  \BibitemOpen
  \bibfield  {author} {\bibinfo {author} {\bibfnamefont {K.}~\bibnamefont {Yasumoto}},\ }\href {https://www.routledge.com/Electromagnetic-Theory-and-Applications-for-Photonic-Crystals/Yasumoto/p/book/9780849336775} {\emph {\bibinfo {title} {Electromagnetic theory and applications for photonic crystals}}}\ (\bibinfo  {publisher} {CRC press},\ \bibinfo {year} {2018})\BibitemShut {NoStop}%
\bibitem [{\citenamefont {Snyder}\ \emph {et~al.}(1983)\citenamefont {Snyder}, \citenamefont {Love} \emph {et~al.}}]{snyder1983optical}%
  \BibitemOpen
  \bibfield  {author} {\bibinfo {author} {\bibfnamefont {A.~W.}\ \bibnamefont {Snyder}}, \bibinfo {author} {\bibfnamefont {J.~D.}\ \bibnamefont {Love}}, \emph {et~al.},\ }\href {https://doi.org/10.1364/JOSAA.3.000378} {\emph {\bibinfo {title} {Optical waveguide theory}}},\ Vol.\ \bibinfo {volume} {175}\ (\bibinfo  {publisher} {Chapman and {H}all London},\ \bibinfo {year} {1983})\BibitemShut {NoStop}%
\end{thebibliography}%

\clearpage
\newpage

\onecolumngrid
\appendix

\section{Detailed derivation of the Green's function}

For completeness, we here give a comprehensive description for the calculation of the Green's function, following the references~\cite{white2002multipole,fussell2004three,yasumoto2018electromagnetic}. As the system is invariant in the $z$-direction, the $z$-components of the electric and magnetic fields are sufficient to reconstruct the remaining components~\cite{snyder1983optical}. With Eq.~\eqref{eq:global-field} as our starting point, we more explicitly write the source term outside $(+)$ of all the cylinders as
\begin{equation}
    \label{app-eq:global-exp}
    \widetilde{G}^{V}_{zu} =\frac{i}{4} D_u^V\left\{H_0^{(1)}\left(k_\rho \rho'\right)\right\} 
    +\sum_{l=1}^{N} \sum_{m=-\infty}^{\infty} B_{um}^{V l+} H_m^{(1)}\left(k_\rho \rho_l\right) e^{i m \phi_l},
\end{equation}
where the cylindrical coordinates $(\rho',\phi')$ of the source are taken with respect to the observation point. We have now written $\widetilde{G}^{V}_{zu}$ for the $z$-component of the Fourier-transformed electric ($V = E$) or magnetic ($V = H$) field with source dipole orientation $u \ (=x,y,z)$. The differential operators $D_{u}^{V}$~\cite{yasumoto2018electromagnetic} may be found by, e.g., considering the Fourier transform of Eq.~\eqref{eq:green-de} with $\textbf{G} \to \textbf{G}_{0}, n(\textbf{r}) \to 1,$ but we omit their expression here.  To apply boundary conditions at each cylinder interface we consider the local field expansion,
\begin{equation}
    \label{app-eq:local-exp}
    \widetilde{G}^{Vl+}_{zu}=\sum_{m=-\infty}^{\infty}\left[A_{um}^{V l+} J_m\left(k_\rho \rho_l\right)+B_{um}^{V l+} H_m^{(1)}\left(k_\rho \rho_l\right)\right] e^{i m \phi_l},
\end{equation}
valid for $\bm{\rho}$ in the annulus $\mathcal{A}_{l}$ external to and containing only the $l$-th fiber. Whilst Eq.~\eqref{app-eq:local-exp} appears to contain twice the number of unknowns as Eq.~\eqref{app-eq:global-exp}, the demand for consistency $\widetilde{G}^{Vl+}_{zu} = \widetilde{G}^{V}_{zu}$ in $\mathcal{A}_{l}$ allows us to eliminate the additional unknowns in the following. We denote the source within the local coordinates about cylinder $m$ as $\bm{\rho}' - \bm{c}_{l} = (\rho'_{l},\phi_{l}').$ Within the region $\rho'_{l} > \rho_{l}$ we may use Graf's addition theorem
\begin{align}
        &H_{m}^{(1)}(k_{\rho}\rho')e^{im\phi'} =  \sum_{m'}H_{m'-m}^{(1)}(k_{\rho}\rho_{l}')J_{m'}(k_{\rho}\rho_{l})e^{im'\phi_{l}}e^{-i(m'-m)\phi_{l}'}.
       \label{app-eq:graf}
\end{align}
Using \eqref{app-eq:graf} and the explicit form of $D_{u}^{V}$ to expand the source term in \eqref{app-eq:global-exp} in this region then yields
\begin{equation}
    \label{eq:source_terms}
    \frac{i}{4} D_u^V\left\{H_0^{(1)}\left(k_\rho \rho'\right)\right\} = \frac{i}{4}\sum_{m=-\infty}^{\infty}K^{Vl}_{um}J_m(k_{\rho}\rho_{l})e^{im\phi_{l}},
\end{equation}
for the source terms $K_{um}^{Vl}$, with
\begin{align}
        &K_{xm}^{El} = \frac{i
        \beta k_{\rho}}{2k^2}\left(H_{m-1}^{ls}e^{-i(m-1)\phi_{ls}} - H_{m+1}^{ls}e^{-i(m+1)\phi_{ls}}\right)\\
        &K_{ym}^{El} = \frac{
        \beta k_{\rho}}{2k^2}\left(H_{m-1}^{ls}e^{-i(m-1)\phi_{ls}} + H_{m+1}^{ls}e^{-i(m+1)\phi_{ls}}\right)\\
        &K_{zm}^{El} = -\left(1 - \frac{\beta^2}{k^2}\right)H_{m}^{ls}e^{-im\phi_{ls}}\\
        &K_{xm}^{Hl} = \left(\frac{1}{Z_{0}}\right)\frac{k_{\rho}}{2k_{0}}\left(H_{m-1}^{ls}e^{-i(m-1)\phi_{ls}} + H_{m+1}^{ls}e^{-i(m+1)\phi_{ls}}\right)\\
        &K_{ym}^{Hl} = \left(\frac{1}{Z_{0}}\right)\frac{k_{\rho}}{2ik_{0}}\left(H_{m-1}^{ls}e^{-i(m-1)\phi_{ls}} - H_{m+1}^{ls}e^{-i(m+1)\phi_{ls}}\right)\\
        &K_{zm}^{Hl} = 0
\end{align}
for free-space impedance $Z_{0} = \sqrt{\epsilon_0 / \mu_0}$. An alternate usage of Graf's addition theorem
\begin{align}
\label{eq:app-Graf}
    &H_m^{(1)}\left(k_\rho \rho_{l}\right) e^{i m  \phi_{l}} = 
    \sum_{m
'=-\infty}^{\infty}
H_{m'-m}^{(1)}\left(k_\rho \rho_{ll'}\right) e^{i (m-m') \phi_{ll'}}  J_m^{(1)}\left(k_\rho \rho_{l'}\right) e^{i m \phi_{l'}},
\end{align}
in the region $\rho_{ll'} > \rho_{l'}$ and for $\rho_{ll'} = |\bm{c}_{l'} - \bm{c}_{l}|, \phi_{ll'} = \text{arg}[\bm{c}_{l'} - \bm{c}_{l}]$
relates the cylindrical field about one center to the cylindrical fields about another center. In the intersection of the regions $\rho_{l}' > \rho_{l}$ and $\rho_{ll'} > \rho_{l'}$, we may thus equate Eqs.~\eqref{app-eq:local-exp} and \eqref{app-eq:global-exp}, and use Eq.~\eqref{eq:app-Graf} to express the result purely in terms of $\rho_{l}$ and $\phi_{l}$. Matching identically at each order $m$ relates source terms to coefficients in the local expansion,
\begin{equation}
    \label{eq:app-graf-matching}
    A_{um}^{V l+}=K_{u m}^{V l}+\sum_{l' \neq l}^{N} \sum_{m'=-\infty}^{\infty} S_{m m'}^{l l'} B_{u m'}^{V l'+},
\end{equation}
for

\begin{equation}
    \label{eq:app-relation_two}
   S_{m m'}^{l l'} = H_{m'-m}^{(1)}\left(k_\rho \rho_{ll'}\right) \\ e^{i (m-m') \phi_{l'l}}. 
\end{equation}
We may now use boundary conditions for the $l$-th fiber to eliminate one set of coefficients, obtaining the other in terms of the known source terms. Whilst \textit{outside} the $l$-th cylinder we can write the field as in Eq.~\eqref{app-eq:local-exp}, \textit{inside} the cylinder we can write 
\begin{equation}
\label{app-eq:local_exp_inner}
    \tilde{G}^{Vl-}_{zu}=\sum_{m=-\infty}^{\infty}A_{um}^{V l-} J_m\left(k_\rho \rho_l\right) e^{i m \phi_l},
\end{equation}
absent the Hankel functions that are diverging for $\rho_{l} = 0$.
We match 
\begin{equation}
    \tilde{G}_{\mu u}^{Vl-} = \tilde{G}_{\mu u}^{Vl+}, 
    \label{app-eq:boundary}
\end{equation}
at the boundary for $\mu = z, \phi_{l}$, noting that the latter can be performed via the relation
\begin{align}
    \tilde{G}_{\phi_{l}u}^{El\pm} &= \frac{i}{k_{\rho}^{\pm 2}} \bigg[ \frac{\beta}{\rho_{l}} \frac{\partial}{\partial \phi_{l}}{\tilde{G}_{zu}^{El\pm}} -
    k\frac{\partial}{\partial \rho_{l}}{\tilde{G}_{zu}^{Hl\pm}} \bigg] \\
    \tilde{G}_{\phi_{l}u}^{Hl\pm} &= \frac{i}{k_{\rho}^{\pm 2}} \bigg[ \frac{\beta}{\rho_{l}} \frac{\partial}{\partial \phi_{l}}{\tilde{G}_{zu}^{El\pm}} +
    kn_{\pm}^2\frac{\partial}{\partial \rho_{l}}{\tilde{G}_{zu}^{Hl\pm}} \bigg],
\end{align}
where we define $k_{\rho}^{-} = \sqrt{k^2 n_{-}^2 - \beta^2}$ and write $k_{\rho}^{+} = k_{\rho}$. We input Eq.~\eqref{app-eq:local-exp} and \eqref{app-eq:local_exp_inner} and match the equations \eqref{app-eq:boundary} also at each order $m$ to eliminate $A^{Vl-}_{um}$ and obtain the relation
\begin{equation}
\label{eq:app-scattering-matrix}
    \begin{bmatrix}
       B^{El+}_{um} \\
       B^{Hl+}_{um}
    \end{bmatrix} =
    \begin{bmatrix}
       R^{EEl}_{m} & R^{EHl}_{m} \\ 
       R^{HEl}_{m} & R^{HHl}_{m}
    \end{bmatrix}
    \begin{bmatrix}
       A^{El+}_{um} \\
       A^{Hl+}_{um}
    \end{bmatrix},
\end{equation}
involving the scattering matrix elements $R_{m}^{VV'l}.$ Inverting Eqs.~\eqref{eq:app-scattering-matrix} and substituting the $A_{um}^{Vl+}$ into Eq.~\eqref{eq:app-graf-matching} finally gives a linear system expressing the unknown scattered terms $B^{Vl+}_{um}$ in terms of the known source terms $K^{Vl}_{um}.$ To solve numerically we truncate at a value $m$ where the resulting expression for $\tilde{G}^{V}_{zu}$ has sufficiently converged in Fourier space. To obtain the Green's function in real space we invert the Fourier transform using the integral contour detailed in~\cite{fussell2004three}. 

\section{Comprehensive study of figures of merit -- two ONFs}

\begin{figure}[h!]
		\centering
		\includegraphics[width=0.33\linewidth]{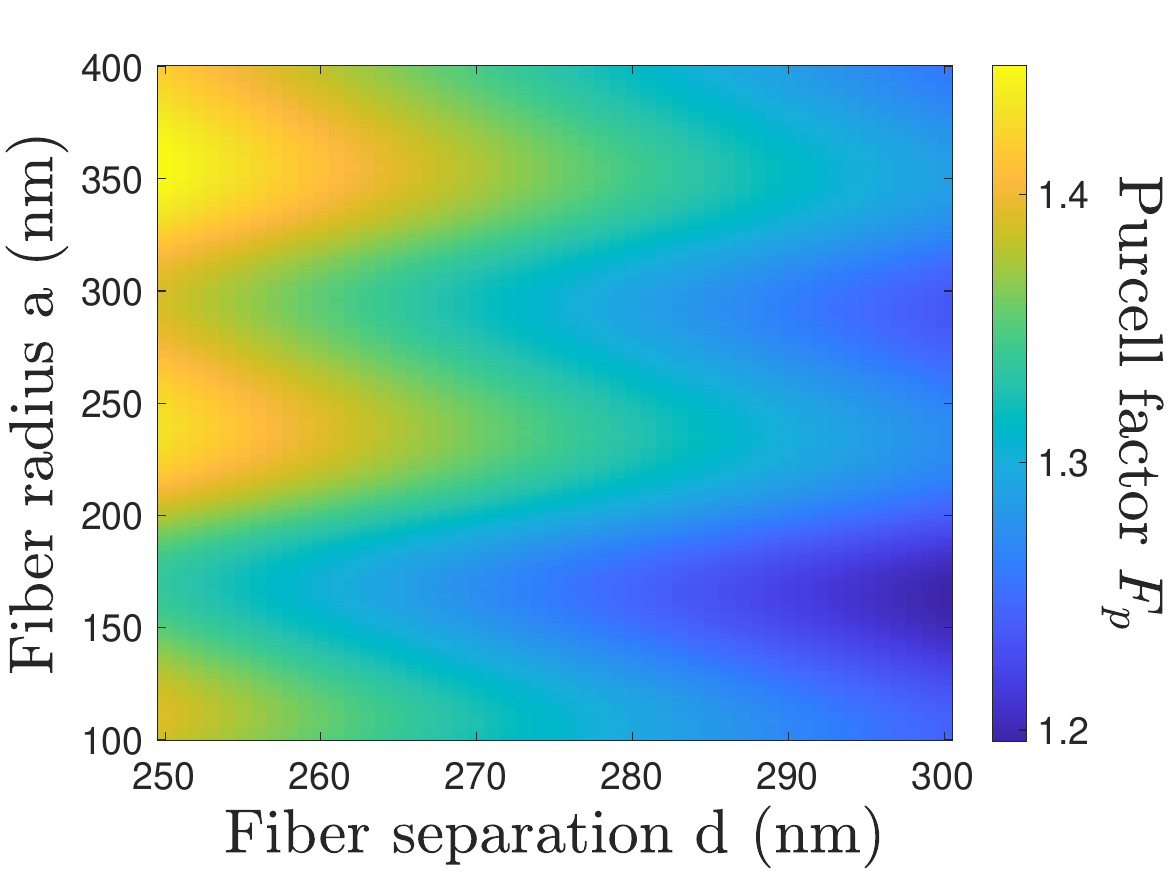}
        \includegraphics[width=0.33\linewidth]{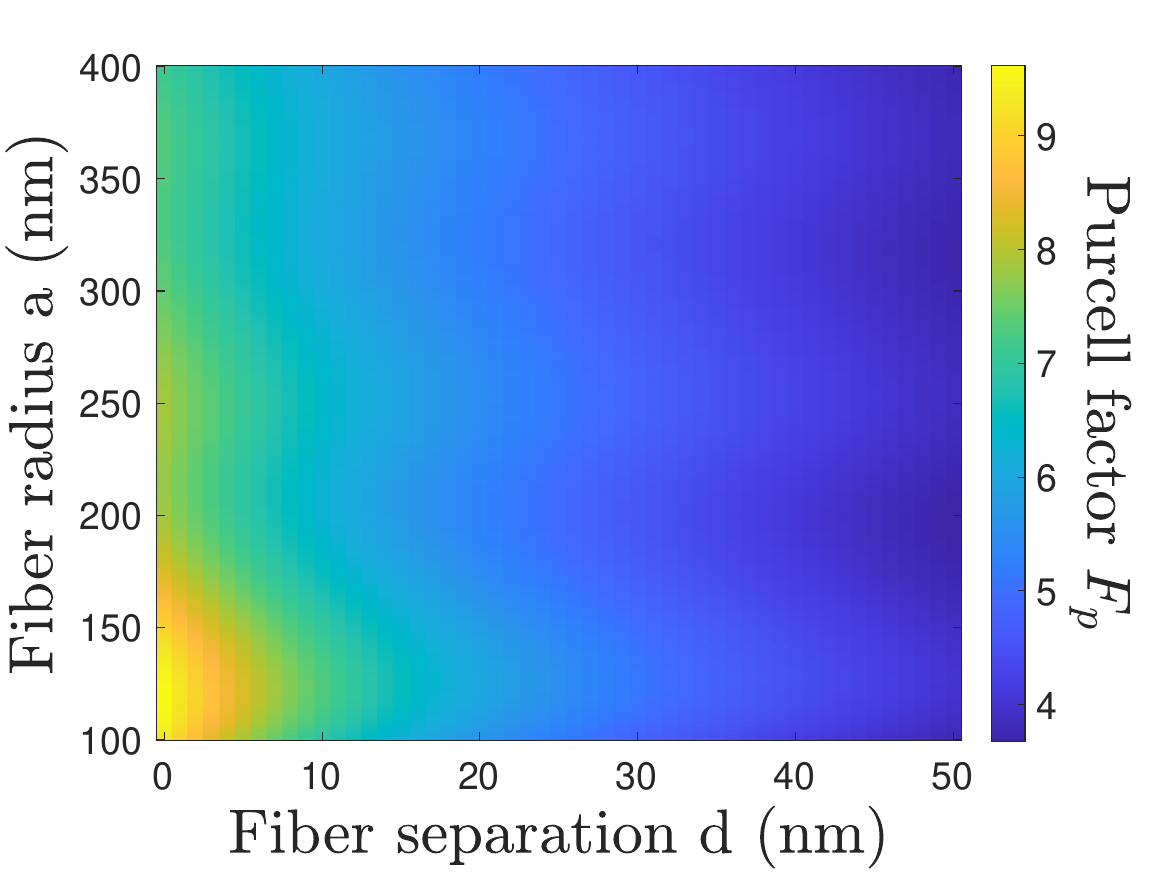}
		\caption{Purcell factor of an $x$-oriented emitter at the center of the two-ONF system in the region (left) $250$ nm $\leq a \leq 300$ nm (right) $0$ nm $\leq a \leq 50 $ nm, as the ONF separation varies.}
		\label{fig:single-atom PF A}
\end{figure}

Here we give more details into the figures of merit explored in the main text. To get clearer picture of the Purcell factor of an $x$-oriented emitter at the center of the two ONFs (see Fig.~\ref{fig:single-atom main}(c)) we show a more detailed plot in the ranges of larger $250$ nm $\leq a \leq 300$ nm and shorter separations $0$ nm $ \leq a \leq 50$ nm  in Fig.~\ref{fig:single-atom PF A}. One can see that the Purcell factor oscillates for increasing ONF radii as higher order modes begin to dominantly couple to the emitter. Notably, the Purcell factor may be as large as 1.4 for an atom-surface separations of $125$nm, the value is already comparable to the Purcell factor for an emitter at the surface of a single ONF~\cite{fam_2005}. 

\begin{figure}[h!]
	\centering
    \includegraphics[width=0.32\linewidth]{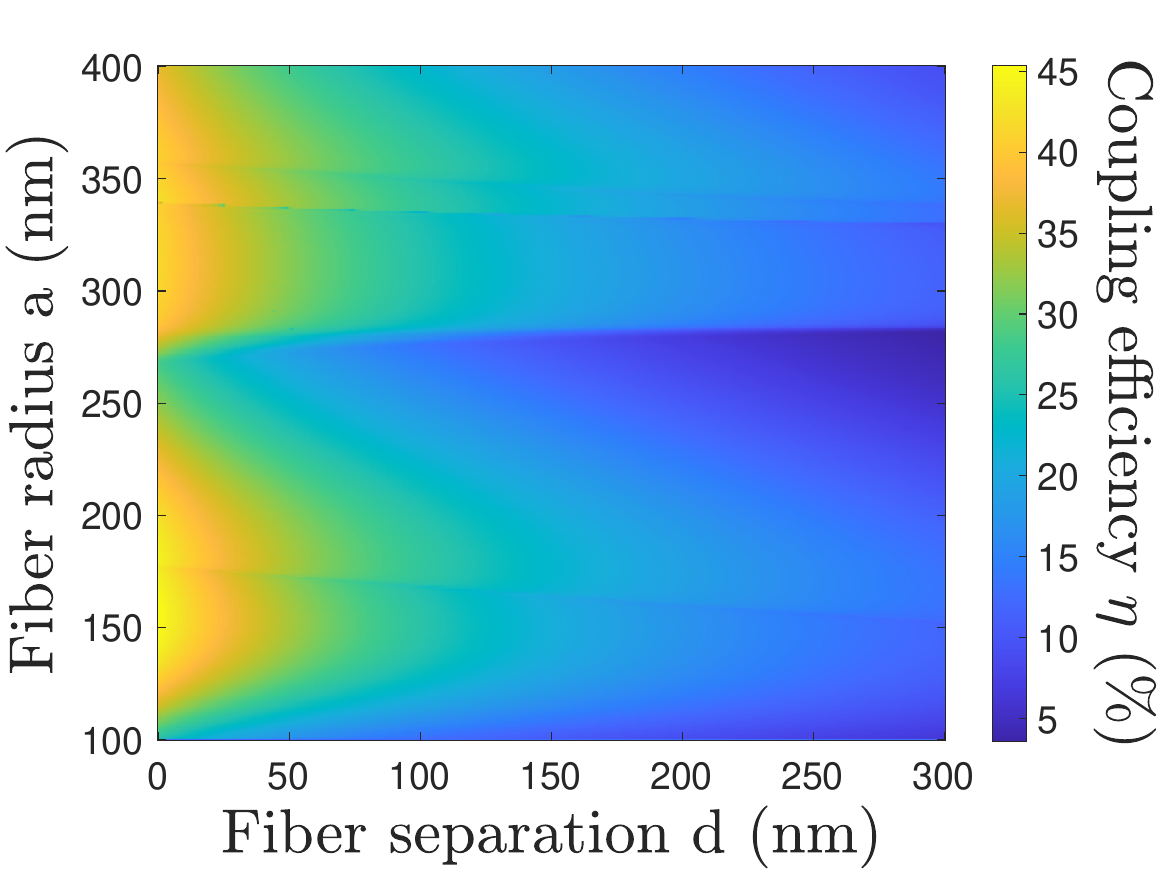}
    \includegraphics[width=0.32\linewidth]{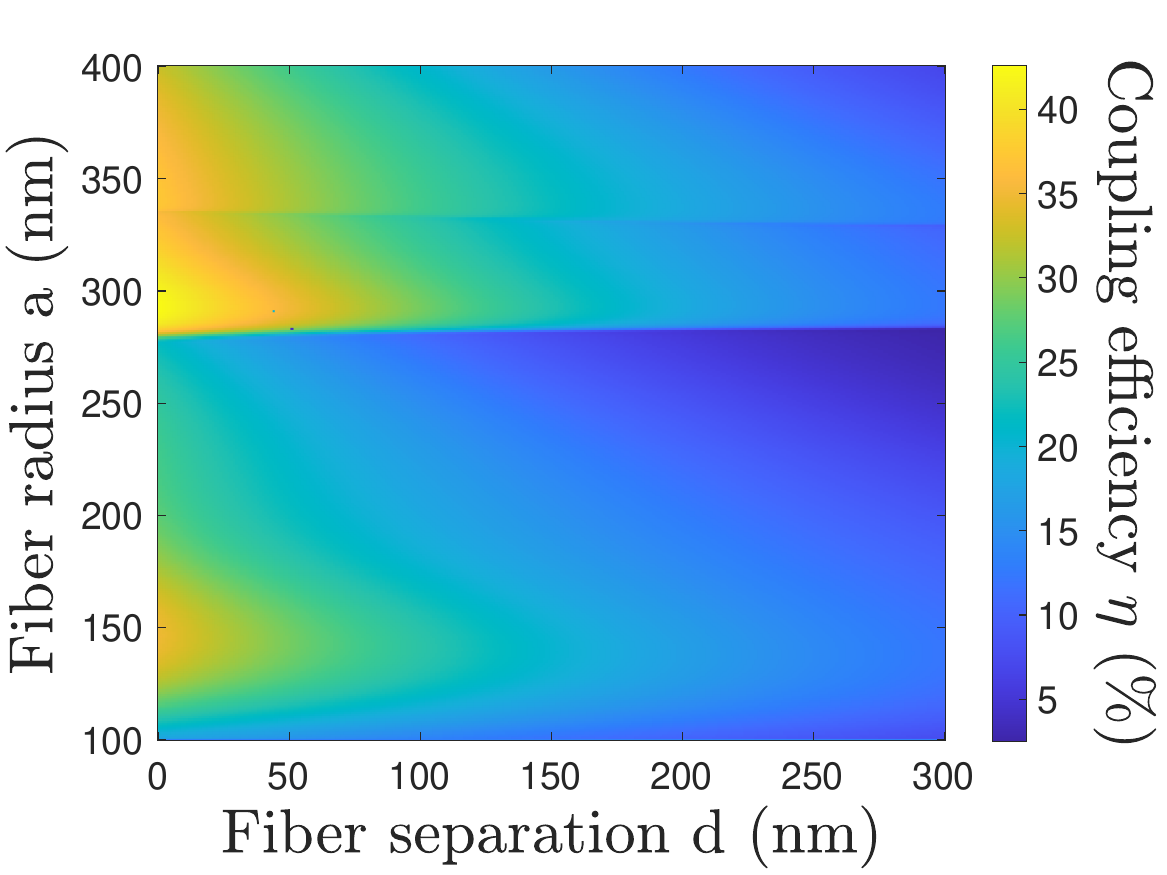}
    \includegraphics[width=0.32\linewidth]{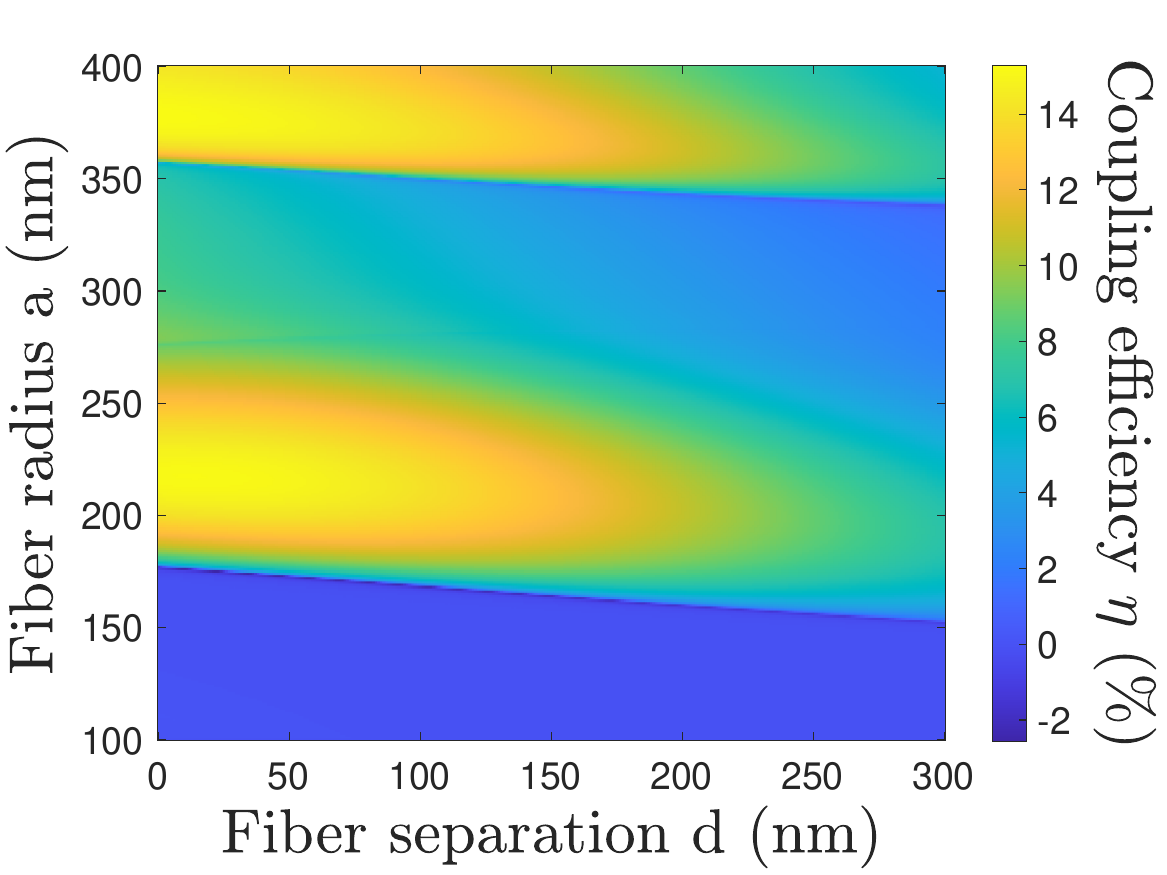}
    \includegraphics[width=0.32\linewidth]{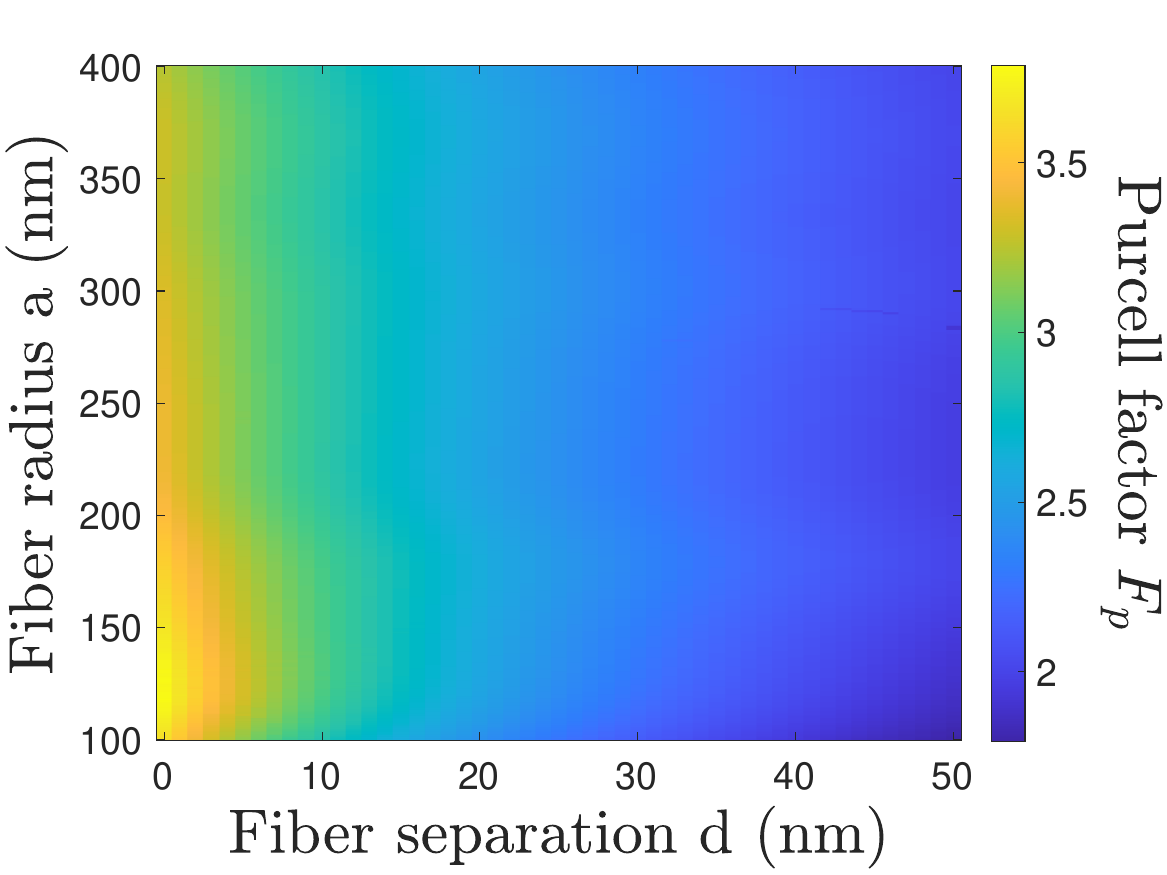}
    \includegraphics[width=0.32\linewidth]{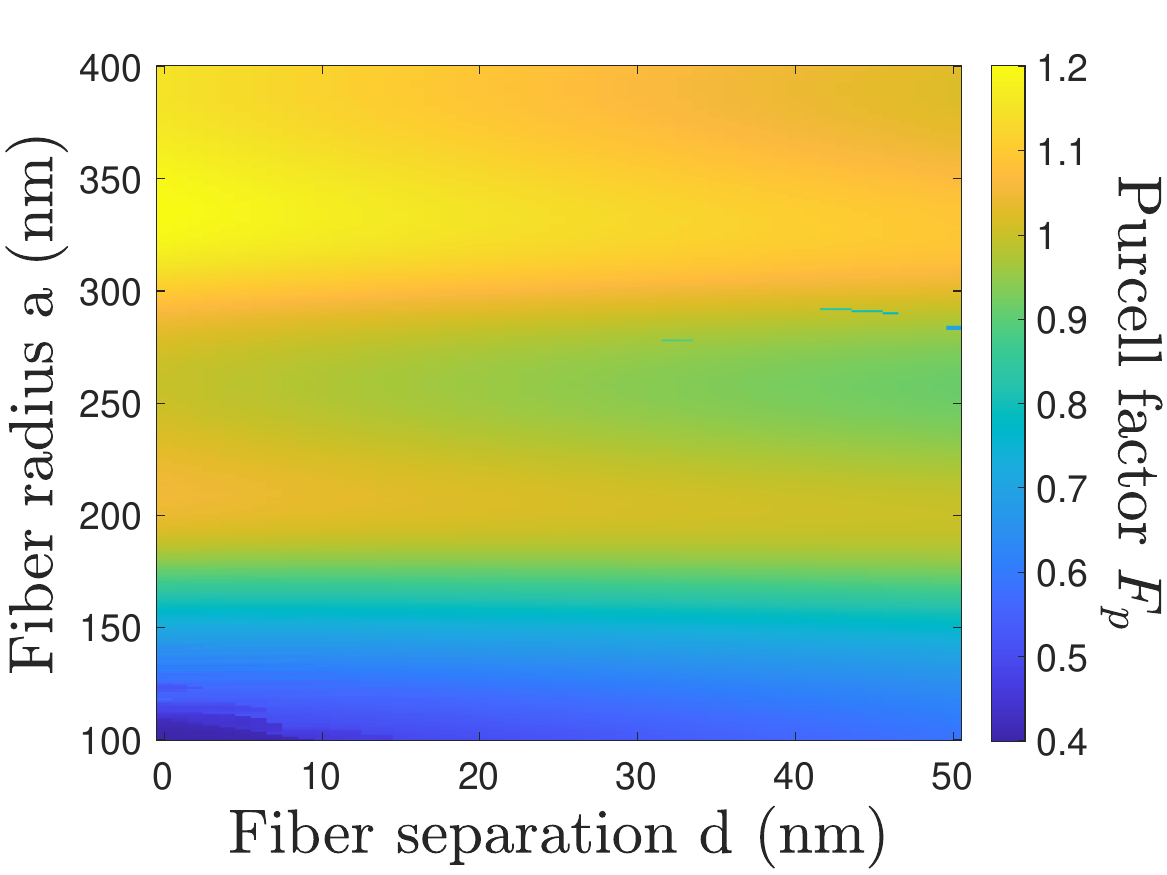}
    \includegraphics[width=0.32\linewidth]{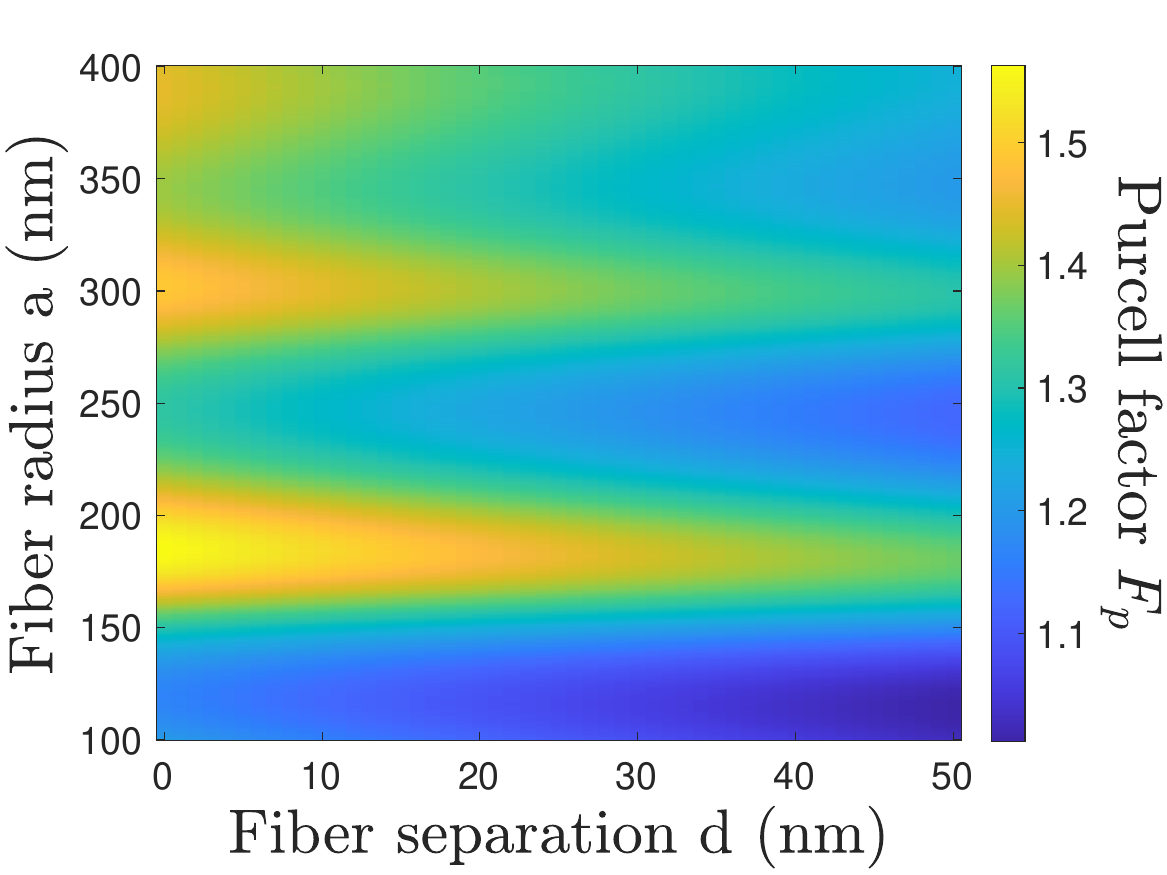}
	\caption{Coupling efficiencies (top) and Purcell factor (bottom) of emitters at the center of the two-fiber system as the fiber radii and separation vary, for the case of (left) averaged dipole orientation (middle) a $y$-oriented dipole, and (right) a $z$-oriented dipole.}
    \label{fig:single-atom rand-oriented A}
\end{figure}

\begin{figure}[h!]
		\centering
        \includegraphics[width=0.33\linewidth]{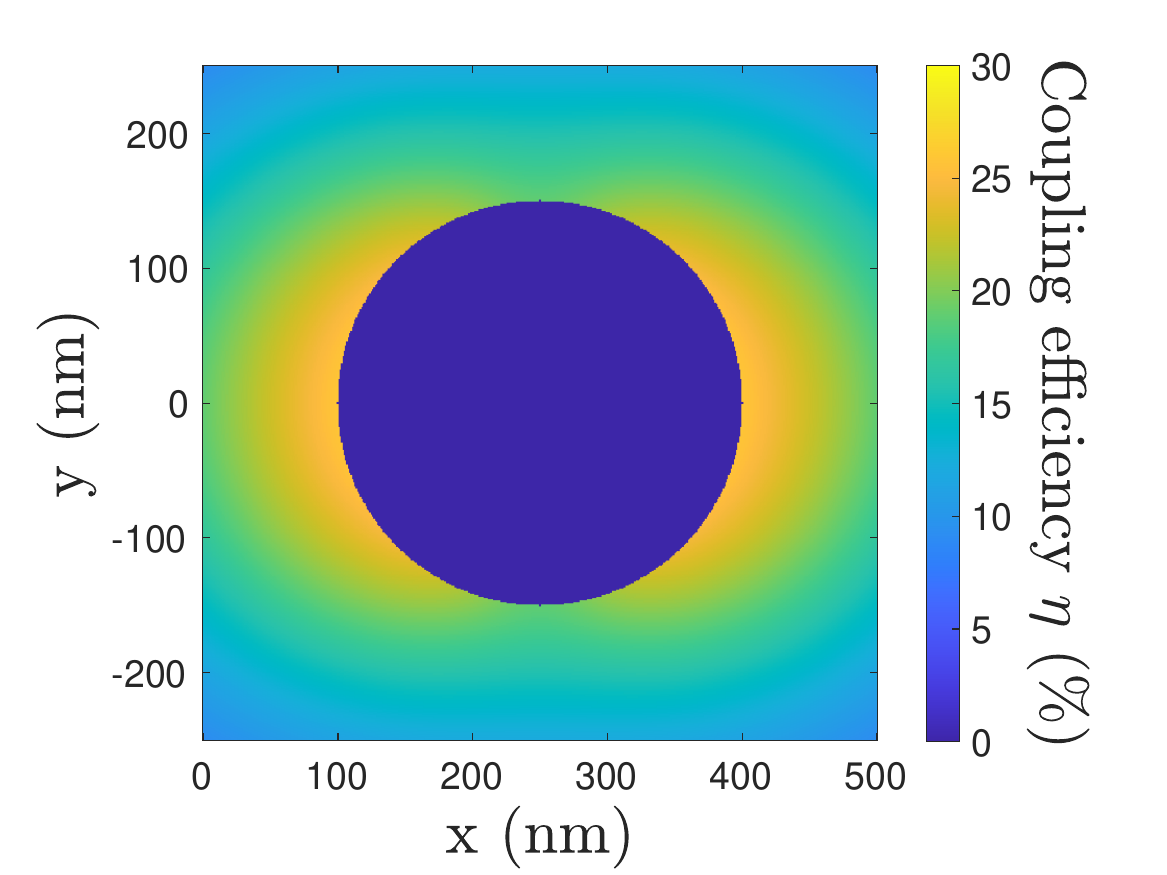}
        \includegraphics[width=0.33\linewidth]{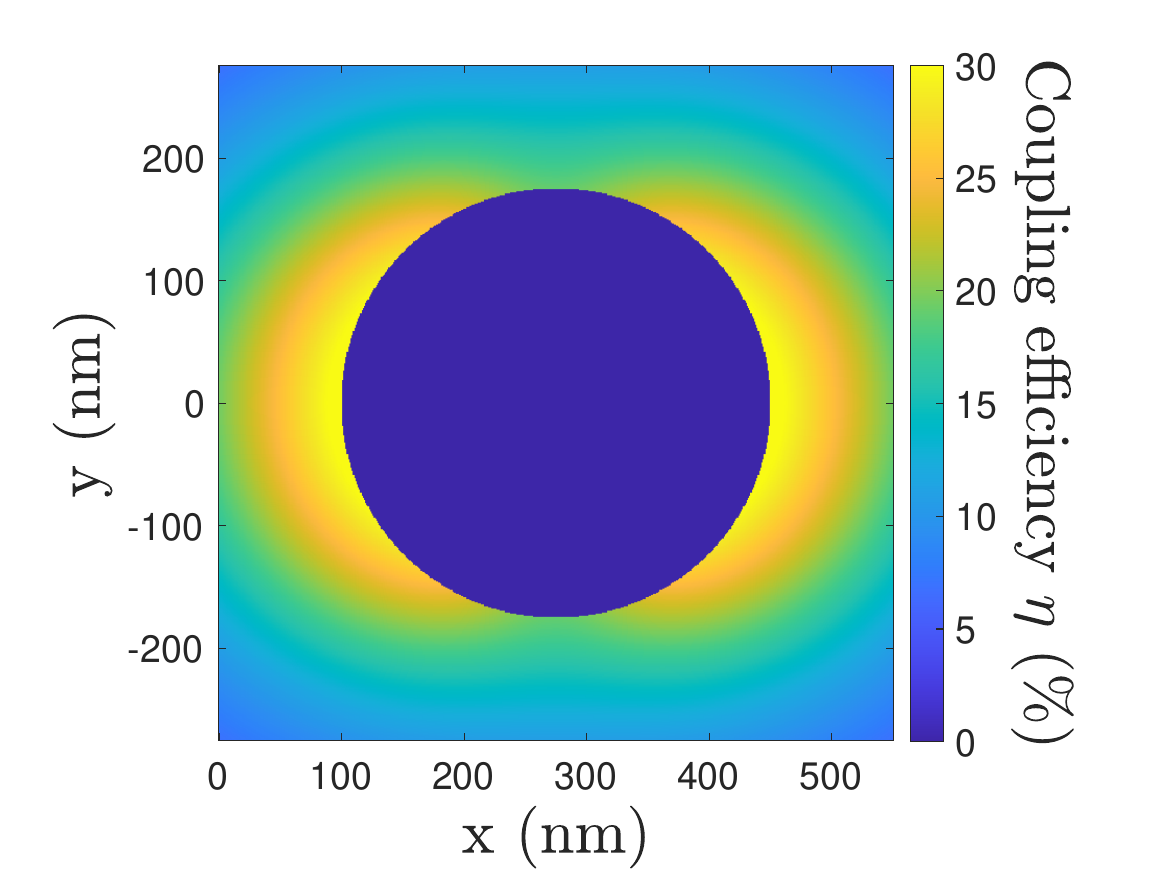}
		\caption{Coupling efficiency of an $x$-oriented emitter with a single ONF of (left) radius $a= 150$ nm (right) $a=175$ nm for varying transverse emitter position.}
		\label{fig:single-atom single-fiber A}
\end{figure}

As the dipole orientation may not be specified in practice, figures of merit are often averaged uniformly over all dipole orientations. In Fig.~\ref{fig:single-atom rand-oriented A}, we plot the dipole-averaged coupling efficiency and Purcell factor of an emitter at the center of the two fibers. The weaker coupling of $z$- and $y$-oriented dipoles to guided modes (also presented in Fig.~\ref{fig:single-atom rand-oriented A}) reduces the averaged coupling efficiency below that of $x$-oriented dipoles. Nonetheless, the coupling efficiency and Purcell factor remain as large as $45\%$ and $3.6$ respectively for surface-bound emitters. These values may be more than twice as large as those in the optimal single-ONF scenario~\cite{yalla2012efficient}.
To allow for a clearer comparison of the coupling efficiency between the single-ONF and two-ONF configurations, we also plot $\eta$ for an x-oriented emitter in the vicinity of a single ONF of two different radii whilst varying the transverse position of the emitter (Fig.~\ref{fig:single-atom single-fiber A}). The two lobes of large coupling efficiency correspond to regions of high intensity for the quasi-$x$ polarized mode of the single-ONF. 

\begin{figure}[h!]
		\centering
        \raisebox{-8mm}{\includegraphics[width=0.35\linewidth]{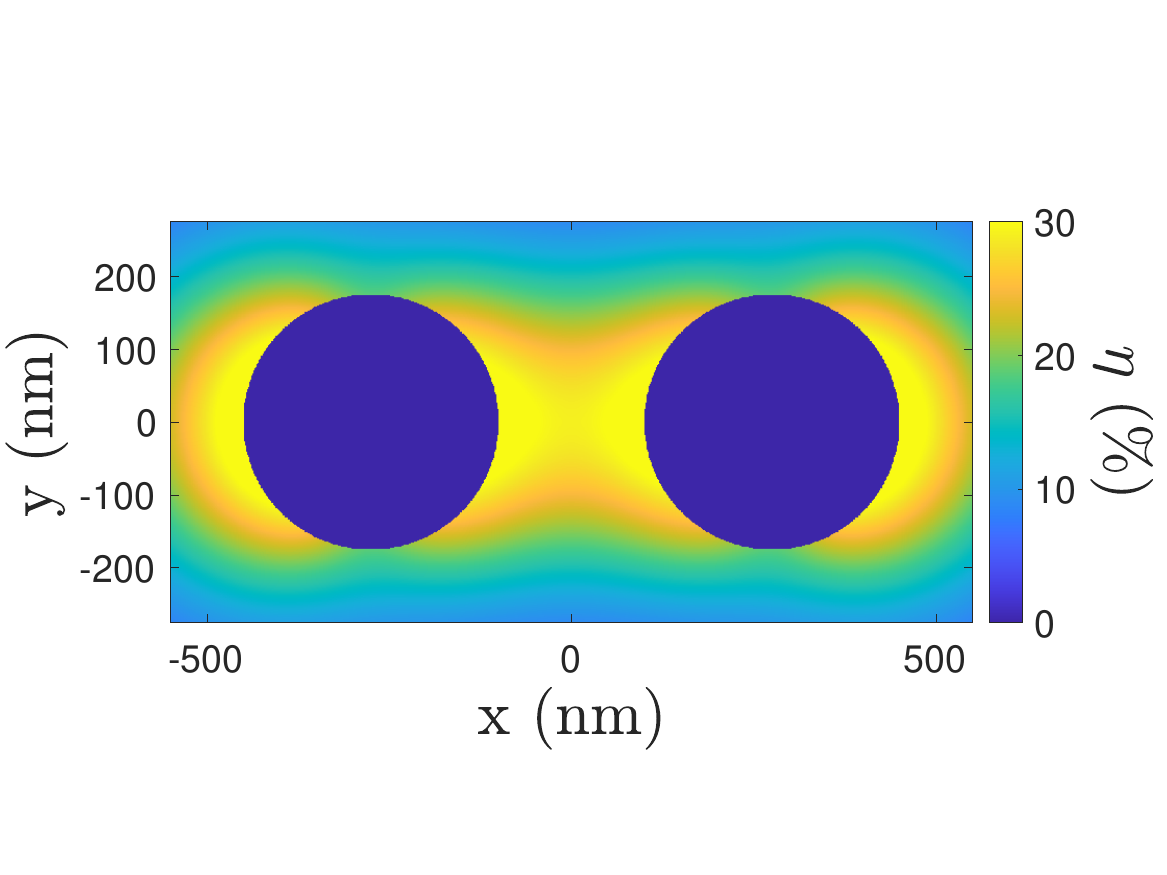}
        \includegraphics[width=0.35\linewidth]{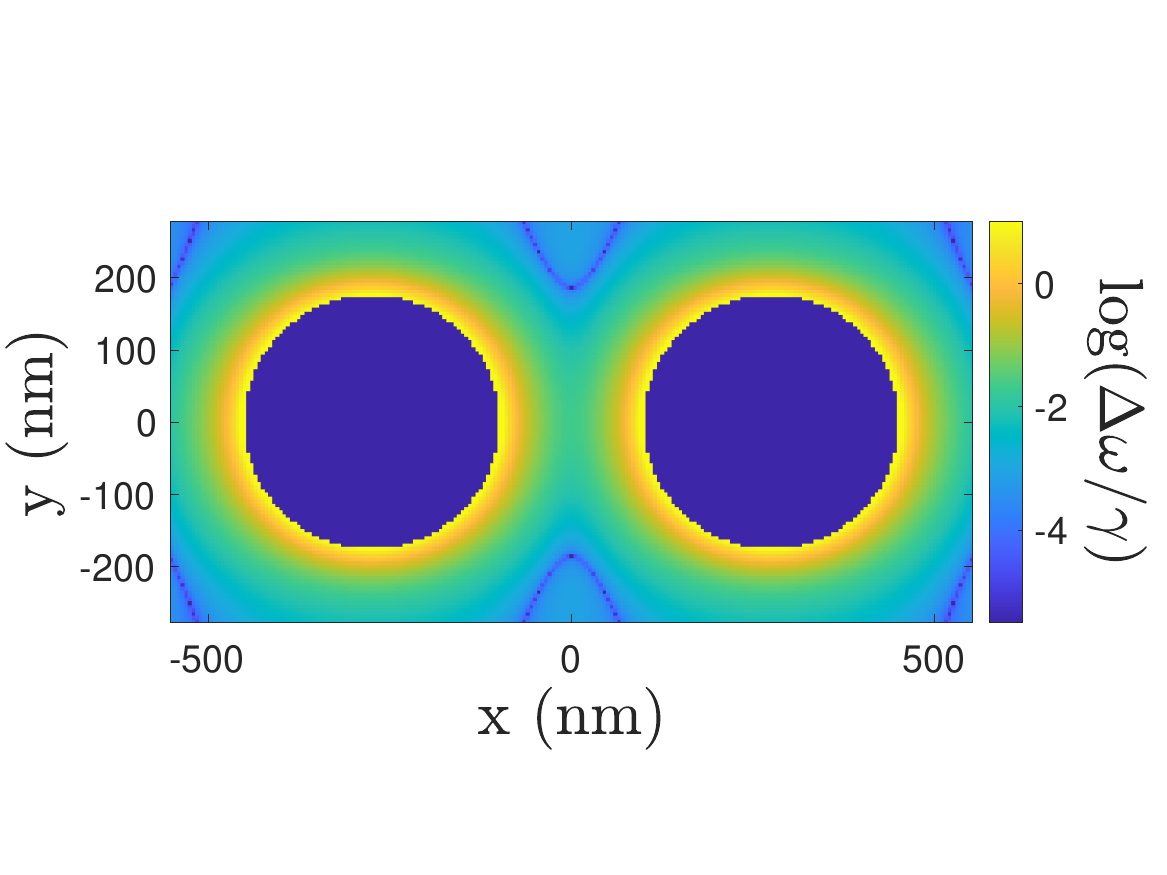}}
        \includegraphics[width=0.25\linewidth]{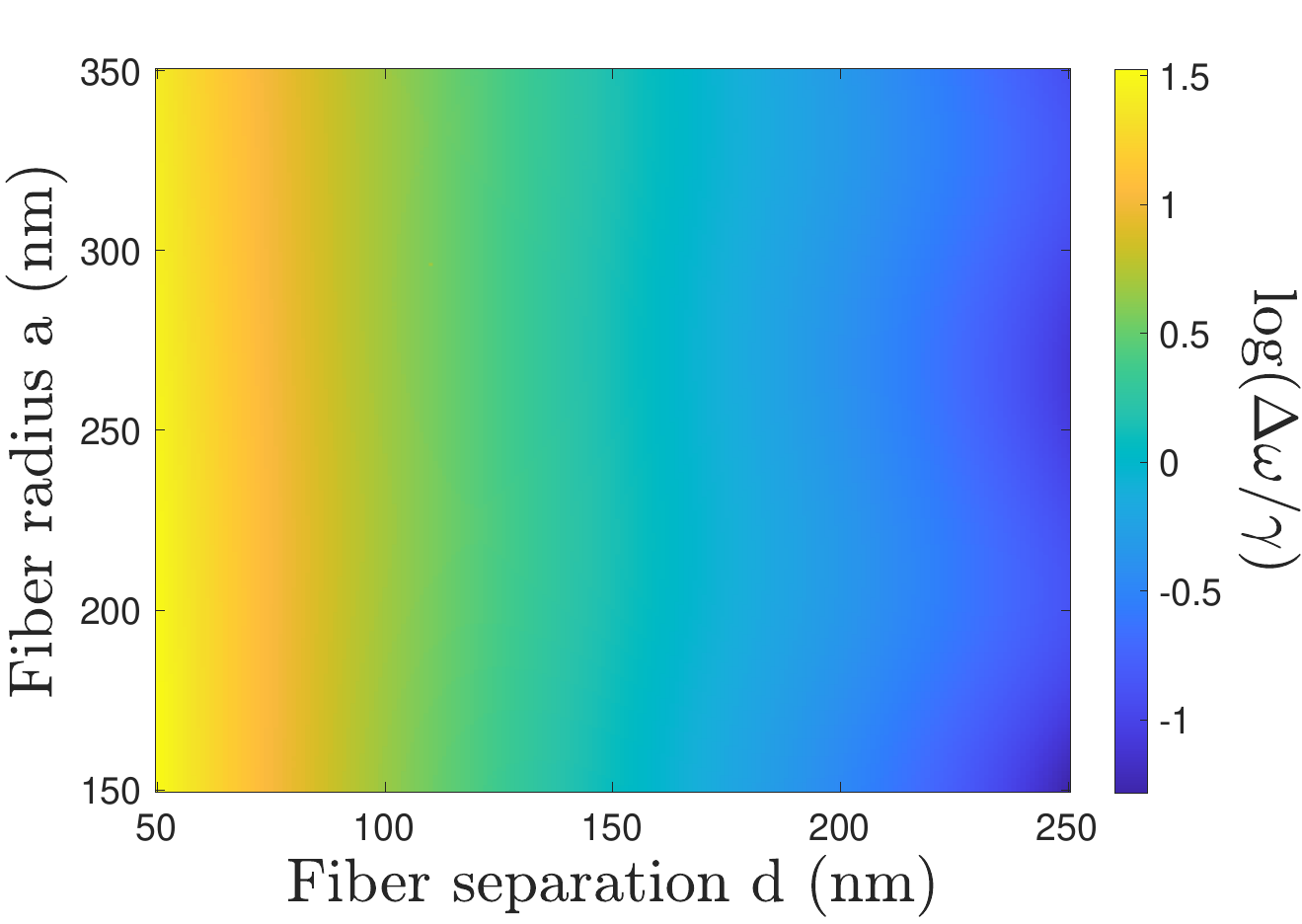}
		\caption{(Left) Coupling efficiency and (middle) normalized Lamb shift (log-plot) of an $x$-oriented emitter for varying transverse position in the system of two ONFs with radii $a=175$ nm and separation $d=200$ nm. (Right) Normalized Lamb shift (log-plot) of an $x$-oriented emitter at the center of the two-fiber system as a function of fiber radii and separation.}  
		\label{fig:single-atom positionvary and freq shifts A}
\end{figure}

Since in the two-ONF system higher-order modes may already be present for radii $a \approx 200$ nm~\cite{le2021spatial},  we here briefly present the coupling efficiency and Lamb shift for $a=175$ nm and $d = 200$ nm, where four guided modes exist. The second TE-like mode features high intensity lobes on the outer sides of the two ONFs, which increases the coupling efficiency in these regions also (Fig.~\ref{fig:single-atom positionvary and freq shifts A}, left), whilst the coupling efficiency is slightly reduced in the central region when compared to Fig.~\ref{fig:single-atom main}(d) of the main text. 
The Lamb shift (Fig.~\ref{fig:single-atom positionvary and freq shifts A}, middle), on the other hand, is less sensitive to cutoffs and does not differ greatly from Fig.~\ref{fig:single-atom main}(e) of the main text. For completeness, Fig.~\ref{fig:single-atom positionvary and freq shifts A} (right) displays the shift of an $x$-oriented emitter in the center of two ONFs as $a$ and $d$ are varied. Once $d \lessapprox 75$ nm, the shift becomes appreciable, compared to the free-space atomic linewidth, for all considered values of $a$. 

\section{Robustness of coupling efficiency}
The previous results suggest a robustness of our geometry with respect to the transverse position of the emitter and to small variations in the two-ONF system, which we further highlight in the following.

\begin{figure}[h!]
   	\centering
        \raisebox{-1mm}{\includegraphics[width=0.24\linewidth]{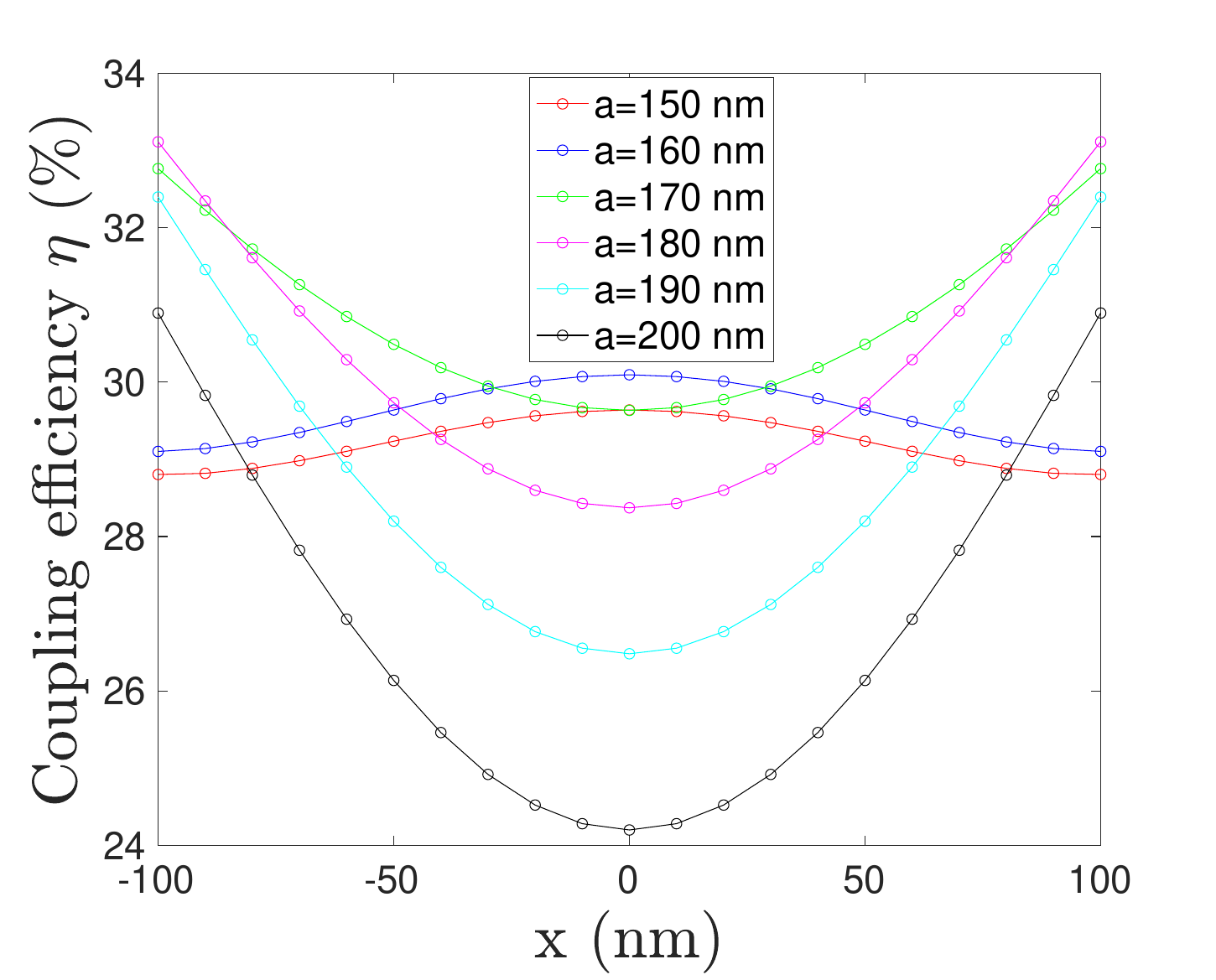}}
          \includegraphics[width=0.24\linewidth]{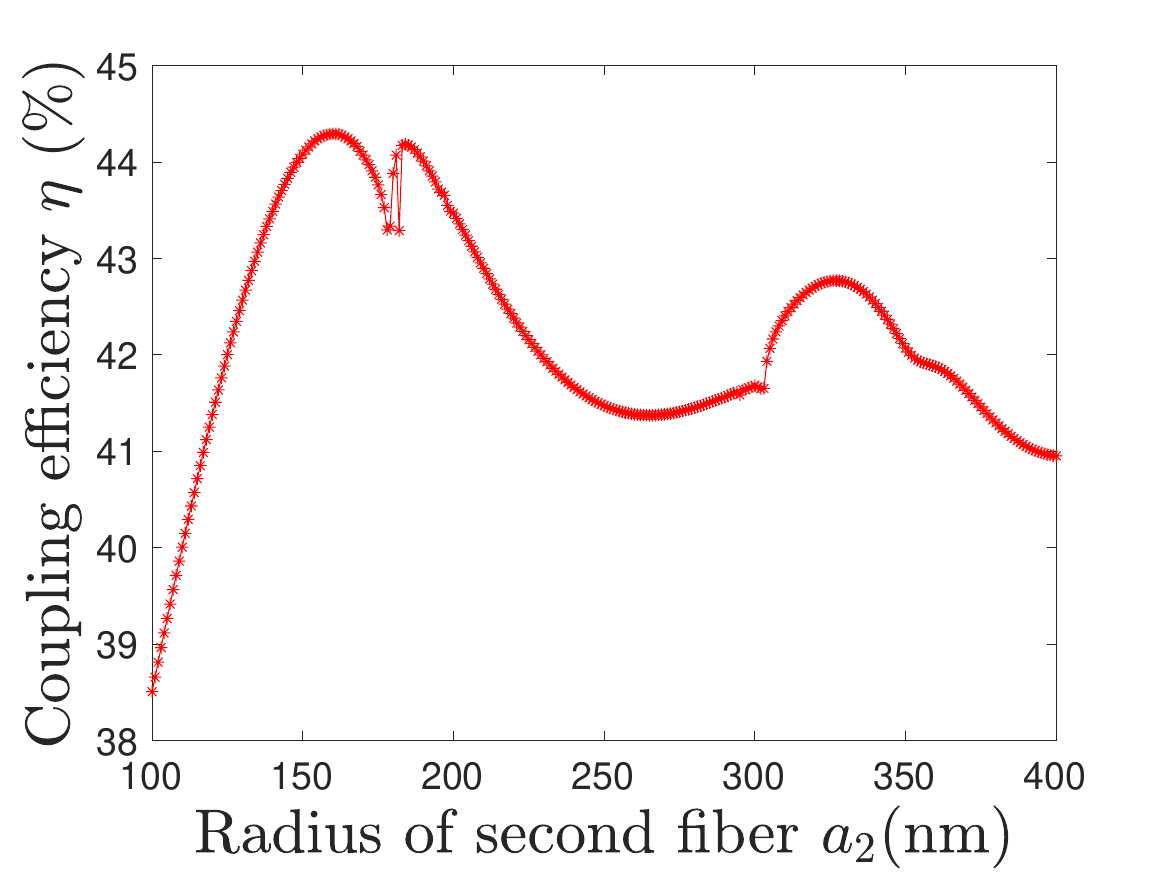}
        \includegraphics[width=0.24\linewidth]{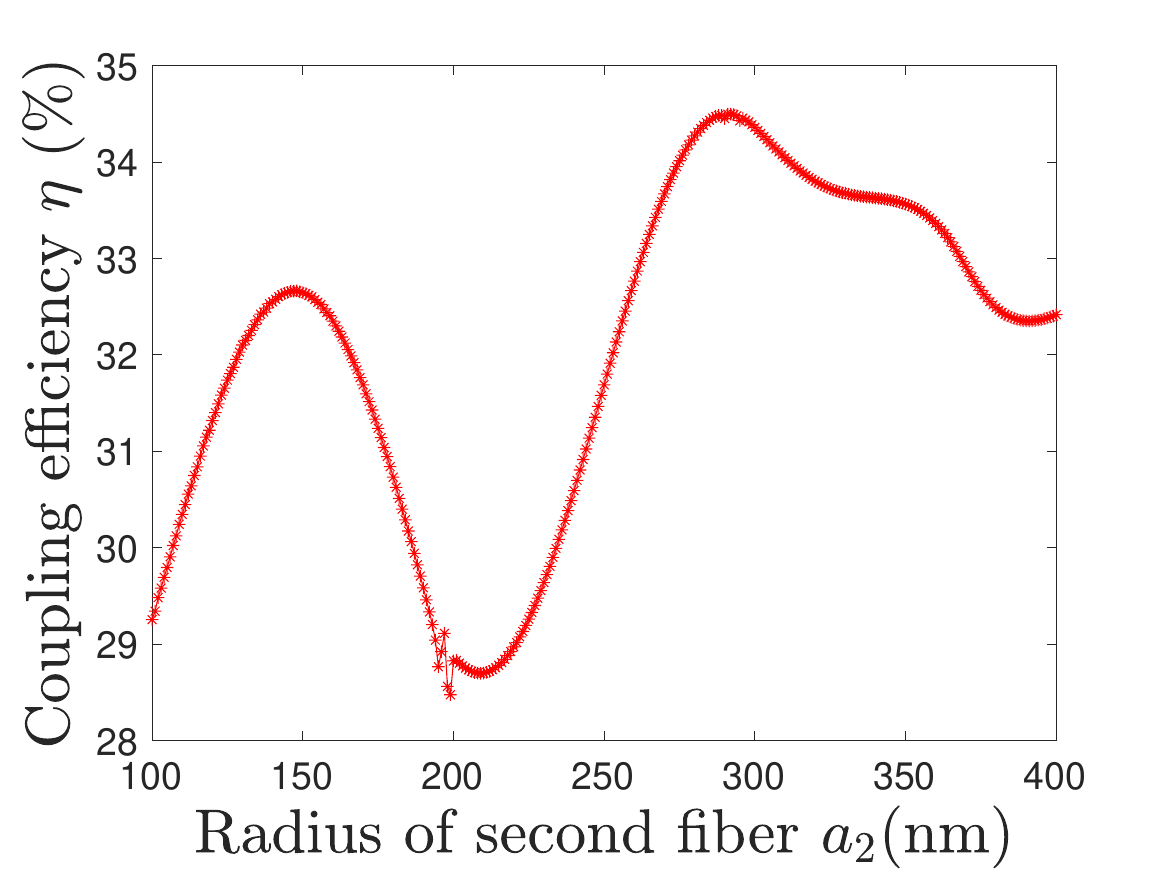}
        \includegraphics[width=0.24\linewidth]{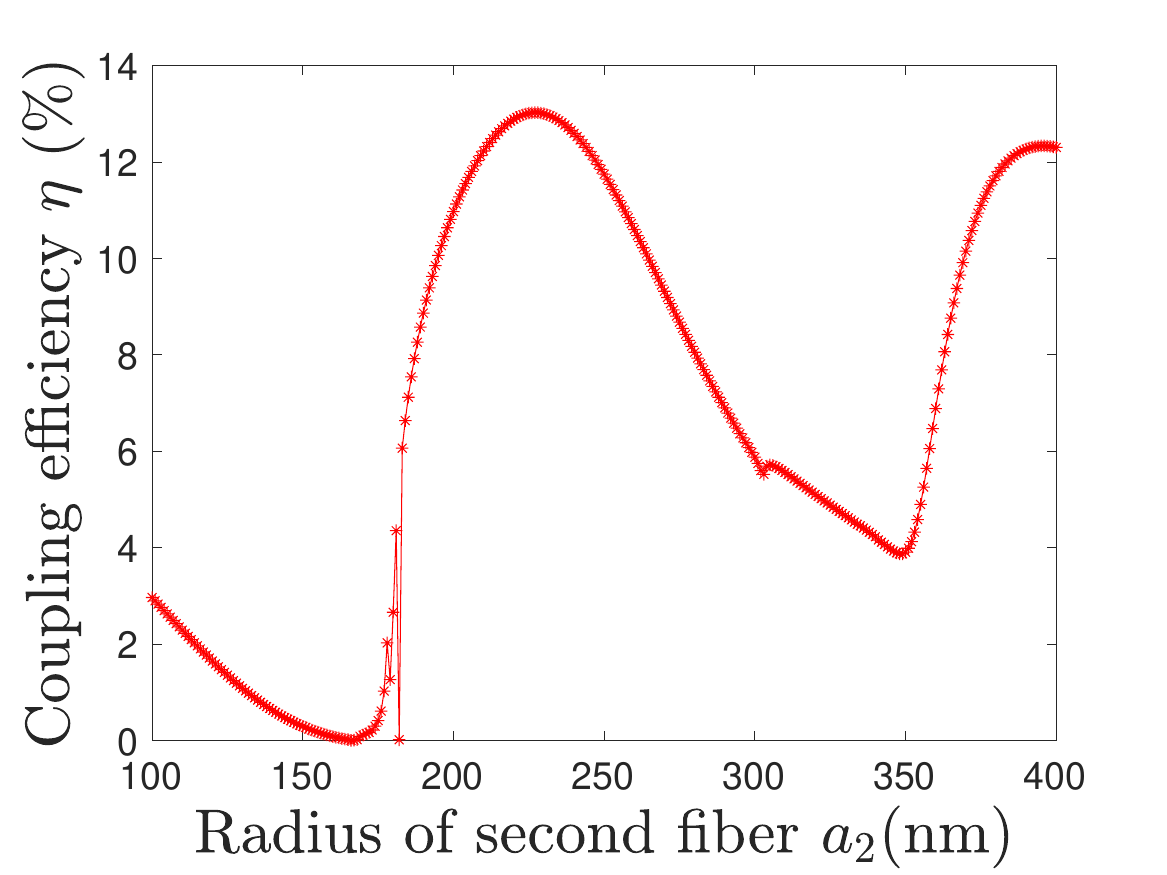}

		\caption{(Left) Variation of coupling efficiency for an $x$-polarized emitter with varying position on the $x$-axis between two identical ONFs, with $a = 175$ nm and $d = 200$ nm. (Middle-left to far-right) Coupling efficiency for the emitter at the center of two nonidentical and touching ($d=0$) ONFs, where the radius $a_1 = 175$ nm is fixed for the first fiber, whilst the radius $a_2$ of the second fiber varies for (middle-left) an emitter with random dipole orientation (middle-right) a $y$-oriented emitter (far-right) a $z$-oriented emitter.}
		\label{fig:vary-rad}
\end{figure}

Fig.~\ref{fig:vary-rad} (left) shows the coupling efficiency for emitters with their transverse position varying along the $x$-axis and only relatively small variations of $\eta$ occur within a $50$ nm distance from the center, independently of the fiber-radii.
To account for possible differences in ONF radii owing to fabrication imperfections, we show in Fig.~\ref{fig:vary-rad} (middle-left to right) the variation in coupling efficiencies with the radius of a one fiber, while the other one is fixed, for dipoles other than the $x$-orientation considered in Fig.~\ref{fig:single-atom positionvary and freq shifts A}(e) of the main text. Away from cutoffs, variations of only a few percent typically occur for experimentally relevant~\cite{Fatemi:17} changes in the second fiber radius on the order of 10s of nanometers. It is worth noting that the variation in coupling efficiency is much greater near the guided mode cutoffs, which could allow coupled emitters to act as sensitive probes of fiber radii.

\section{Coupling efficiency for increasing ONF number}

\begin{figure}[h!]
   	\centering
    \includegraphics[width=0.33\linewidth]{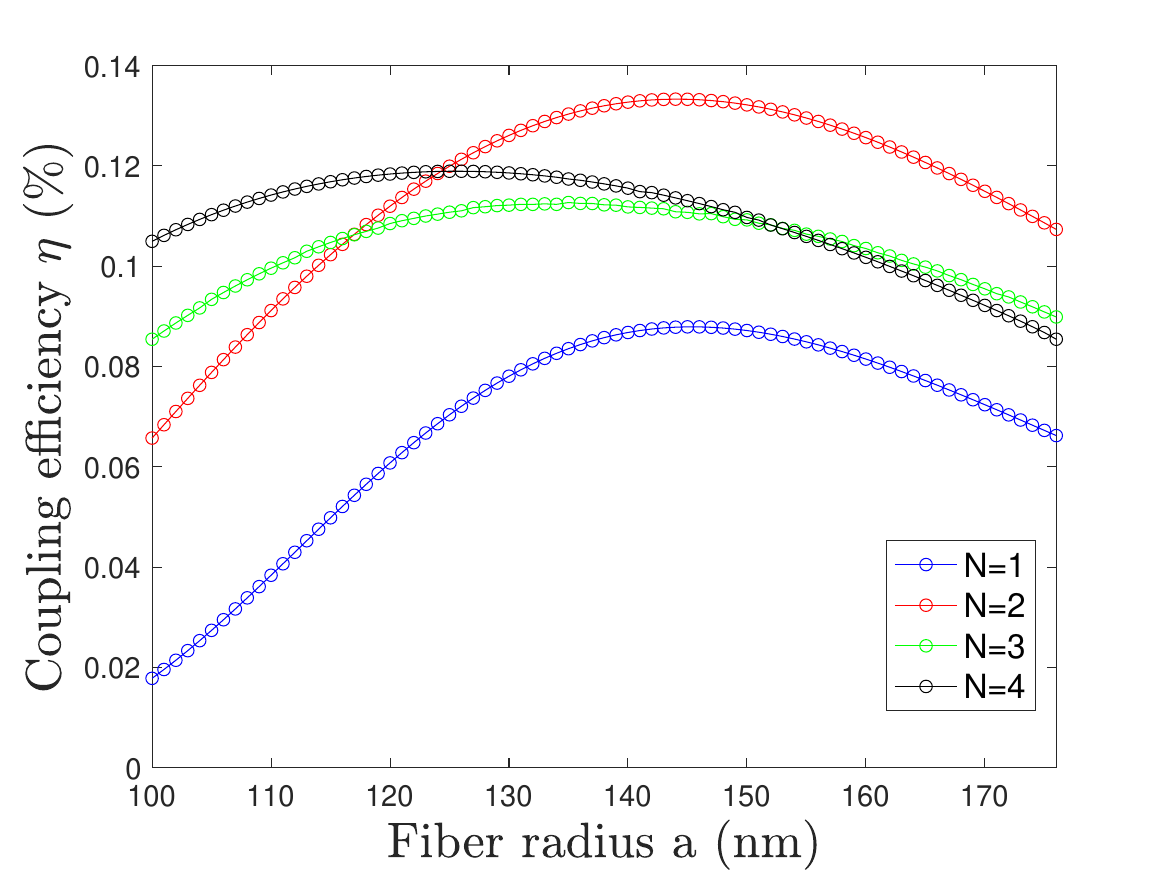}
    \caption{Variation of coupling efficiency for an emitter at the center of $N$ fibers with an emitter-surface distance $d/2=250$ nm, as the radius of the fibers are varied.}
	\label{fig:multiple-fibers}
\end{figure}

Whilst the two-ONF system is demonstrated in the main text to show figures of merit, most notably coupling efficiency, improved over the single-fiber system, we here show that a two-ONF geometry can also outperform increasingly complex geometries of more ONFs that might naively be expected to confine the field further. In Fig.~\ref{fig:multiple-fibers}, we plot the coupling efficiency for an $x$-polarized emitter at the center of $N$ identical ONFs for $N = 1,2,3,4$, with centers $\bm{\rho}_{l} = (a+d/2)(\cos\frac{2\pi l}{N},\sin\frac{2\pi l}{N})$, for emitter-surface distance $d/2 = 250$ nm and varying fiber-radii $a$. When optimizing coupling efficiency over radii, the coupling efficiency increases first for $N = 2,$ and then decreases again. We have checked separately for $N = 5,6$ that the radii-optimized coupling efficiency remains below that of the simplest hybrid system of two ONFs. This result suggests that the majority of field confinement for the TE-like mode can already be achieved for $N=2$, and that exploiting complex in-plane geometries in an attempt to increase coupling efficiency may not be worth the additional difficulties that will be encountered in their fabrication.

\section{Verification of off-diagonal Green's function elements}
\begin{figure}[h!]
    \centering
    \includegraphics[width=0.5\linewidth]{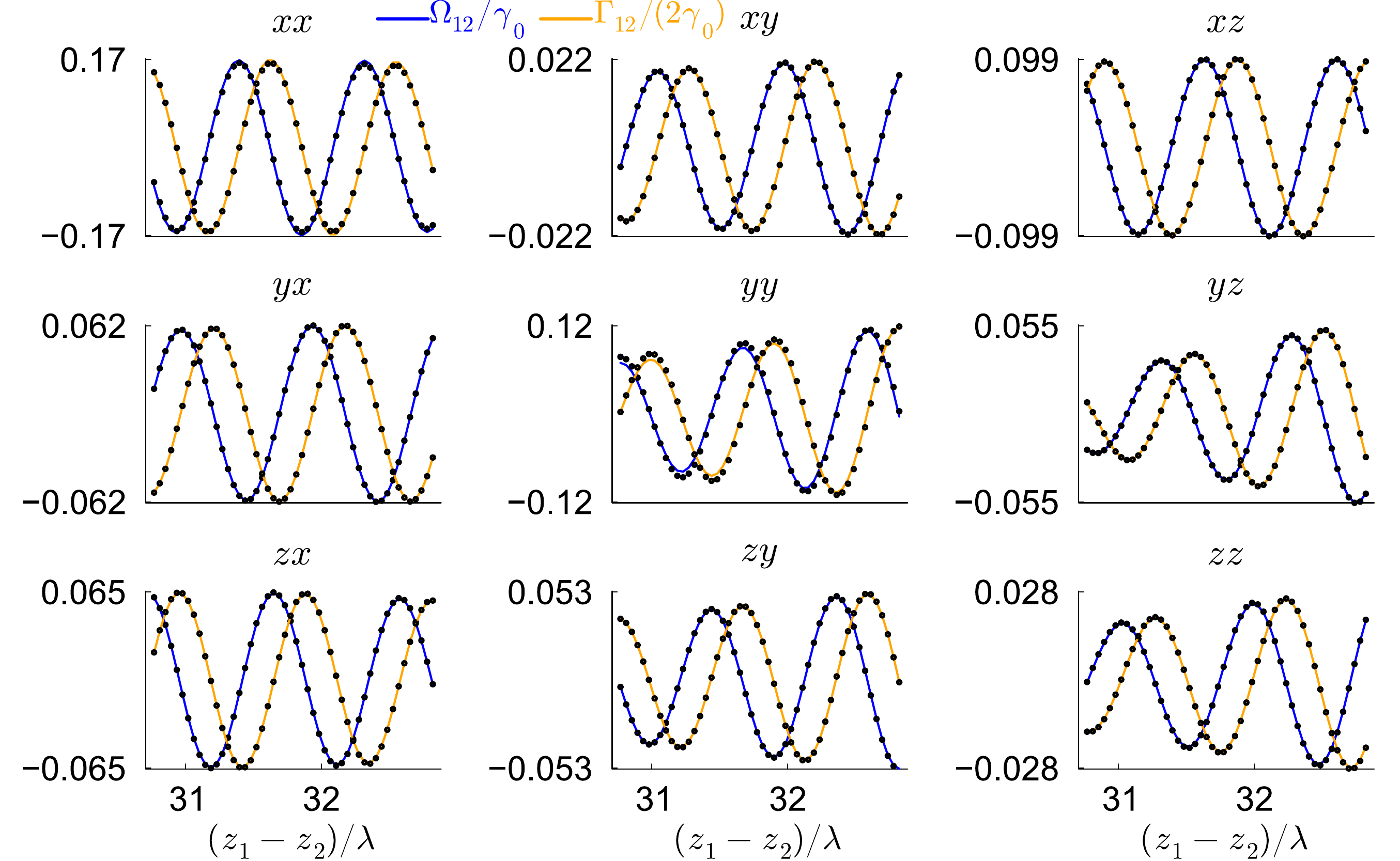}
    \caption{Comparison of the coupling elements Eq.~\eqref{eq:om} and Eq.~\eqref{eq:gam} obtained using the exact expression \eqref{app-eq:global-exp} (solid lines) and the approximation \eqref{eq:long-range} (scatter points) of the Green's function for $(a,d,\lambda) = (200,100,800)$ nm and $\bm{\rho} = (-400,-300)$ nm, $\bm{\rho}' = (-50,-100)$ nm. $xx$ denotes e.g., the Green's function element $G_{xx}.$}
    \label{fig:app:verif-two}
\end{figure}

In order to verify the Green's function elements between two distinct atoms at different positions, we compare with the asymptotic expression \eqref{eq:long-range} for large on-axis separations. In the specific case of two ONFs, we use the electric field mode functions obtained separately in~\cite{le2021spatial}.   We set $(a,d,\lambda) = (200,100,800)$ nm and choose off-axis in-plane positions $\bm{\rho} = (-400,-300)$ nm and $\bm{\rho}' = (-50,-100)$ nm. The full Green's function \eqref{app-eq:global-exp} and the approximation \eqref{eq:long-range} observe excellent agreement for large separations  and for all dipole orientations as can be seen in Fig.~\ref{fig:app:verif-two}. Residual error can be attributed to far-field $O(\lambda|z_1 - z_2|^{-1})$ decay of radiation mode contributions in the full expression.

\section{Calculation of concurrence}
For completeness we here present the concurrence $C$, as defined in \cite{wooters1998entanglement}
\begin{equation}
    C = \text{min}[0,\lambda_{1} - \lambda_{2} - \lambda_{3} - \lambda_{4}].
\end{equation}
Here, $\lambda_{j}$ are the square roots of the real eigenvalues, sorted in descending order, of the operator $\rho\tilde{\rho}$ for $\tilde{\rho} = \sigma_{1y}\sigma_{2y}\rho^{*}\sigma_{1y}\sigma_{2y}.$

\end{document}